\documentclass[prx,reprint,twocolumn,showpacs,superscriptaddress]{revtex4-2}
\usepackage{amsmath,amssymb,bm,mathrsfs,graphicx, braket, times, mathtools}
\usepackage[colorlinks=true,citecolor=blue,linkcolor=blue]{hyperref}
\usepackage{longtable}
\usepackage{multirow}
\usepackage[all,cmtip]{xy}
\usepackage[normalem]{ulem}
\usepackage[usenames,dvipsnames]{color}
\usepackage{tikz-cd}
\usepackage[compat=1.0.0]{tikz-feynman}
\usepackage{bbding}

\usepackage[utf8]{inputenc}
\usepackage[english]{babel}
\usepackage{amsthm}
\usepackage{wasysym}

\renewcommand{\vec}[1]{\mathbf{#1}}
\newcommand{\vechat}[1]{\hat{\mathbf{#1}}}

\begin{document}
\title{Symmetric Jordan-Wigner transformation in higher dimensions}

\author{Hoi Chun Po}
\affiliation{Department of Physics, Hong Kong University of Science and Technology, Clear Water Bay, Hong Kong, China}
\affiliation{Department of Physics, Massachusetts Institute of Technology, Cambridge, Massachusetts, USA}

\begin{abstract}
The Jordan-Wigner transformation is traditionally applied to one dimensional systems, but recent works have generalized the transformation to fermionic lattice systems in higher dimensions while keeping locality manifest. These developments could aid the theoretical or even experimental studies of strongly correlated electronic problems through their bosonic counterparts. In this work, we develop a scheme for higher-dimensional Jordan-Wigner transformation which keeps all relevant symmetries manifest on the bosonic side. Our approach connects the discussion of exact lattice bosonization to the familiar notions of fractionalized partons like spinons and chargeons, and works for spin-$1/2$ fermions---like the physical electrons---on four-coordinated lattices. The construction is applied to fermions defined on the square, kagome and diamond lattices, and we provide explicit expressions for the bosonized versions of well-known models of strongly correlated electrons, like the Hubbard and $t$-$J$ models.
\end{abstract}

\maketitle

\tableofcontents

\section{Introduction}

Particle statistics is a fundamental aspect of quantum mechanics. While bosonic operators localized at different points in space commute, fermionic operators anticommute. The sign difference in particle exchange statistics leads to opposite generic behaviors, in that bosons tend to condense into the lowest available modes and behave collectively, whereas fermions exhibit Pauli exclusion and fill up an extensive number of single-particle states.
The dichotomy between bosons and fermions, however, is in fact not as strict as it might appear. 
In one spatial dimension, the Jordan-Wigner (JW) transformation \cite{JW} provides an exact mapping between one-dimensional fermions and hardcore bosons or, equivalently, spin-$1/2$ magnetic moments. The general insight is that the anti-commutation relation among fermionic operators could be enforced on the bosonic side at the cost of introducing a non-local JW string. Importantly, the non-local portion of the string cancels in the Hamiltonian, which ensures the locality of the Hamiltonian is preserved.
While the generalization of such locality-preserving, statistics-transmuting transformation to higher than one spatial dimension is nontrivial, much progress has already been made from field-theory \cite{PhysRevLett.71.3622, BURGESS199418, Kopietz1997, PASQUIER1998719, PhysRevB.58.16262,   Gaiotto2016,  SEIBERG2016395, PhysRevX.6.031043, JHEP032016131, Kapustin2017, Bhardwaj2017, PhysRevLett.120.016602,  radicevic2019spin,  SENTHIL20191, 10.21468/SciPostPhys.7.1.007,Thorngren2020, huang2021nonabelian}, lattice-system \cite{Wosiek:1981mn, cmp/1103942539, PhysRevLett.63.322, PhysRevB.43.3786, WENG200067,  PhysRevLett.86.1082, Dobrov_2003, PhysRevLett.95.176407, Verstraete_2005, PhysRevB.75.144401, PhysRevLett.98.087204, PhysRevB.76.193101, Chen_2008, Cobanera2011, PhysRevB.86.085415, PhysRevB.98.075119, Minami2016, CHEN2018234, PhysRevB.100.245127, MINAMI2019465, PhysRevResearch.2.033527, PhysRevD.102.114502, Clifford,  PhysRevB.102.245118, Meier_2021}, and quantum-simulation \cite{BRAVYI2002210, Seeley2012, setia2018bravyi, setia2019superfast, PhysRevB.104.035118, derby2021compact} perspectives. The key to achieving such statistics transmutation is the introduction of gauge fields and their associated anyonic excitations, which provide a natural framework for fermions to emerge in a system with microscopic bosonic degrees of freedom \cite{Wen2004}, say in the process of gauging the fermion parity to arrive at a bosonic description \cite{PhysRevB.86.115109, PhysRevB.90.035451, PhysRevB.100.115147}. 

Such gauge fields may also emerge naturally in quantum many-body systems, which are exemplified by the strongly correlated electrons in complex quantum materials. Indeed, the intriguing phenomenologies of high-temperature superconductors \cite{RevModPhys.78.17} and heavy fermion compounds \cite{Si1161} have led to the invention of the slave particle formalism, in which the fermionic electron could be fractionalized into partons with different particle statistics \cite{Barnes_1976, Barnes_1977, PhysRevB.29.3035, PhysRevLett.105.057201,  Castelnovo2012, Savary_2016,CHEN20173, PhysRevX.8.011012, PhysRevResearch.2.023172}. Such parton description shares the same spirit as the mentioned generalizations of the JW transformation, in that the microscopic particle statistics does not have to agree with the the emergent ones. However, unlike the JW discussion which provides a universal mapping from fermionic to bosonic systems, parton frameworks are usually developed with reference to the expected physical phenomena one wishes to capture \cite{PhysRevX.8.011012, PhysRevResearch.2.023172}: for parton theories the emphasis is not on the exact rewriting of the underlying physical degrees of freedom, but rather how they could provide a physically transparent description of the system in certain regions of the parameter space.

In this work, we revisit the higher-dimensional JW problem from the lens of parton construction. In particular, we will always refer to ``JW transformation'' as an exact procedure for bosonizing a lattice fermionic system which is applicable to all possible Hamiltonians defined on the lattice \footnote{In one dimension, the JW transformation proceeds in both ways, namely, one could also femrionize a bosonic system. In higher dimensions, certain bosonic problems may also admit a femrionic description, but this is not generic. Our focus here is the bosonization of generic femrionic systems.}. 
On the one hand, the exact nature of JW transformation is desirable for numerical or quantum simulations, in which one attempts to obtain unbiased answers on questions like the physical nature and symmetry-breaking pattern in a quantum many-body system. In this regard, the JW transformation serves as a universal shortcut for the development of fermionic simulation algorithm \cite{PhysRevA.81.010303, PhysRevB.80.165129, PhysRevB.81.165104, PhysRevB.101.155105}, since one could simply apply bosonic algorithms to the transformed Hamiltonian \cite{RevModPhys.77.259, itensor}. 
The existing discussions on higher-dimensional JW transformation, however, tend to focus on the consistency and global properties of the theory, and spinless fermions are typically considered \cite{CHEN2018234, PhysRevB.100.245127,  PhysRevResearch.2.033527, PhysRevD.102.114502, Clifford}. Electrons, however, are spin-$1/2$ fermions, and to explore the plethora of symmetry-breaking phases that could arise in a quantum many-body system, it is important to understand how symmetries are represented in the bosonized description. Addressing this question is one of the major motivations for the present work.

On the other hand, important conceptual insights on strongly correlated quantum phases have been gleaned from the parton framework. An enticing feature of the parton description is that all the relevant quantum numbers of the excitations are made explicit. Such fractionalized excitations could have decoupled dynamics and this leads to phenomena like spin-charge separation.
Indeed, in spin liquids although the electronic charges are non-dynamical, spinons, which carry the spin but not the charge of the electrons, could emerge as the low-energy quasiparticles \cite{Wen2004}; alternatively, one could also imagine charge liquids in which the low-energy quasiparticles are chargeons/holons and carry the charge but not the spin of the electrons \cite{Ribhu2008}.

Here, we also consider the fractionalization of an electron into spinons and chargeons/holons. Yet, in contrast to the conventional schemes both species of partons are fermionic. Our scheme is motivated by the group of transformations associated with electrons occupying a single orbital, which naturally points to a way for performing JW transformation for single-band electronic models on four-coordinated lattices.
With the explicit quantum numbers assignment for the spinons and chargeons/holons, all the internal unitary symmetries are automatically manifest after JW transformation. We further show that the internal anti-unitary symmetries and crystalline symmetries are represented naturally. Curiously, in our construction the anti-unitary particle-hole transformation is implemented as a time-reversal on Dirac spinors, which is reminiscent of the proposal in the context of composite Fermi liquid in half-filled lowest Landau level \cite{PhysRevX.5.031027}.

As noted, the higher-dimensional JW problem has already been considered from many different angles in the literature. In the following we briefly comment on the relation between the present approach and some of the most relevant existing works. 
Starting from the electron fractionalization perspective, our scheme bears resemblance to that proposed in Ref.\ \onlinecite{PhysRevLett.105.057201}, in that spinons and chargeons/holons have the same particle statistics. Here, however, instead of considering bosonic spinons and chargeons/holons together with a fermionic, neutral Majorana parton, we use only femrionic partons and the fermionic statistics of the physical electrons is emergent.
From the JW perspective, our approach is inspired by the recent discussions in Ref.\  \onlinecite{CHEN2018234, PhysRevB.100.245127,  PhysRevResearch.2.033527, PhysRevD.102.114502, Clifford}, especially on issues pertaining to the global aspects of the construction. On the bosonic side, however, we associate operators to the original physical sites \cite{PhysRevD.102.114502, Clifford} instead of plaquettes \cite{CHEN2018234, PhysRevB.100.245127,  PhysRevResearch.2.033527}. 
Also, as mentioned we would consider spin-$1/2$ fermions (i.e., electrons) instead of spinless fermions, and we focus on the interplay between symmetries and JW transformation.
Furthermore, we also provide a prescription for identifying the bosonic counterparts of fermion-odd operators, like a single creation or annihilation operator, at the cost of attaching a (non-abelian) JW string.

Let us conclude the introduction by outlining our general approach to the higher-dimensional JW problem. As a first step, in Sec.\ \ref{sec:dot} we consider a single site with two complex fermion modes, corresponding to a spin-$1/2$ electron in a single orbital. We will discuss how the on-site terms in the Hamiltonian can be expressed in terms of fermionic spinon and chargeon/holon, and why this perspective helps keep the internal symmetries manifest. Next, in Sec.\ \ref{sec:square} we  specialize to the case of a square lattice and discuss how to consistently patch up the individual sites and enforce fermionic statistics throughout the lattice. We  also provide a concrete recipe for writing down a general JW transformed Hamiltonian, and in the process the non-abelian JW string emerges. The more subtle global properties arising from the periodic boundary conditions are discussed in Sec.\ \ref{sec:global}, in which we also address how the fermion-odd operators could be bosonized. These preparations allow us to address the symmetry properties of the JW transformed Hamiltonian in Sec.\ \ref{sec:sym}, and we would see that the internal unitary and anti-unitary symmetries, together with spatial symmetries like translation, $C_4$ rotation and mirror, are all kept manifest. In fact, they all take the simple form of the exponentiation of bilinears in the fermionc partons. To better bridge our results with numerical and quantum simulation, we also discuss in Sec.\ \ref{sec:qubit} how the JW transformed Hamiltonian can be rewritten in terms of qubits. Sec.\ \ref{sec:models} is then devoted to providing explicit expressions for physically interesting models defined on the square lattice, including the spinful and spinless Hubbard model and the $t$-$J$ model. The procedure can also be carried out for other four-coordinated lattices, as we demonstrate in Sec.\ \ref{sec:others} for the two-dimensional kagome and three-dimensional diamond lattices. We then conclude in Sec.\ \ref{sec:discussion} with a discussion on some future directions.

\section{An anecdote about a spinful dot \label{sec:dot}}
Let us begin by considering a single site with $n$ complex fermion modes defined by annihilation operators $\hat f^{i}$, $i = 1,\dots, n$. Physically, the $n$ modes can account for both orbital and spin degrees of freedom, and $n$ is a general integer here. Define the Majorana fermion operators
\begin{equation}\begin{split}\label{eq:}
\hat \gamma^{2j-1} \equiv \hat f^{j}  + \hat f^{j \dagger}   ;~~~
\hat \gamma^{2j} \equiv i ( \hat f^{j}  - \hat f^{j \dagger}),
\end{split}\end{equation}
which satisfy the anti-commutation relation $\{ \hat \gamma^\alpha,\hat \gamma^\beta\} = 2 \delta^{\alpha \beta}$ for $\alpha, \beta = 1,\dots, 2n$. We will be focusing on the Hermitian fermion bilinears 
\begin{equation}\begin{split}\label{eq:theta_0}
\hat \theta_0^{\alpha \beta} \equiv \frac{i}{2} \hat \gamma^{\alpha} \hat \gamma^\beta~~~{\rm for}~~~\alpha \neq \beta,
\end{split}\end{equation}
where the subscript ``$0$'' emphasizes that this is the original representation defined using the physical degrees of freedom. Later on we will consider other representations of the same algebra, for which we will drop the subscript. 
The commutators for the fermion bilienars correspond to that of the Lie algebra $\mathfrak{so}(2n)$:
\begin{equation}\begin{split}\label{eq:}
[\hat \theta_0^{\alpha \beta}, \hat \theta_0^{\gamma \delta}] = i \left(
\delta^{\alpha\delta} \hat \theta_0^{\beta \gamma} 
- \delta^{\alpha\gamma} \hat \theta_0^{\beta \delta} 
+ \delta^{\beta\gamma} \hat \theta_0^{\alpha \delta} 
- \delta^{\beta \delta} \hat \theta_0^{\alpha\gamma}
\right).
\end{split}\end{equation}
Upon exponentiation, one finds, for any fixed $\alpha, \beta$,
\begin{equation}\begin{split}\label{eq:Spin_sign}
e^{-i \xi \hat \theta_0^{\alpha \beta}} = \cos \frac{\xi}{2} + \sin\frac{\xi}{2} \hat \gamma^\alpha \hat \gamma^\beta,
\end{split}\end{equation}
and we make two observations here: first, by setting $\xi = 2\pi$ we get $e^{-i 2\pi \hat \theta_0^{\alpha \beta}} = -1$, which reveals that the Lie group we obtained is in fact ${\rm Spin}(2n)$ instead of ${\rm SO}(2n)$ \cite{KITAEV20062}; second, when $\xi = \pi$ we arrive back at the fermion bilinear $e^{-i \pi\hat \theta_0^{\alpha \beta}}   = \hat \gamma^\alpha \hat \gamma^\beta$. 

These observations highlight that the bilinear $\hat \gamma^\alpha \hat \gamma^\beta$ could be viewed as a group element of ${\rm Spin}(2n)$. This also demonstrates the physical significance of the $-1$ phase distinguishing ${\rm Spin}(2n)$ from ${\rm SO}(2n)$.  For instance, consider the equality
\begin{equation}\begin{split}\label{eq:}
(\hat \gamma^1 \hat \gamma^2 )(\hat \gamma^1 \hat \gamma^3) = 
- (\hat \gamma^1 \hat \gamma^3)(\hat \gamma^1 \hat \gamma^2 ),
\end{split}\end{equation}
for which the $-1$ sign comes from the anti-commutation between the Majorana operators.
From our discussion above, the equality can also be rewritten as
\begin{equation}\begin{split}\label{eq:}
e^{- i \pi \hat \theta_0^{12}} e^{- i \pi \hat \theta_0^{13}} = 
- e^{- i \pi \hat \theta_0^{13}} e^{- i \pi \hat \theta_0^{12}},
\end{split}\end{equation}
which can be viewed as a particular relation between elements in the group ${\rm Spin}(2n)$.  In this sense, the negative sign differentiating between  ${\rm Spin}(2n)$ and  ${\rm SO}(2n)$ is integral to the fermionic statistics.

More generally, for any set of distinct $ \{ \alpha_1,\alpha_2,\dots ,\alpha_{2l}\}$, we can write
\begin{equation}\begin{split}\label{eq:Exp_h}
\hat \gamma^{\alpha_1}\hat \gamma^{\alpha_2} \cdots \hat \gamma^{\alpha_{2l}} = 
e^{-i \pi\hat \theta_0^{\alpha_1 \alpha_2}} \cdots e^{-i \pi\hat \theta_0^{\alpha_{2l-1} \alpha_{2l}}},
\end{split}\end{equation}
which is also an element of ${\rm Spin}(2n)$. Note that such operators involving an even number of fermionic operators preserve the global fermion parity, and are possible terms in a Hamiltonian. We will refer to such operators as ``fermion-even.'' 
Eq.\ \eqref{eq:Exp_h} shows that any fermion-even operator can be viewed as the sum (over $\mathbb C$) of a collection of group elements of ${\rm Spin}(2n)$. 
In other words, one could view the generators of the fermion-even operator algebra as elements of ${\rm Spin}(2n)$, and in this perspective the algebraic relations follow from group multiplications in ${\rm Spin}(2n)$.

\begin{figure}[h]
\begin{center}
{\includegraphics[width=0.25 \textwidth]{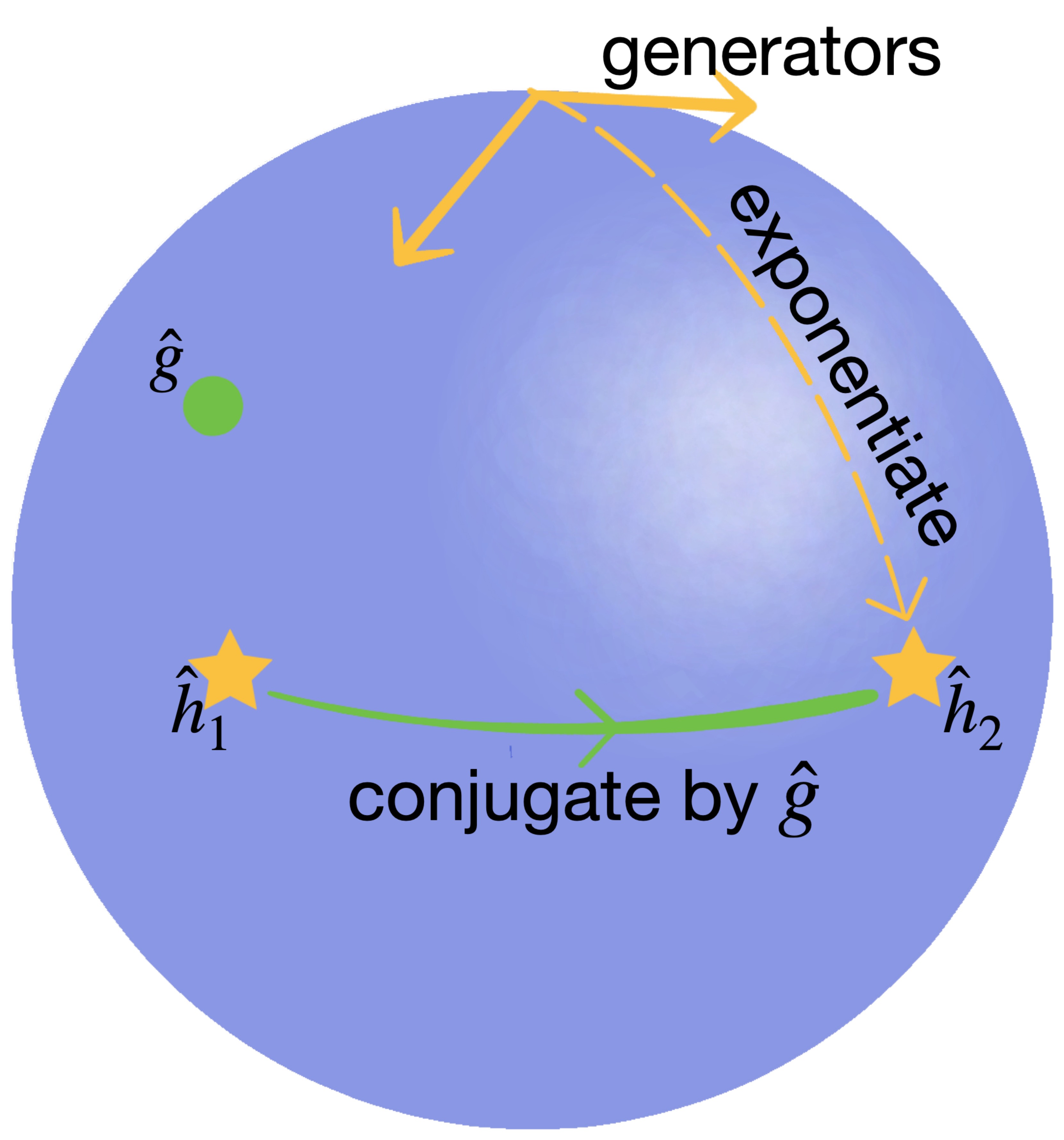}} 
\caption{Symmetries, Hamiltonian, and group multiplication. A symmetry operation $\hat g$ generally relates different terms in the Hamiltonian, say $\hat g \hat h_1 \hat g^\dagger = \hat h_2$. 
Up to a coefficient, we could view $\hat g$, $\hat h_1$ and $\hat h_2$ as elements of a Lie group which can be obtained by exponentiating the generators. For instance, let $\hat \gamma^{i}$ for $i=1,2,3,4$ be Majorana fermions, then we may take $\hat g = e^{ \frac{\pi}{4} \hat \gamma^1 \hat \gamma^3}$, $\hat h_1 = i \hat \gamma^{1} \hat \gamma^{2}$, and $\hat h_2 = i \hat \gamma^{2} \hat \gamma^{3}$ as an example with the relation $e^{ \frac{\pi}{4} \hat \gamma^1 \hat \gamma^3} (\hat \gamma^1 \hat \gamma^2) e^{- \frac{\pi}{4} \hat \gamma^1 \hat \gamma^3}  = \hat \gamma^2 \hat \gamma^3$ being interpreted as a statement on group multiplication in ${\rm Spin}(4)$.
\label{fig:lie_group}
 }
\end{center}
\end{figure}

So far, we have considered a canonical representation of ${\rm Spin}(2n)$ arising from the $2n$ Majorana operators $\{ \hat \gamma^\alpha: \alpha = 1,\dots, 2n \}$. As argued, any representation of ${\rm Spin}(2n)$ will also provide a representation of the fermion-even algebra, and hence a representation of any possible Hamiltonian. 
An advantage of this point of view is that symmetries will remain manifest in the alternative representation (Fig.\ \ref{fig:lie_group}). To see why, consider an internal unitary symmetry $g$ of the Hamiltonian represented by $\hat U_g \equiv e^{- i \sum_{\alpha <\beta} L_g^{\alpha \beta} \hat \theta_0^{\alpha \beta}}$, where $L^{\alpha\beta}_g$ are  $6$ real coefficients. If $\hat h$ is a term in the Hamiltonian, then $\hat U_g \hat h \hat U_g^\dagger$ is also a term in the Hamiltonian \footnote{We implicitly assumed we write the Hamiltonian as an expansion over a unitary operator basis. For instance, while complex fermion hopping terms like $\hat f^\dagger \hat f' + {\rm h.c.}$ is not unitary in general, we may expand the term using Majorana bilinears which are unitary.}. 
By viewing $\hat h$ as a group element of ${\rm Spin}(2n)$ (up to the coefficient), the symmetry relations are automatically guaranteed by the group multiplication even when we switch to a different representation of ${\rm Spin}(2n)$ by replacing the generators $\hat \theta_0^{\alpha \beta} \mapsto \hat \theta^{\alpha \beta}$. In addition, the representation of the internal unitary symmetries in the problem follows immediately from our specification of $\hat \theta^{\alpha \beta}$.

Up to this point, we have kept $n$, the number of complex fermion modes in the problem, generic. We now specialize to the case of $n=2$, which corresponds physically to a single orbital possibly occupied by electron(s) with spin(s) up and/ or down. More explicitly, we consider the fermion operators
\begin{equation}\begin{split}\label{eq:}
\hat f^{\uparrow} = \frac{1}{2} \left( \hat \gamma^1 - i \hat \gamma^2 \right); ~~~
\hat f^{\downarrow} = \frac{1}{2} \left( \hat \gamma^3 - i \hat \gamma^4 \right).
\end{split}\end{equation}
This $n=2$ problem is rather special, since in this case the relevant group is ${\rm Spin}(4)$, which enjoys an exceptional isomorphism
\begin{equation}\begin{split}\label{eq:}
{\rm Spin}(4) \simeq  {\rm SU_s}(2)\times {\rm SU_c}(2),
\end{split}\end{equation}
where the subscripts ``s '' and ``c'' respectively denote spin and charge. This interpretation could be rationalized as follows: the four-dimensional site Hilbert space separates into two sectors labeled by the fermion parity. The even-parity sector consists of the vacuum state $|0 \rangle$ and the doubly occupied state $\hat f^{\uparrow \dagger} \hat f^{\downarrow \dagger} | 0 \rangle$, both of which are spin-singlets; in contrast, the odd-parity sector is spanned by the two singly occupied states $\hat f^{\uparrow \dagger} | 0 \rangle$ and $\hat f^{\downarrow \dagger} |0 \rangle$, which transforms as a spin-1/2. Unitary transformations generated by the fermion bilinears do not mix the two sectors, and one could analyze the rotations in the two sectors independently. The ${\rm SU}(2)$ transformation within the odd-parity (i.e., singly occupied) sector is simply the familiar spin rotation. In contrast, the ${\rm SU}(2)$ transformations within the even-parity sector rotate between states with different particle number, and are thus designated as a ``charge'' rotation. Correspondingly, the singly occupied states are singlet under ${\rm SU_c}(2)$, an observation which has played a prominent role in the gauge structure of the parton constructions for certain spin liquid states \cite{PhysRevB.38.745, PhysRevB.38.2926, RevModPhys.78.17}.

An explicit form of the generators could be chosen as
\begin{equation}\begin{split}\label{eq:fermion_bare}
\hat \sigma^i_0 =& \frac{1}{2} 
\left(
\begin{array}{cc}
\hat f^{\uparrow \dagger} & \hat f^{\downarrow \dagger}
\end{array}
\right)
\tau^i
\left(
\begin{array}{c}
\hat f^{\uparrow } \\ \hat f^{\downarrow }
\end{array}
\right);\\
\hat \chi^i_0 =& \frac{1}{2} 
\left(
\begin{array}{cc}
\hat f^{\uparrow \dagger} & \hat f^{\downarrow }
\end{array}
\right)
\tau^i
\left(
\begin{array}{c}
\hat f^{\uparrow } \\ \hat f^{\downarrow \dagger}
\end{array}
\right),
\end{split}\end{equation}
where $\tau^i$ for $i=1,2,3$ denote the Pauli matrices, and one could check that $[\hat \sigma^i_0 , \hat \chi^j_0]=0$ as they should. $\hat \sigma_0^i$ and $\hat \chi_0^i$ respectively generate the ${\rm SU_s}(2)$ and ${\rm SU_c}(2)$ rotations.
Given the stated isomorphism, $\hat \sigma_0^i$ and $\hat \chi_0^i$ also provide another basis for the generators of ${\rm Spin}(4)$. More explicitly, the transformation is given by
\begin{equation}\begin{split}\label{eq:Spin_trans}
\begin{array}{ll}
\hat \theta_0^{23} = \hat \chi^1_0 + \hat \sigma^1_0; &
\hat \theta_0^{14} = \hat \chi^1_0 - \hat \sigma^1_0;\\
\hat \theta_0^{13} = -(\hat \chi^2_0 + \hat \sigma^2_0); &
\hat \theta_0^{24} = \hat \chi^2_0 - \hat \sigma^2_0;\\
\hat \theta_0^{12} = -(\hat \chi^3_0 + \hat \sigma^3_0); &
\hat \theta_0^{34} =  -\hat \chi^3_0 + \hat \sigma^3_0.
\end{array}
\end{split}\end{equation}

With these preparations we are now ready to switch to an alternative representation of ${\rm Spin}(4)$, which will serve as an anchor for our subsequent discussion on JW transformation. 
Let us construct the familiar representations of the two ${\rm SU}(2)$ factors independently using two pairs of complex fermion modes, i.e., 
we introduce four complex fermion modes with annihilation operators $\hat u$, $\hat d$, $\hat c$ and $\hat h$, which we would refer to as ``partons.''  Define 
\begin{equation}\begin{split}\label{eq:sigma_chi}
\hat \sigma^i \equiv & \frac{1}{2} 
\left(
\begin{array}{cc}
\hat u^{\dagger} & \hat d^{ \dagger}
\end{array}
\right)
\tau^i
\left(
\begin{array}{c}
\hat u \\ \hat d
\end{array}
\right);\\
\hat \chi^i \equiv & \frac{1}{2} 
\left(
\begin{array}{cc}
\hat c^{ \dagger} & \hat h^{\dagger}
\end{array}
\right)
\tau^i
\left(
\begin{array}{c}
\hat c \\ \hat h
\end{array}
\right).
\end{split}\end{equation}
Note the factor of $1/2$ in the definitions above, which is different from the (perhaps more popular) convention in which $\hat \sigma^i$ stands for the Pauli operator. Also, notice the absence of the subscript $0$. Here, $\hat \sigma^i$ should be interpreted as the generators for spin rotations, and similarly $\hat \chi^i$ are generators for the charge rotations.

From Eq.\ \eqref{eq:sigma_chi}, it is suggestive that $\hat u$ and $\hat d$ can be interpreted as fermionic spinons carrying the original spin-1/2 of the electron but not its electric charge. In contrast, $\hat c$ and $\hat h$ can respectively be viewed as fermionic chargon and holon, which are spinless and carry electric charges of $\pm 1$.
Note that the current fractionalization scheme is different from a conventional one, in that both the spinon and the chargon/holon are fermions. As we will see later, once we go beyond the single-site problem and consider a lattice, the proper statistics is enforced by nontrivial constraints which involve both species of partons.

The partons define a $2^4 = 16$-dimensional Hilbert space, into which we embed the original $4$-dimensional physical Hilbert space of the site. Our next task is to see how we could obtain a faithful representation of ${\rm Spin}(4)$, and hence a general fermionic Hamiltonian defined on our single dot, using this enlarged Hilbert space. First, we simply replace $\hat \sigma_0\mapsto \hat \sigma$  and $\hat \chi_0 \mapsto \hat \chi$ in Eq.\ \eqref{eq:Spin_trans} to obtain $\hat \theta^{\alpha \beta}$ defined using the partons. Now, notice that
\begin{equation}\begin{split}\label{eq:}
e^{-i 2 \pi \hat \theta^{12}} = & e^{i 2 \pi (\hat \chi^3 + \hat \sigma^3)} = \hat \Gamma;\\
\hat \Gamma \equiv & e^{i \pi 
(\hat u^\dagger \hat u + \hat d^\dagger \hat d + \hat c^\dagger \hat c + \hat h^\dagger \hat h)},
\end{split}\end{equation}
where $\hat \Gamma$ is the fermion parity operator in the parton Hilbert space and satisfies $\hat \Gamma^2 = 1$.
More generally one can check that $e^{-i 2 \pi \hat \theta^{\alpha \beta}} = \hat \Gamma$ for any $\alpha \neq \beta$.  
From the discussion surrounding Eq.\ \eqref{eq:Spin_sign}, we know that fermionic statistics follows from $e^{-i 2 \pi \hat \theta^{\alpha \beta}} \mapsto -1$, and so we restrict ourselves to the $8$-dimensional subspace of the parton Hilbert space in which $\hat \Gamma\mapsto -1$. 
To avoid possible confusion with the physical fermion parity, we will refer to this subspace as ``$\Gamma_-$,'' and correspondingly the subspace with $\hat \Gamma \mapsto +1$ will be referred to as ``$\Gamma_+$.''
The necessity to restrict our description to the $\Gamma_-$ subspace implies we are freezing the parton fermion parity operator in our description. In this sense, although we have introduced fermionic partons, our construction is eventually bosonic in nature.
We remark that the $\Gamma_-$ subspace is still twice as large as the physical one, which is only four-dimensional, i.e., our enlarged description contains an extra qubit per site. 
We will see in the next section that this extra qubit is locally consumed once we carry the procedure over from a single dot to a lattice.

Let us finish up our single-dot discussion by considering the possible Hamiltonian. Physically, there are $6$ possible, linearly independent Hermitian fermion bilinears $i \hat \gamma^{\alpha} \hat\gamma^{\beta}$ for $1\leq \alpha <\beta\leq 4$, as well as one quartic term $ \hat \gamma^1\hat \gamma^2\hat \gamma^3\hat \gamma^4$. As in Eq.\ \eqref{eq:Exp_h}, in our prescription we reconcile terms in the Hamiltonian by a ``$\pi$-rotation'' using the generators we have used for the representation of ${\rm Spin}(4)$. Furthermore, it will be convenient to ensure the term is Hermitian. Since in our current scheme we have 
$e^{-i 2 \pi \hat \theta^{\alpha \beta}} = \hat \Gamma$, we define $\sqrt{\hat \Gamma}$ as the diagonal matrix which takes value $+1$ in the $\Gamma_+$ subspace, and $i$ in the $\Gamma_-$one. With this convention, we have $\left(\sqrt{\hat \Gamma}\right)^2 = \hat \Gamma$ and $\left(\sqrt{\hat \Gamma}\right)^\dagger = \hat \Gamma \sqrt{\hat \Gamma}$.

Recall we have to restrict our attention to the $\Gamma_-$ subspace to obtain a faithful representation of ${\rm Spin}(4)$. In this subspace, we have $\sqrt{\hat \Gamma} \mapsto i$, and it will be natural for us to utilize $\sqrt{\hat \Gamma}$ in lieu of the imaginary unit for any term in the Hamiltonian. More explicitly, we replace the term $i \hat \gamma^\alpha \hat \gamma^\beta $ in the Hamiltonian by 
\begin{equation}\begin{split}\label{eq:h_map}
i \hat \gamma^\alpha \hat \gamma^\beta  \mapsto \hat \Theta^{\alpha \beta} \equiv
\sqrt{\hat \Gamma} e^{- i \pi \hat \theta^{\alpha \beta}},
\end{split}\end{equation}
where $\alpha \neq \beta$.
One can check that the right-hand side is also Hermitian. We also remark that, previously, we have focused on the case when $\alpha < \beta$. When $\alpha > \beta$ we simply define $\hat \theta^{\alpha \beta} = - \hat \theta^{ \beta \alpha} $, and correspondingly we have
\begin{equation}\begin{split}\label{eq:}
 \hat \Theta^{\alpha \beta} = \hat\Gamma  \hat \Theta^{\beta \alpha }.
\end{split}\end{equation}

One can verify that $\hat \Theta^{\alpha \beta}$ acts in the same way as the generating fermion bilinear (up to the coefficient) in the $\Gamma_-$ subspace:
\begin{equation}\begin{split}\label{eq:Gamma_odd}
\left. \hat \Theta^{\alpha \beta} \right|_{\hat \Gamma\mapsto -1}
= \left. 2 \hat \theta^{\alpha \beta} \right|_{\hat \Gamma\mapsto -1}.
\end{split}\end{equation}
For brevity, in the following we will use the decoration ``$\cdot$'' to indicate that a relation is valid only when restricted to the $\Gamma_-$ subspace, viz.\ $\hat \Theta^{\alpha \beta} \dot = 2 \hat \theta^{\alpha \beta}$.
We remark that this simplification is nontrivial, given the exponential of a fermion bilinear is generally a sum of fermion-even terms. 
To understand its origin, we introduce the notion of ``roots of parity'' in Appendix \ref{app:RoP}. While the properties of roots of parity turn out to be quite useful in constructing our bosonic description, it is a technical trick in nature and so we relegate the detailed discussion to an Appendix.

\begin{figure}[h]
\begin{center}
{\includegraphics[width=0.45 \textwidth]{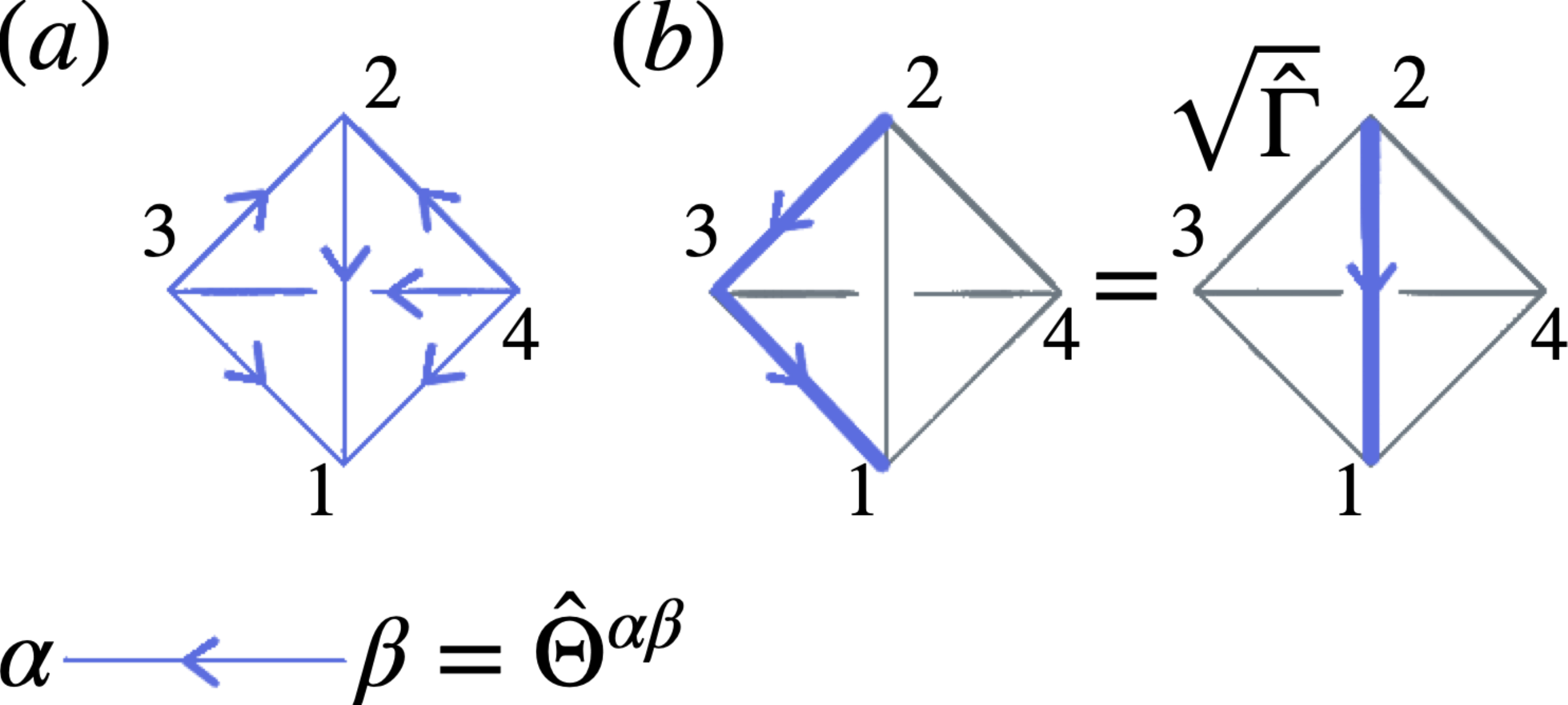}} 
\caption{Graphical representation of the bosonized operators. (a) The four Majorana operators defined on a single site are represented by labeled vertices, and an arrow connecting the $\beta$ vertex to the $\alpha$ one corresponds to the operator $\hat \Theta^{\alpha \beta}$, which is the bosonized version of the original fermion bilinear term $i \hat \gamma^\alpha \hat \gamma^\beta$. The indicated arrows correspond to $\hat \Theta^{\alpha \beta}$ with $\alpha < \beta$; reversal of the arrows gives $\hat \Theta^{\beta \alpha} = \hat \Gamma \hat \Theta^{\alpha \beta}$. (b) The arrows can be composed, which translates into the operator relation $\hat \Theta^{13} \hat \Theta^{32} = \sqrt{\hat \Gamma} \hat \Theta^{12}$. This can be compared with the original relation $(i \hat \gamma^1 \hat \gamma^3)(i \hat \gamma^3 \hat \gamma^2) = i (i \hat \gamma^{1}\hat \gamma^2)$.
\label{fig:dot}
 }
\end{center}
\end{figure}

As promised, one can verify that, when restricted to the $\Gamma_-$ subspace, the relation between the operators $\hat \Theta^{\alpha \beta}$ follows that between $i \hat \gamma^\alpha \hat\gamma^\beta$.  
For instance, one can verify that when $\alpha,\beta,\gamma$ are all distinct, we have (no summation on repeated indices)
\begin{equation}\begin{split}\label{eq:Theta_comm}
\hat \Theta^{\alpha \beta} \hat \Theta^{\beta \gamma}  = 
\sqrt{\hat \Gamma} \hat \Theta^{\alpha \gamma} 
= \hat \Gamma \hat \Theta^{\beta \gamma}  \hat \Theta^{\alpha \beta}.
\end{split}\end{equation}
Similarly, $ [\hat \Theta^{\alpha \beta} , \hat \Theta^{ \gamma \delta}  ] = 0$ when $\alpha,\beta,\gamma,\delta$ are all distinct.
It is helpful to introduce a graphical representation as a mnemonic device for the algebraic relations between $\hat \Theta^{\alpha \beta}$ \cite{CHEN2018234, PhysRevB.100.245127,  PhysRevResearch.2.033527, PhysRevD.102.114502, Clifford}.
As shown in Fig.\ \ref{fig:dot}, we represent each operator $\hat \Theta^{\alpha \beta}$ as an arrow pointing from $\beta$ to $\alpha$. When two arrows share a common vertex, we say they ``collide,''  and the associated operators anti-commute in the $\Gamma_-$ subspace. Non-colliding arrows commute.

We can now provide explicit expressions for the possible terms in the Hamiltonian using the parton operators. First, consider the electron number operator $\hat n \equiv \sum_{\sigma = \uparrow, \downarrow} \hat n^\sigma$ with $\hat n^\sigma \equiv \hat f^{\sigma \dagger}\hat f^\sigma$. 
We can identify its correspondence in the enlarged Hilbert space by first re-expressing it using the Majorana operators, and then mapping each of the terms according to Eq.\ \eqref{eq:h_map}:
\begin{equation}\begin{split}\label{eq:n_map}
\hat n 
= & 1 - \frac{1}{2} \left(i \hat \gamma^1 \hat \gamma^2 + i \hat \gamma^3 \hat \gamma^4 \right) 
\mapsto  1 - \frac{1}{2}\left(\hat \Theta^{12} + \hat \Theta^{34}\right).
\end{split}\end{equation}
Furthermore, by Eq.\ \eqref{eq:Gamma_odd}, the mapping can be written as
\begin{equation}\begin{split}\label{eq:}
\hat n 
~\dot \mapsto ~
1 + 2 \hat \chi^3 =  \hat c^\dagger c + \hat h \hat h^\dagger.
\end{split}\end{equation}
Similarly the on-site interaction term can be expressed as 
\begin{equation}\begin{split}\label{eq:interaction_1}
(\hat n -1)^2
~\dot \mapsto ~
\left( \hat c^\dagger c - \hat h^\dagger  \hat h\right)^2.
\end{split}\end{equation}
We remark, however, that this is not the only way we could represent the interaction term. For instance, by recasting
\begin{equation}\begin{split}\label{eq:U_map}
(\hat n -1)^2
= \frac{(i \hat \gamma^1 \hat \gamma^2) (i \hat \gamma^3 \hat \gamma^4) + 1}{2} \mapsto  \frac{\hat \Theta^{12} \hat \Theta^{34}+1}{2},
\end{split}\end{equation}
one finds
\begin{equation}\begin{split}\label{eq:interaction_2}
(\hat n -1)^2
~\dot \mapsto ~
\frac{1}{2} \left[ \left( \hat c^\dagger c - \hat h^\dagger  \hat h\right)^2 - 
\left( \hat u^\dagger u - \hat d^\dagger  \hat d \right)^2 + 1 \right].
\end{split}\end{equation}
Eqs.\ \eqref{eq:interaction_1} and \eqref{eq:interaction_2} could be reconciled by noticing
\begin{equation}\begin{split}\label{eq:}
\left( \hat c^\dagger c - \hat h^\dagger  \hat h\right)^2 + \left( \hat u^\dagger u - \hat d^\dagger  \hat d \right)^2 \dot =  1,
\end{split}\end{equation}
and one could also view it as alternative way to specify the $\Gamma_-$ subspace.

Following the same procedure, we can also write down the operators corresponding to the other fermion bilinears. The usual spin operators $\hat S^{i}$ follow naturally: $\hat S^i \dot \mapsto \hat \sigma^i$ \footnote{This may appear to be trivially guaranteed; but we emphasize that $\hat S^i$ represents a term in the Hamiltonian, whereas $\hat \sigma^i$ is interpreted as a symmetry generator. The fact that the two agree in the $\Gamma$-odd subspace is a special property of the current problem with $4$ complex partons (Appendix \ref{app:RoP}).}. We can also consider the on-site pairing 
\begin{equation}\begin{split}\label{eq:}
\frac{1}{2} \left(
\hat f^{\uparrow \dagger} \hat f^{\downarrow \dagger}
+ \hat f^{\downarrow}\hat f^{\uparrow}\right) 
\dot \mapsto \hat \chi^1 = \hat c^\dagger h + h^\dagger c,
\end{split}\end{equation}
which can be physically interpreted by viewing the spin-singlet pairing as the conversion of a spinless holon into a chargon and vice versa.

\section{Connecting the dots \label{sec:square}}
So far, we have detailed an elaborated scheme to represent the four-dimensional Hilbert space defined by two complex fermions $\hat f^\uparrow, \hat f^\downarrow$ using an enlarged Hilbert space of dimension $8$ (i.e., the $\Gamma_-$ subspace of the parton Hilbert space). If our full problem was defined by only two complex fermions, the preparation was clearly unnecessary: one could readily take two qubits, which define a four-dimensional Hilbert space, and write down operators corresponding to all fermion-even operators in the original problem. In other words, our preparation pays off only when we consider fermions living on a lattice, which requires us to also bosonize terms which simultaneously change the physical fermion parity on multiple sites, e.g., a hoping term $\hat f^{\uparrow \dagger}_{\vec r} \hat f^\uparrow_{\vec r'}$. In this section, we will consider a two-dimensional square lattice and discuss how we could identify operators in the enlarged Hilbert space corresponding to such hopping terms. We remark that our general approach in this section could be traced to the corresponding discussions in Refs.\ \onlinecite{CHEN2018234, PhysRevB.100.245127,  PhysRevResearch.2.033527, PhysRevD.102.114502, Clifford}.

\begin{figure}[h]
\begin{center}
{\includegraphics[width=0.45 \textwidth]{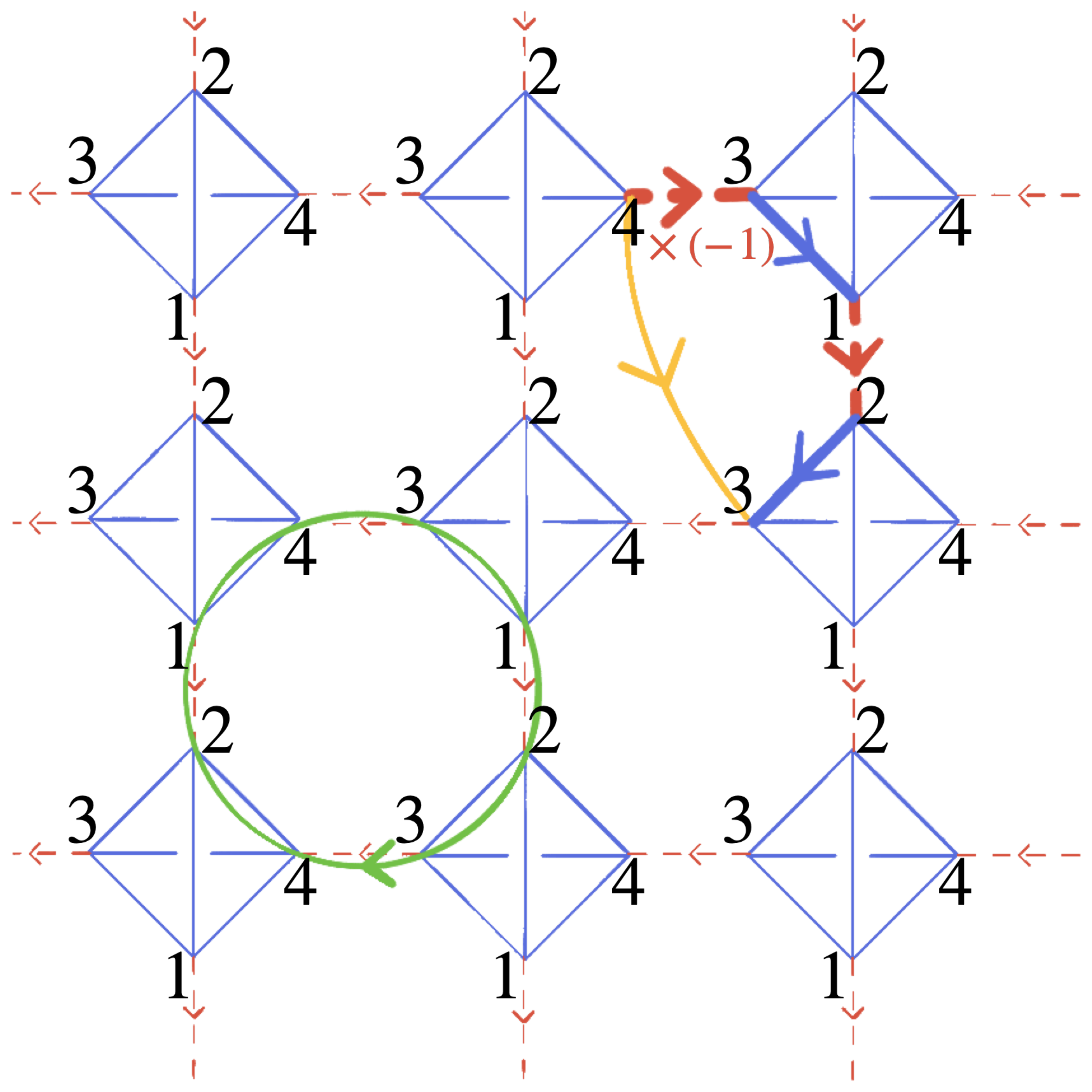}} 
\caption{Fermionic algebra on the square lattice. 
We again impose the rule that arrows anti-commute in the physical Hilbert space if they sharing a common point, and commute otherwise. 
For instance, a dashed arrow pointing from $\beta$ at site $\vec r+\vechat x$ to $\alpha$ at site $\vec r$ could correspond to the fermionic process $i \hat \gamma^{\alpha}_{\vec r} \hat \gamma^{\beta}_{\vec r+\vechat x}$. Although we only define operators for the arrows shown, which are either on-site or between nearest neighbors, any other terms can be decomposed into products of the operators defined, say by viewing the curved arrow as the product of the thickened ones. When a dashed arrow is reversed the operator acquires a negative sign. Importantly, closed loops, like the green arrow, should reduce to the identity in the physical Hilbert space, and hence they give rise to operator constraints. We remark that we do not have to assign an orientation to the arrows within a site, since the sign of the term on the bosonic side automatically follows from the relation $\hat \Theta^{\alpha \beta}_{\vec r} \dot = - \hat \Theta^{\beta\alpha }_{\vec r}$.
\label{fig:square}
 }
\end{center}
\end{figure}

To this end, it is helpful to resort back to our graphical representation for the operator algebra (Fig.\ \ref{fig:dot}). 
Similar in spirit to the constructions in Refs.\ \onlinecite{PhysRevD.102.114502, Clifford}, we simply arrange the sites into a square lattice, and then draw oriented arrows connecting suitable vertices between nearest-neighbor sites (Fig.\ \ref{fig:square}). Heuristically, we again think of each arrow as symbolizing the fermion bilinear $i \hat \gamma^\alpha_{\vec r} \hat \gamma^\beta_{\vec r'}$
\footnote{As a cautionary remark, for reasons detailed in Sec.\ \ref{sec:global} it is not generally possible to make such an identification for all arrows on the lattice. The more accurate statement here is as follows: to each arrow we attach an operator, and the algebra of the operators is the same as that obtained by replacing each arrow by $i \hat \gamma^\alpha_{\vec r} \hat \gamma^\beta_{\vec r'}$.
}. 
More concretely, for each arrow connecting two vertices (on the same or nearest-neighbor sites), we attach a Hermitian operator acting on the enlarged Hilbert space. Furthermore, upon restriction to the $\Gamma_-$-subspace of each of the sites, two arrows anti-commute if they share a common point, and commute otherwise. Importantly, the arrows generate the entire fermion-even algebra, and so any fermion-even term in the original problem can be decomposed into sums and products of the defined arrows (which may or may not share any common point) \cite{CHEN2018234, PhysRevB.100.245127,  PhysRevResearch.2.033527, PhysRevD.102.114502, Clifford}. For instance, a next-nearest-neighbor bilinear can be decomposed into shorter arrows connecting only vertices on the same or nearest-neighbor sites (Fig.\ \ref{fig:square}).

From the discussion above, we see that the main task in lattice bosonization is to identify operators in the enlarged Hilbert space which could furnish the algebraic relations as stipulated on the arrows. While we have detailed how that could be achieved for a single site in the preceding section, we would now need to identify operators that could correspond to the dashed arrows connecting nearest-neighbor sites. Without any further enlargement of the Hilbert space, it will be most natural to represent an arrow connecting two sites $\vec r$ and $\vec r'$ by a bilocal operator on the two respective sites. More explicitly, we consider \footnote{One could alternatively assert that if we reverse an arrow connecting sites $\vec r$ and $\vec r'$, then we attach the term $\hat \Gamma_{\vec r}$ to the operator. Once we restrict to the $\Gamma_-$-subspace of all the sites the two definitions become equivalent.}
\begin{equation}\begin{split}\label{eq:link}
\raisebox{-.25\height}{\includegraphics[width=0.08\textwidth]{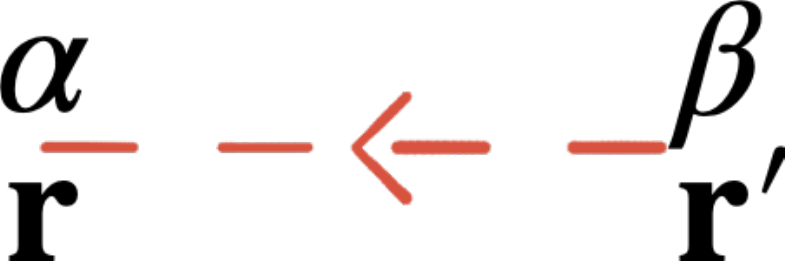}} \,
=
\left\{
\begin{tabular}{cl}
$\hat \Lambda^{\alpha \alpha}_{\vec r} \hat \Lambda^{\beta \beta}_{\vec r'}$ & 
if the orientation of the arrow\\ 
 ~ &  agrees with that in Fig.\ \ref{fig:square};\\
 \\
$- \hat \Lambda^{\alpha \alpha}_{\vec r} \hat \Lambda^{\beta \beta}_{\vec r'} $& 
 \text{if reversed.}\\ 
\end{tabular}
\right.\\
\end{split}\end{equation}
Here, the double superscript $\alpha \alpha$ and $\beta \beta$ should be viewed as indicating the diagonal elements of a $4\times 4$ matrix, as we will make explicit later.
We also require $\hat \Lambda^{\alpha \alpha}_{\vec r} \hat \Lambda^{\beta \beta}_{\vec r'}$ to be Hermitian.
We can then propose the identification of the operators
\begin{equation}\begin{split}\label{eq:proposal}
i \hat \gamma^{\alpha}_{\vec r} \hat \gamma^{\beta}_{\vec r'} 
~\dot \longleftrightarrow~
\raisebox{-.25\height}{\includegraphics[width=0.08\textwidth]{dashed_arrow.pdf}} ,\end{split}\end{equation}
for nearest neighbors $\vec r$, $\vec r'$ and matched values of $\alpha, \beta$ as indicated in Fig.\ \ref{fig:square}.
We caution, however, that this proposal may not be globally consistent. We defer the discussion of global consistency requirements to Sec.\ \ref{sec:global}; in the following we first study the local implications of the proposal.

The operator $\hat \Lambda^{\alpha \alpha}_{\vec r}$ is defined on the site Hilbert space at $\vec r$, and should commute with $\hat \Gamma_{\vec r}$ given we need to impose the $\Gamma_-$-subspace constraint. This implies $[\hat \Lambda^{\alpha \alpha}_{\vec r}, \hat \Lambda^{\beta \beta}_{\vec r'} ] = 
[\hat \Lambda^{\alpha \alpha}_{\vec r}, \hat \Theta^{\beta \gamma}_{\vec r'} ]  = 0$ when $\vec r \neq \vec r'$.
As such, the (anti-)commutation relation of the dashed arrow  $\pm \hat \Lambda^{\alpha \alpha}_{\vec r} \hat \Lambda^{\beta \beta}_{\vec r'}$ with the arrow $\hat \Theta^{\alpha \beta}_{\vec r}$ is carried solely by $\Lambda^{\alpha \alpha}_{\vec r}$.
More concretely, we require
\begin{equation}\begin{split}\label{eq:Lambda_rule}
\{ \hat \Lambda^{11}_{\vec r} , \hat \Theta^{12} _{\vec r}\}
\dot = \{ \hat \Lambda^{11}_{\vec r} , \hat \Theta^{13} _{\vec r}\} &
\dot = \{ \hat \Lambda^{11}_{\vec r} , \hat \Theta^{14} _{\vec r}\}
\dot = 0;\\
 [ \hat \Lambda^{11}_{\vec r} , \hat \Theta^{23} _{\vec r}]
\dot = [ \hat \Lambda^{11}_{\vec r} , \hat \Theta^{24} _{\vec r}] &
\dot = [ \hat \Lambda^{11}_{\vec r} , \hat \Theta^{34} _{\vec r}]
\dot = 0,
\end{split}\end{equation}
and similarly for $\hat \Lambda^{22}_{\vec r}$, $\hat \Lambda^{33}_{\vec r}$, and $\hat \Lambda^{44}_{\vec r}$. 
In addition, it follows from definitions that we also require $[\hat \Lambda^{\alpha \alpha}_{\vec r}, \hat \Lambda^{\beta \beta}_{\vec r} ]\dot = 0$.
As shown in Appendix \ref{app:RoP}, the conditions on $\hat \Lambda^{\alpha \alpha}_{\vec r}$ can be solved by writing 
\begin{equation}\begin{split}\label{eq:RoP_form}
\hat \Lambda^{\alpha \alpha}_{\vec r} = \sqrt{\hat \Gamma} e^{- i \pi \hat \lambda^{\alpha \alpha}_{\vec r}},
\end{split}\end{equation}
for some parton bilinear $\lambda^{\alpha \alpha}_{\vec r}$, i.e., it takes the same form as that of $\hat \Theta^{\alpha \beta}$ in Eq.\ \eqref{eq:h_map}. The operators $\hat \Lambda^{\alpha \alpha}$ also enjoy the same simplification in the $\Gamma_-$-subspace as in Eq.\ \eqref{eq:Gamma_odd}, and 
the explicit expressions for $\hat \lambda^{\alpha \alpha}$ are provided in Appendix \ref{app:RoP}. For the purpose of the current session, we will be content with the assertion that the operators $\hat \Lambda^{\alpha \alpha}$ satisfying the required conditions exist; their concrete forms in terms of the parton operators are provided in Appendix \ref{app:RoP}. Here, we merely note that the four operators $\hat \Lambda^{\alpha \alpha}$ we found are not independent; rather they satisfy
\begin{equation}\begin{split}\label{eq:Lambda_fuse}
\hat \Lambda^{11}_{\vec r} \hat \Lambda^{22}_{\vec r}\hat \Lambda^{33}_{\vec r}\hat \Lambda^{44}_{\vec r}=-1.
\end{split}\end{equation}
This point will be relevant for our subsequent discussion in Sec.\ \ref{sec:global}.

Next, we discuss how the JW string arises when we consider processes with range beyond the nearest-neighbor. For concreteness, consider the process indicated by the yellow arrow in Fig.\ \ref{fig:square}, which can be reconciled with the bilinear $i \hat \gamma^{3}_{\vec r}  \hat \gamma^{4}_{\vec r- \vechat x+\vechat y}$. As shown in Fig.\ \ref{fig:square}, we can decompose the ``long'' arrow into segments of smaller arrows. Algebraically, on the fermionic side we have the identity
\begin{equation}\begin{split}\label{eq:gamma_break}
&i \hat \gamma^{3}_{\vec r} i \hat \gamma^{4}_{\vec r- \vechat x+\vechat y}\\
= & (-i)^3(i \hat \gamma^{3}_{\vec r} \hat \gamma^{2}_{\vec r})
(i \hat \gamma^{2}_{\vec r} \hat \gamma^{1}_{\vec r +\vechat y})
(i \hat \gamma^{1}_{\vec r +\vechat y} \hat \gamma^{3}_{\vec r +\vechat y})
(i \hat \gamma^{3}_{\vec r +\vechat y} \hat \gamma^{4}_{\vec r- \vechat x+\vechat y}).
\end{split}\end{equation}
Translating to the enlarged bosonic description, we note that each of the bilinear encapsulated in parentheses could be reconciled with a defined operator, and so the right-hand side can be identified with
\begin{equation}\begin{split}\label{eq:bosonized_break}
&  \sqrt{\hat \Gamma_{\vec r}}^{\dagger } 
\left( \sqrt{\hat \Gamma_{\vec r+\vechat y}}^{\dagger }\right)^2 
\hat \Theta^{32}_{\vec r} 
(\hat \Lambda^{22}_{\vec r} \hat \Lambda^{11}_{\vec r +\vechat y})
\hat \Theta^{13}_{\vec r+\vechat y} 
(-\hat \Lambda^{33}_{\vec r +\vechat y} \hat \Lambda^{44}_{\vec r- \vechat x+\vechat y})\\
= & -
\left(\sqrt{\hat \Gamma_{\vec r}}^\dagger \hat \Theta^{32}_{\vec r} \hat \Lambda^{22}_{\vec r}  \right)
\left(\hat \Gamma_{\vec r+\vechat y} 
\hat \Lambda^{11}_{\vec r +\vechat y} \hat \Theta^{13}_{\vec r+\vechat y}
\hat \Lambda^{33}_{\vec r +\vechat y} \right )
\hat \Lambda^{44}_{\vec r- \vechat x+\vechat y},
\end{split}\end{equation}
notice that the reversal of a dashed arrow leads to a sign of $-1$, but that of the simple arrow is taken care of by the relation $\hat \Theta^{32}_{\vec r} = \hat \Gamma_{\vec r} \hat \Theta^{23}_{\vec r} $. Also, the replacement
\begin{equation}\begin{split}\label{eq:}
(-i)^3 \dot \mapsto   \sqrt{\hat \Gamma_{\vec r}}^{\dagger }  \left( \sqrt{\hat \Gamma_{\vec r+\vechat y}}^{\dagger }\right)^2 
\end{split}\end{equation}
is motivated by the following observation: one factor of $-i$ arises in Eq.\ \eqref{eq:gamma_break} whenever we insert the identity $(\hat \gamma_{\vec r}^\alpha)^2 = 1$ to achieve the arrow decomposition. Graphically, the ``insertion points'' can be reconciled with the internal vertices in the chain of arrows (Fig.\ \ref{fig:square}). In the bosonized description, we simply replace a $(-i)$ factor by $\sqrt{\hat \Gamma_{\vec r}}^\dagger$ at each of the internal vertices concerned.

In Eq.\ \eqref{eq:bosonized_break}, we have grouped the operators by the sites they act on. Although we have only studied the decomposition of a specific term $i \hat \gamma^{3}_{\vec r} i \hat \gamma^{4}_{\vec r- \vechat x+\vechat y}$, similar patterns of operators appear when we bosonize an arbitrary fermion bilinear (and correspondingly, any fermion-even term by taking a product of bilnears). This motivates the definition of the following operators for $\alpha \neq \beta$ (no summation on repeated indices):
\begin{equation}\begin{split}\label{eq:Lambda_Phi}
\hat \Lambda^{\alpha \beta}_{\vec r} \equiv & \sqrt{\hat \Gamma_{\vec r}}^\dagger \hat \Theta^{\alpha \beta}_{\vec r}\hat \Lambda^{\beta \beta}_{\vec r};\\
\hat \Phi^{\alpha \beta}_{\vec r} \equiv &
\sqrt{\hat \Gamma_{\vec r}}^\dagger 
\hat \Lambda^{\alpha \alpha}_{\vec r}
\hat \Lambda^{\alpha \beta}_{\vec r}
=\hat \Gamma_{\vec r} \hat \Lambda^{\alpha \alpha}_{\vec r}
\hat \Theta^{\alpha \beta}_{\vec r}\hat \Lambda^{\beta \beta}_{\vec r}.
\end{split}\end{equation}
Note that $\hat \Phi^{\alpha \beta}_{\vec r} = \hat \Gamma_{\vec r} \hat \Phi^{\beta \alpha }_{\vec r}$. Also, both $\hat \Lambda^{\alpha \beta}_{\vec r}$ and $\hat \Phi^{\alpha \beta}_{\vec r} $ are defined such that they are Hermitian and square to $1$. In fact, they are both roots of parity (Appendix \ref{app:RoP}) and can be expressed in the form Eq. \eqref{eq:RoP_form} with suitable parton bilinears $\hat \lambda^{\alpha \beta}_{\vec r}$ and $\hat \phi^{\alpha \beta}_{\vec r}$.

The transformation properties of $\hat \Lambda^{\alpha \beta}_{\vec r}$ plays a key role in our analysis of the symmetry manifestation in the bosonized model, and we defer that discussion to Sec.\ \ref{sec:sym}. Here, we simply note the following relation between the $\hat \Phi$ operators: if $\alpha, \beta, \gamma$ are all distinct, we have
\begin{equation}\begin{split}\label{eq:Phi_prod}
\hat \Phi^{\alpha \beta}_{\vec r} \hat \Phi^{ \beta \gamma}_{\vec r} 
= \sqrt{\hat \Gamma_{\vec r}}^\dagger \hat \Phi^{\alpha \gamma}_{\vec r}.
\end{split}\end{equation}
By squaring this, one sees that $\hat \Phi^{\alpha \beta}_{\vec r} \hat \Phi^{ \beta \gamma}_{\vec r}  = \hat \Gamma_{\vec r}
\hat \Phi^{ \beta \gamma}_{\vec r} \hat \Phi^{\alpha \beta}_{\vec r} $. Furthermore, we have $[\hat \Phi^{\alpha \beta}_{\vec r}, \hat \Phi^{ \gamma \delta}_{\vec r}] = 0$ when $\alpha,\beta,\gamma,\delta$ are all distinct. In other words, $\hat \Phi$'s realize a (conjugated) copy of the algebra furnished by $\hat \Theta$, as discussed around Eq.\ \eqref{eq:Theta_comm}. Also, we note that $[\hat \Theta^{\alpha \beta}_{\vec r}, \hat \Phi^{ \gamma \delta}_{\vec r'}] = 0$ for all $\vec r, \vec r'$ (i.e., even when $\vec r = \vec r'$).

Using the $\hat \Lambda$ and $\hat \Phi$ operators, we may simply write
\begin{equation}\begin{split}\label{eq:}
i \hat \gamma^{3}_{\vec r} i \hat \gamma^{4}_{\vec r- \vechat x+\vechat y}
\dot \longleftrightarrow - \hat \Lambda^{32}_{\vec r} \hat \Phi^{13}_{\vec r+\vechat y} \hat \Lambda^{44}_{\vec r - \vechat x + \vechat y}.
\end{split}\end{equation}
More generally, one could apply the same procedure to an arbitrary fermion bilinear $i \hat \gamma^{\alpha}_{\vec r}\hat \gamma^{\beta}_{\vec r'}$. Schematically, the bosonized operator takes the form $\pm \hat \Lambda^{\alpha *}_{\vec r} \hat \Phi^{**}_{\#} \cdots \hat \Phi^{**}_{\#}\hat \Lambda^{\beta *}_{\vec r'}$, where $*$ denotes indices which are determined by the relative position of $\vec r$ and $\vec r'$, and $\#$ denote sites traversed by a straightened arrow connecting $\vec r'$ to $\vec r$. From this generic form, we see that the operators $\hat \Phi^{\alpha \beta}$ play the role of a non-abelian JW string.
The indices are determined as follows:
\begin{enumerate}
\item Assuming $\vec r \neq \vec r'$, draw a straightened arrow connecting $\vec r'$ to $\vec r$ using only nearest-neighbor links.
\item The second index of the first term $\hat \Lambda^{\alpha *}_{\vec r}$ depends on the direction the arrow enters the site $\vec r$: 
(south, north, west, east) = $(1,2,3,4)$.
For instance, if the arrow enters $\vec r$ from the north, the first term is $\hat \Lambda^{\alpha 2}_{\vec r}$.
\item Similarly, we determine the second index of the last term $\hat \Lambda^{\beta *}_{\vec r'}$ depending on the direction from which the arrow leaves, e.g., if it leaves from the west, then the last term is $\hat \Lambda^{\beta 3}_{\vec r'}$.
\item For each site $\vec x$ traversed by the arrow, we append $\hat \Phi^{**}_{\vec x}$ depending on the directions the arrow enters and leaves the site. For instance, if the arrow comes in from the east and exits from the south, we insert the operator $\hat \Phi^{14}_{\vec x}$ to the string.
\item The overall sign of $\pm 1$ is given by the number of times a nearest-neighbor link (i.e., a dashed arrow in Fig.\ \ref{fig:square}) is flipped. With our convention, the sign is given by ${\rm sign}(r_x'-r_x)^{|r_x'-r_x|} \times {\rm sign}(r_y'-r_y)^{|r_y'-r_y|}$, with the understanding that the associated sign is $+1$ if $r_x'=r_x$ or $r_y'=r_y$.
\end{enumerate}

Crucially, in the prescription above we have not specified how the straightened arrow should be chosen. Generally speaking, different choices of the arrows could, for example, traverse different sites. In the original fermionic problem, different straightening simply corresponds to different ways of inserting the identity operator. Yet, in the bosonized description different straightening gives rise to different patterns of the JW strings, which are not {\it a priori} identical. The choice of straightening is clearly unphysical, and so, correspondingly, we have to impose constraints on the enlarged Hilbert space to ensure that all the possible JW strings associated with the same physical process are equivalent in the restricted Hilbert space. In other words, we demand that all closed loops in Fig.\ \ref{fig:square} evaluate to the identity in the restricted Hilbert space. Barring homologically nontrivial loops (with $\mathbb Z_2$ coefficients), for the square lattice it suffices to ensure the condition is satisfied for each of the plaquettes formed by the nearest-neighbor links \cite{CHEN2018234, Clifford}. More explicitly, for each $\vec r$ we consider
\begin{equation}\begin{split}\label{eq:}
&(i \hat \gamma^{4}_{\vec r} \hat \gamma^{3}_{\vec r + \vechat x})
(i \hat \gamma^{3}_{\vec r + \vechat x} \hat \gamma^{2}_{\vec r + \vechat x})
(i \hat \gamma^{2}_{\vec r + \vechat x} \hat \gamma^{1}_{\vec r + \vechat x + \vechat y})\times \\
&~~
(i \hat \gamma^{1}_{\vec r + \vechat x + \vechat y} \hat \gamma^{3}_{\vec r + \vechat x + \vechat y})
(i \hat \gamma^{3}_{\vec r + \vechat x + \vechat y}\hat \gamma^{4}_{\vec r + \vechat y})
(i \hat \gamma^{4}_{\vec r + \vechat y} \hat \gamma^{1}_{\vec r + \vechat y})\times\\
&~~~~
(i \hat \gamma^{1}_{\vec r + \vechat y} \hat \gamma^{2}_{\vec r})
(i \hat \gamma^{2}_{\vec r}\hat \gamma^{4}_{\vec r})\\
\dot \longleftrightarrow & 
(\hat \Lambda^{44}_{\vec r} \hat \Lambda^{33}_{\vec r+\vechat x})
\hat \Theta^{32}_{\vec r + \vechat x}
(\hat \Lambda^{22}_{\vec r + \vechat x} \hat \Lambda^{11}_{\vec r+\vechat x + \vechat y}) \hat \Theta^{13}_{\vec r+\vechat x + \vechat y} \times\\
&~~(-\hat \Lambda^{33}_{\vec r+\vechat x + \vechat y} \hat \Lambda^{44}_{\vec r + \vechat y})
\hat \Theta^{41}_{\vec r + \vechat y}
(- \hat \Lambda^{11}_{\vec r + \vechat y} \hat \Lambda^{22}_{\vec r}) 
\hat \Theta^{24}_{\vec r}\\
=&   \hat \Phi^{24}_{\vec r }
(\hat \Gamma_{\vec r+\vechat x} \hat \Phi^{32}_{\vec r + \vechat x})
(\hat \Gamma_{\vec r+\vechat x+\vechat y} \hat \Phi^{13}_{\vec r + \vechat x+\vechat y})
(\hat \Gamma_{\vec r+\vechat y} \hat \Phi^{41}_{\vec r + \vechat y})\\
\dot =& - \hat \Phi^{24}_{\vec r }  \hat \Phi^{32}_{\vec r + \vechat x} \hat \Phi^{13}_{\vec r + \vechat x+\vechat y} \hat \Phi^{41}_{\vec r + \vechat y}.
\end{split}\end{equation}
Since in the original fermionic problem the product of operators simply gives the identity, we impose the constraint
\begin{equation}\begin{split}\label{eq:plaquette}
\hat C_{\vec r^\vee} \equiv
\hat \Phi^{24}_{\vec r }  \hat \Phi^{32}_{\vec r + \vechat x} \hat \Phi^{13}_{\vec r + \vechat x+\vechat y} \hat \Phi^{41}_{\vec r + \vechat y} \overset{c}{=} -1
\end{split}\end{equation}
for each plaquette $\vec r^\vee$ with the bottom-left corner being $\vec r$. The notation $\overset{c}{=}$ reads as ``the left-hand side is constrained to equal to the right-hand side in the physical subspace.'' In particular, the $\Gamma_-$ condition is always enforced.
One can also verify that the different constraints commute with each other in the $\Gamma_-$-subspace.

Locally, the constraints filter out half of the states in the $\Gamma_-$-subspace on each of the site. This consumes the additional qubit mentioned in Sec.\ \ref{sec:dot}, and brings the dimension of the physical site Hilbert space down to four, which matches with that of the original fermionic problem. However, as is typical for such local plaquette constraints, say in the toric code, the local constraints may not be globally independent: on topologically nontrivial spaces like the torus, there can be additional independent constraints associated with the homologically nontrivial loops \cite{CHEN2018234, Clifford}. We will next explore what modifications are required to account for such global constraints.

\section{Global affairs \label{sec:global}}
To address global concerns in constructing the JW transformation, we focus our attention to a square lattice defined on a two-dimensional torus of size $L_x \times L_y$. For definiteness, we write the coordinates as $\vec r = (r_x, r_y)$ with $r_{x,y} \in [0,L_{x,y}-1]$, and identify $r_{x,y} = L_{x,y}$ with $r_{x,y} = 0$. One could also investigate how the topology of a more general manifold could impact the JW transformation procedures. We will refrain from these more general problems and stay with the torus geometry; instead, we refer the interested readers to Refs.\ \onlinecite{CHEN2018234, PhysRevB.100.245127,  PhysRevResearch.2.033527, PhysRevD.102.114502, Clifford} which provide a detailed discussions on issues such as global consistency of the transformation and the relation with a lattice analog of the spin structure.

Consider a general site $\vec r$ and a nontrivial loop passing through $\vec r$ along the $\vechat y$ direction. Taking the product of all the arrows along the loop, we define
\begin{equation}\begin{split}\label{eq:}
\hat {\mathcal V}_{\vec r}
 \equiv& (-1)^{L_y-1}
\hat \Lambda^{22} _{\vec r}
\hat \Lambda^{11}_{\vec r+\vechat y}
\hat \Theta^{12}_{\vec r + \vechat y}
\cdots 
\hat \Lambda^{22}_{\vec r+(L_y-1) \vechat y} 
\hat \Lambda^{11}_{\vec r}
\hat \Theta^{12}_{\vec r} \\
\dot =& (-1)^{L_y-1} \hat \Phi^{12}_{\vec r}
(-\hat \Phi^{12}_{\vec r + \vechat y})
\cdots 
(-\hat \Phi^{12}_{\vec r+(L_y-1) \vechat y} )\\
=&  \prod_{n=0}^{L_y-1} \hat \Phi^{12}_{\vec r+ n \vechat y}.
\end{split}\end{equation}
Note that, in the $\Gamma_-$ subspace, the definition of $\hat {\mathcal V}_{\vec r}$ is independent of the choice of the base point along the loop.
Analogously, we also define for the horizontal loop 
\begin{equation}\begin{split}\label{eq:}
\hat {\mathcal H}_{\vec r}
 \equiv& (-1)^{L_x-1}
\hat \Lambda^{44} _{\vec r}
\hat \Lambda^{33}_{\vec r+\vechat x}
\hat \Theta^{34}_{\vec r + \vechat x}
\cdots 
\hat \Lambda^{44}_{\vec r+(L_x-1) \vechat x} 
\hat \Lambda^{33}_{\vec r}
\hat \Theta^{34}_{\vec r} \\
\dot =&  \prod_{m=0}^{L_x-1} \hat \Phi^{34}_{\vec r+ m \vechat x}.
\end{split}\end{equation}
From the commutation relation among the $\hat \Phi$ operators, we also have $[\hat {\mathcal V}_{\vec r}, \hat {\mathcal V}_{\vec r'}] = [\hat {\mathcal H}_{\vec r}, \hat {\mathcal H}_{\vec r'}]  = [\hat {\mathcal V}_{\vec r}, \hat {\mathcal H}_{\vec r'}] =0$. 

With the proposed operator identification in Eq.\ \eqref{eq:proposal}, we have
\begin{equation}\begin{split}\label{eq:gamma_loop}
& (i \hat \gamma^{2}_{\vec r} \hat \gamma^{1}_{\vec r + \vechat y}) (i \hat \gamma^{1}_{\vec r+\vechat y} \hat \gamma^{2}_{\vec r + \vechat y})\cdots 
(i \hat \gamma^{2}_{\vec r+(L_y-1) \vechat y} \hat \gamma^{1}_{\vec r }) (i \hat \gamma^{1}_{\vec r} \hat \gamma^{2}_{\vec r })\\
\dot \longleftrightarrow
& (-1)^{L_y-1}\hat {\mathcal V}_{\vec r}.
\end{split}\end{equation}
The left-hand side evaluates to $i^{2L_y}=(-1)^{L_y}$, and so we have the constraint
\begin{equation}\begin{split}\label{eq:}
\hat {\mathcal V}_{\vec r} \dot = \prod_{n=0}^{L_y-1} \hat \Phi^{12}_{\vec r+ n \vechat y} 
\overset{c}{=}& -1
\end{split}\end{equation}
and similarly also
\begin{equation}\begin{split}\label{eq:}
\hat {\mathcal H}_{\vec r} \dot =\prod_{m=0}^{L_x-1} \hat \Phi^{34}_{\vec r+ m \vechat x} \overset{c}{=}& -1.
\end{split}\end{equation}
Now, let us consider the product
\begin{equation}\begin{split}\label{eq:VH_prod}
& \hat {\mathcal V}_{\vec 0}\hat {\mathcal V}_{\vechat x}\cdots \hat {\mathcal V}_{(L_x-1)\vechat x}
\hat {\mathcal H}_{\vec 0}\hat {\mathcal H}_{\vechat y}\cdots \hat {\mathcal H}_{(L_y-1)\vechat y}\\
\dot =& \prod_{m=0}^{L_x-1} \prod_{n=0}^{L_y-1} 
\hat \Phi^{12}_{m \vechat x+ n \vechat y } 
\hat \Phi^{34}_{m \vechat x+ n \vechat y}\\
\overset{c}{=} & (-1)^{L_x + L_y},
\end{split}\end{equation}
and notice that 
\begin{equation}\begin{split}\label{eq:}
\hat \Phi^{12}_{\vec r}  \hat \Phi^{34}_{\vec r}
= \hat \Lambda^{11}_{\vec r} \hat \Theta^{12}_{\vec r} \hat \Lambda^{22}_{\vec r}
\hat \Lambda^{33}_{\vec r} \hat \Theta^{34}_{\vec r} \hat \Lambda^{44}_{\vec r}
=-\hat \Theta^{12}_{\vec r} \hat \Theta^{34}_{\vec r},
\end{split}\end{equation}
where the last equality follows from Eq.\ \eqref{eq:Lambda_fuse}. On the fermionic side, we have
\begin{equation}\begin{split}\label{eq:}
\hat \Theta^{12}_{\vec r} \hat \Theta^{34}_{\vec r} 
\dot \longleftrightarrow 
(i\hat \gamma^{1}_{\vec r} \hat \gamma^{2}_{\vec r})
(i\hat \gamma^{3}_{\vec r} \hat \gamma^{4}_{\vec r}),
\end{split}\end{equation}
which is simply the on-site parity operator. Taking the product over all the sites in Eq.\ \eqref{eq:VH_prod} then gives the global parity operator $\hat P$ of the physical fermions. So we conclude that, as part of our proposal, we have implicitly demanded
\begin{equation}\begin{split}\label{eq:P_bare}
\hat P \overset{c}{=} (-1)^{L_x L_y + L_x + L_y}.
\end{split}\end{equation}
Clearly, the right-hand side is non-dynamical, and hence for a given torus we could only describe half of the states in the original fermionic Hilbert space with the specified global fermion parity. For instance, in this formulation we must take $L_x$ and $L_y$ to be both even in order to capture the vacuum state with no physical fermions.

The analysis above raises a natural question: in the discussion so far we have only emphasized on representing the fermion-even algebra in the bosonic Hilbert space. In other words, we have made no attempt to identify the fermion-odd operators, like $\hat f^{\dagger }$,  which would flip the global parity $\hat P$. Is it possible for us to modify the identification rule in Eq.\ \eqref{eq:proposal} and achieve this goal? 
To answer this question, we address one subtlety in our operator identification: our prescription requires that to each arrow in Fig.\ \ref{fig:square} we associate an operator, and that the algebra of the operators is dictated by the identifications Eqs.\ \eqref{eq:h_map} and \eqref{eq:proposal}. In the original fermionic problem, however, we could have freely modified any one of the fermion-even operators without changing the algebra by attaching to it a center element of the algebra, i.e., an operator which commutes with all the other elements in the algebra. More explicitly, modifying a dashed arrow from $ i\hat \gamma^{\alpha }_{\vec r} \hat \gamma^{\beta}_{\vec r'}$ to $\pm (i \hat \gamma^{\alpha }_{\vec r} \hat \gamma^{\alpha }_{\vec r'})$ or $\pm \hat P (i \hat \gamma^{\alpha }_{\vec r} \hat \gamma^{\alpha }_{\vec r'})$ would not affect the algebraic relations between the arrows.

In our choice of the orientation of the arrows, we have already made a choice on the sign $\pm 1$ of the operators. Yet, there is still freedom left in the attachment of the global fermion parity. While the identification of the on-site terms is well-motivated from the representation of ${\rm Spin}(4)$ and satisfies the expected relations as operator identities (Fig.\ \ref{fig:dot}), the identification of nearest-neighbor links is more ambiguous. 
Riding on this ambiguity, we consider the modification
\begin{equation}\begin{split}\label{eq:proposal_gen}
i \hat \gamma^{\alpha}_{\vec r} \hat \gamma^{\beta}_{\vec r'} \hat P^{\zeta_{\langle \vec r \vec r' \rangle}} 
~\dot \longleftrightarrow~
\raisebox{-.25\height}{\includegraphics[width=0.08\textwidth]{dashed_arrow.pdf}} ,\end{split}\end{equation}
where $\zeta_{\langle \vec r \vec r' \rangle} = 0,1 \in \mathbb Z_2$ is defined only on the nearest-neighbor links.
With this generalization, the plaquette constraint Eq.\ \eqref{eq:plaquette} becomes
\begin{equation}\begin{split}\label{eq:plaquette_gen}
\hat C_{\vec r^\vee} \overset{c}{=} - \hat P^{\sum_{\langle \vec r \vec r'\rangle \in \partial\vec r^\vee}\zeta_{\langle \vec r \vec r'\rangle}},
\end{split}\end{equation}
where the sum in the exponent is over the four links around the plaquette $\vec r^\vee$.
Similarly, the constraints on $\hat {\mathcal V}_{\vec r}$ and $\hat {\mathcal H}_{\vec r}$ are modified analogously, and Eq.\ \eqref{eq:P_bare} becomes
\begin{equation}\begin{split}\label{eq:P_gen}
\hat P^{\zeta -1} \overset{c}{=} (-1)^{L_x L_y + L_x + L_y},
\end{split}\end{equation}
where
\begin{equation}\begin{split}\label{eq:}
\zeta \equiv \sum_{\langle \vec r \vec r'\rangle} \zeta_{\langle \vec r \vec r'\rangle}
\end{split}\end{equation}
is summed over all the $2 L_x L_y$ nearest-neighbor links on the lattice. 
We remark that this observation is closely related to the use of an Eulerian circuit in Ref.\ \onlinecite{Clifford}.

We can now analyze how Eq.\ \eqref{eq:P_gen} could be satisfied for both $\hat P = \pm 1$. First, for the even-parity sector with $\hat P = 1$, the left-hand side is $+1$ regardless of the value of $\zeta$. This forces us to consider an even-by-even torus such that the right-hand side is also $+1$ \footnote{
We impose this mostly for simplicity; one could also relax our construction and treat the sign in the operator identification also as a variable, and then solve the constraints.
}. Now, in order for our map to cater the possibility of $\hat P = -1$, we need to require $\zeta  = 1 \mod 2$. This condition rules out adopting Eq.\ \eqref{eq:proposal} for all links on the lattice, which would lead to $\zeta = 0$. Nevertheless, we could still adopt that identification almost everywhere except for one particular link, which must come with $\hat P$ in order to realize $\zeta = 1$.

\begin{figure}[h]
\begin{center}
{\includegraphics[width=0.45 \textwidth]{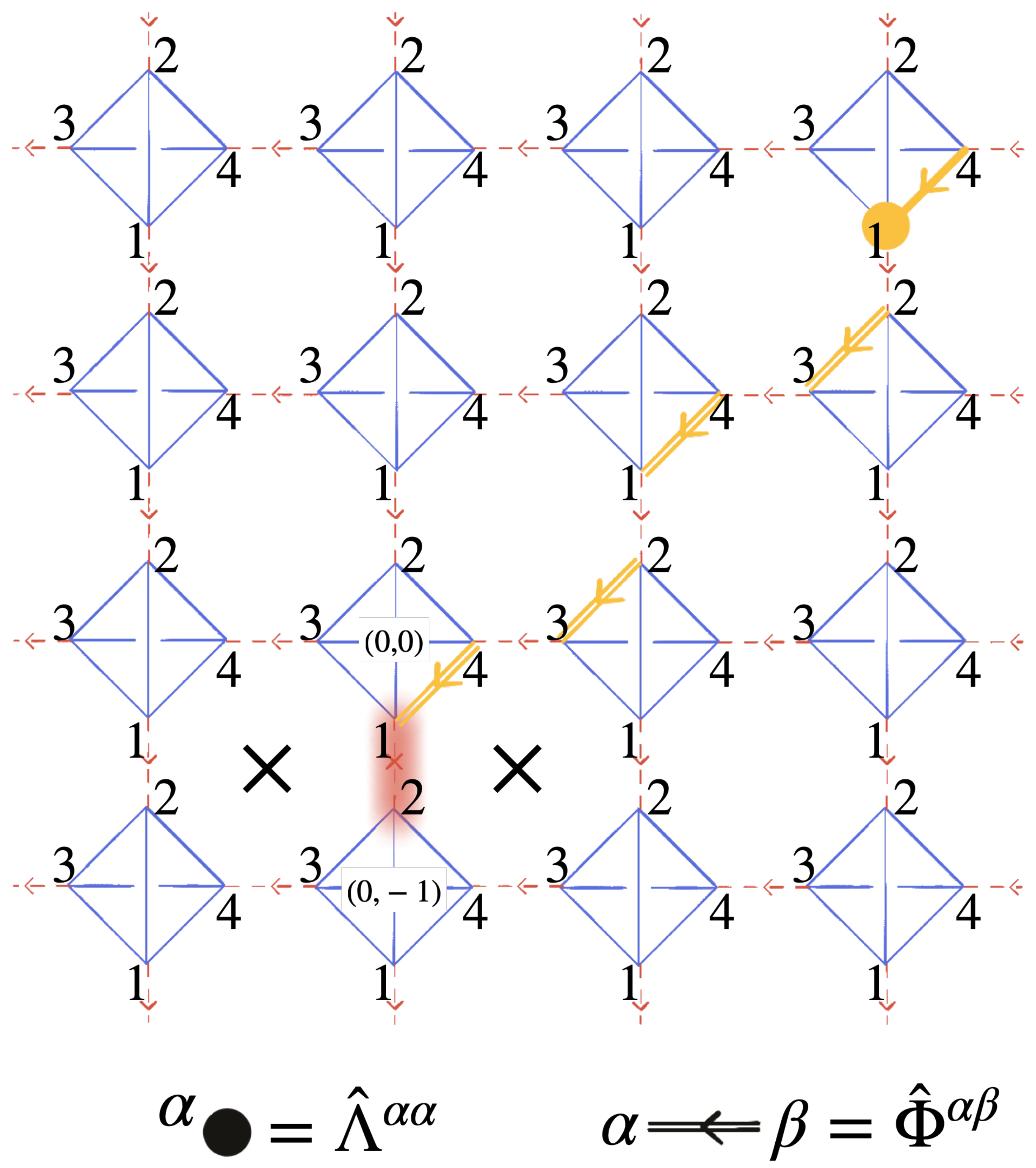}} 
\caption{Decoration by the fermion parity operator. In order to capture both the parity-even and odd subspaces of the original femrionic problem, we have to consider an even-by-even torus and attach the fermion parity operator $\hat P$ to an odd number of the nearest-neighbor link. We can choose to attach it to the link between vertices $1$ on $(0,0)$ and $2$ on $(0,-1)\equiv(0,L_y-1)$ (highlighted dashed line marked $\times$). Correspondingly, the constraints on the two plaquettes marked by a cross are modified to $- \hat P$. With this identification, the fermion-odd operators (up to the coefficients) can also be identified by suitable combinations of the on-site terms with the Jordan-Winger string attached to the origin. The product of operators corresponding to  $\hat \gamma^{4}_{(2,2)}$ are shown as an example (orange).
\label{fig:square_odd}
 }
\end{center}
\end{figure}

To this end, we assign the ``decorated link'' to the one between sites $\vec 0$ and $-\vechat y \equiv (L_y-1)\vechat y$, i.e., we make the modified identification 
\begin{equation}\begin{split}\label{eq:P_insert}
(i \hat \gamma^{2}_{-\vechat y} \hat \gamma^{1}_{\vec 0}) \hat P 
~\dot \longleftrightarrow ~
\hat \Lambda^{22}_{-\vechat y} \hat \Lambda^{11}_{\vec 0}.
\end{split}\end{equation}
Correspondingly, the constraints on the two plaquettes sandwiching the modified link are also modified to $- \hat P$ (Fig.\ \ref{fig:square_odd}). 
Furthermore, the constraints on the loop operators are modified as 
\begin{equation}\begin{split}\label{eq:VH_new}
\hat {\mathcal H}_{\vec r} \overset{c}{=}& -1;\\
\hat {\mathcal V}_{\vec r} \overset{c}{=}& 
\left\{
\begin{array}{cl}
- \hat P & \text{if $\vec r = (0,y)$ for $y\in [0,L-1]$};\\
-1 & \text{otherwise.}
\end{array}
\right.
\end{split}\end{equation}

This modification allows us to identify the bosonic counterparts of the fermion-odd operators.
Consider $\hat \Lambda^{11}_{\vec 0}$. It commutes with all the arrows in the lattice except for the following three:
\begin{equation}\begin{split}\label{eq:}
\Theta^{12}_{\vec 0} ~\dot \longleftrightarrow& ~ i \hat \gamma^{1}_{\vec 0} \hat \gamma^{2}_{\vec 0};\\
\Theta^{13}_{\vec 0} ~\dot \longleftrightarrow& ~ i \hat \gamma^{1}_{\vec 0} \hat \gamma^{3}_{\vec 0}; \&\\
\Theta^{14}_{\vec 0} ~\dot \longleftrightarrow& ~ i \hat \gamma^{1}_{\vec 0} \hat \gamma^{4}_{\vec 0}.
\end{split}\end{equation}
Note that, crucially, $\hat \Lambda^{11}_{\vec 0}$ also commutes with the ``decorated'' arrow identified with $(i \hat \gamma^{2}_{ - \vechat y} \hat \gamma^{1}_{\vec 0}) \hat P $. As such, we see that the algebraic relations between $\hat \Lambda^{11}_{\vec 0}$ and all the other fermion-even operators are consistent with the identification
\begin{equation}\begin{split}\label{eq:fermion_odd}
\hat \Lambda^{11}_{\vec 0} ~\dot \longleftrightarrow~ \hat \gamma^{1}_{\vec 0}.
\end{split}\end{equation}
This then unlocks the bosonization of all the other fermion-odd operators, since they can always be decompsed as the product between $\hat \gamma^1_{\vec 0}$ and another fermion-even operator, both of which we have bosonized already. Intuitively, the bosonized form of the fermion-odd operator $\hat \gamma_{\vec r}^{\alpha}$ comes with a JW string $\hat \Phi \hat \Phi\cdots \hat \Phi$ attached to the origin, together with suitable on-site terms at the site $\vec r$ (Fig.\ \ref{fig:square_odd}). In the constrained, physical Hilbert space the precise shape of the string is immaterial, and generally  the shortest choice possible is given by $|\vec r|$ in the taxicab metric. 
However, when one considers any fermion-even operators the strings cancel out except for the segment(s) contained within the support of the operator. Such cancellation is the same as in the classic one-dimensional JW transformation, and so a local fermionic Hamiltonian (containing only fermion-even terms) is mapped to a local bosonic Hamiltonian---up to the subtlety we explain below.

To faithfully reproduce the fermionic problem, it is necessary for us to impose constraints to the bosonic Hilbert space. 
The entering of the global parity operator in the identification Eq.\ \eqref{eq:proposal_gen} then leads to a subtlety concerning the locality of the transformation. 
While the $\Gamma_-$ condition is manifestly local (and can be taken into account explicitly by a truncation of the site Hilbert space, as we discuss in Sec.\ \ref{sec:qubit}), the plaquette constraints are more subtle. In our current scheme, for most of the plaquettes we can simply add a local term in the bosonized Hamiltonian $K \hat C_{\vec r^\vee}$ with $K\rightarrow + \infty$ to enforce $\hat C_{\vec r^\vee} \overset{c}{=} -1$. Yet, for the two plaquettes $(-\vechat  y)^\vee$ and $(-\vechat  y-\vechat x)^\vee$, the constraint has to be enforced by a non-local term. For instance, the constraint on $(-\vechat y)^\vee$ can be enforced by
\begin{equation}\begin{split}\label{eq:C0P}
K  \hat C_{(-\vechat  y)^\vee}  \hat P= 
K  \hat C_{(-\vechat  y)^\vee} \prod_{\vec r} \hat \Theta^{12}_{\vec r} \hat \Theta^{34}_{\vec r},
\end{split}\end{equation}
which is clearly non-local. Similarly, we could also add the constraints $K \hat {\mathcal H}_{\vec 0}$ and $K \hat {\mathcal V}_{\vechat x}$ to enforce Eq.\ \eqref{eq:VH_new}. 
The last two constraints corresponding to Wilson loop operators on the bosonic side are relatively mild, in the sense that they are non-dynamical and represent topological degeneracy on the torus even without adding a non-local operator to the Hamiltoinan. In contrast, the non-local constraint in Eq.\ \eqref{eq:C0P} is more severe, in that a plaquette term is local on the bosonic side, but its operator value is tied to the non-local, global fermion parity.
As such, we say the JW transformation is only ``graded-local,'' in the sense that locality is manifest only if we consider the two parity sectors independently (and barring the selection of a particular state out of the degenerate ground-state manifold). We comment that such subtleties on locality is not unique to higher dimensions; in fact, the classic one-dimensional JW transformation is also known to display the same subtlety, as we briefly review in Appendix \ref{app:1D}. Along this line, we also remark that the notion of graded-locality is mutual, in the sense that one could also insist that the bosonized Hamiltonian is manifestly local, and that the signs of certain terms in the corresponding fermionic Hamiltonian depend on the fermion parity; this point of view is usually endorsed in discussing the one-dimensional JW transformation.

In the discussion above, we have made an {\it ad hoc} choice of placing the mandated  $\hat P$ operator at the link between $(0,0)$ and $(0,-1)$, which as we have argued is equivalent to picking an origin for the JW string. Given no point is special on the torus, it is natural to ask how we could relate the different choices on the placement of $\hat P$. To this end, let us define a unitary operator
\begin{equation}\begin{split}\label{eq:}
\hat {\mathcal M}(\hat \Phi) \equiv \hat P_+ + \hat P_- \hat \Phi,
\end{split}\end{equation}
where $\hat P_{\pm} \equiv (1 \pm \hat P)/2$ are projectors into the different parity sectors, and $\hat \Phi$ represents any collection of the JW string operators $ e^{i \frac{\pi}{2} m} \hat \Phi^{\alpha_1 \beta_1}_{\vec r_1}\hat \Phi^{\alpha_2 \beta_2}_{\vec r_2}\cdots \hat \Phi^{\alpha_k \beta_k}_{\vec r_k}$ for some integer $m$. The phase is incorporated such that the coordinates $\vec r_{i}$ are not necessarily distinct, and the relation $\hat {\mathcal M}(\hat \Phi_1) \hat {\mathcal M}(\hat \Phi_2) = \hat {\mathcal M}(\hat \Phi_1 \hat \Phi_2)$ holds.
Notice that, with this definition, we have $\hat \Phi^\dagger \hat \Phi=1$ and $\hat \Phi^\dagger = \pm \hat \Phi$.
If $\hat \Phi$ is local, then $\hat {\mathcal M}(\hat \Phi)$ is graded-local since it acts either as the identity or as $\hat \Phi$ in the two subspaces. 

Next, let us consider an operator $\hat O$ which either commutes or anti-commutes with $\hat P$ and $\hat \Phi$ in the $\hat \Gamma_-$ subspace. In practice, such operator $\hat O$ arises when we take products between the $\hat \Theta$ and $\hat \Lambda$. We denote the effective (anti-)commutation by the notation $ \hat O  \hat P \dot = C_{\hat O, \hat P}\hat P \hat O$ with $C_{\hat O, \hat P}=\pm 1$. $C_{\hat O, \hat \Phi } = \pm 1$ is similarly defined.
We can directly evaluate how $\hat O$ transforms under conjugation by $\hat {\mathcal M}(\hat \Phi)$:
\begin{equation}\begin{split}\label{eq:Phi_change}
\hat {\mathcal M}(\hat \Phi) \hat O \hat {\mathcal M}(\hat \Phi)^\dagger \dot = 
\left\{
\begin{array}{ccc}
~& C_{\hat O, \hat P} & C_{\hat O, \hat \Phi} \\
\hat O & +1 & +1\\
\hat P \hat O & +1 & -1\\
\hat O \left[ \frac{\hat \Phi+ \hat \Phi^\dagger}{2} + \hat P \frac{\hat \Phi - \hat \Phi^\dagger}{2}  \right] & -1 & +1\\
- \hat O \left[ \frac{\hat \Phi - \hat \Phi^\dagger}{2} + \hat P \frac{\hat \Phi + \hat \Phi^\dagger}{2}  \right] & -1 & -1
\end{array}
\right.
,
\end{split}\end{equation}
notice that the last two lines can be further simplified using $\hat \Phi^\dagger = \pm \hat \Phi$.
We see that when $\hat O$ commutes with $\hat P$, i.e., those that are fermion-even, it is either unchanged or gets attached to $\hat P$. This is expected since when we move the link at which the mandated $\hat P$ operator is placed, we have to attach $\hat P$ to appropriate terms to maintain the fermionic processes they represent. Furthermore, for $\hat O$ which are fermion-odd and hence anti-commute with $\hat P$, we attach $\hat \Phi$ as the explicit form of the JW string is modified.

To illustrate this more explicitly, let us consider the case of $\hat {\mathcal M} (\hat \Phi_{\vec 0}^{12})$. With the prescriptions Eqs.\ \eqref{eq:P_insert} and \eqref{eq:fermion_odd}, we have
\begin{equation}\begin{split}\label{eq:}
\hat \gamma_{\vechat y}^{1} &= 
- \hat \gamma_{\vec 0}^{1} (i \hat \gamma_{\vec 0}^1 \hat \gamma_{\vec 0}^2)
(i \hat \gamma_{\vec 0}^2 \hat \gamma_{\vechat y}^1)\\
& \dot \longleftrightarrow - \hat \Lambda^{11}_{\vec 0} \hat \Theta^{12}_{\vec 0} \hat \Lambda^{22}_{\vec 0} \hat \Lambda^{11}_{\vechat y} \dot = \hat \Phi^{12}_{\vec 0} \hat \Lambda^{11}_{\vechat y}.
\end{split}\end{equation}
Upon conjugating the right-hand side by $\hat {\mathcal M} (\hat \Phi_{\vec 0}^{12})$, one finds
\begin{equation}\begin{split}\label{eq:}
\hat {\mathcal M} (\hat \Phi_{\vec 0}^{12})
( \hat \Phi^{12}_{\vec 0} \hat \Lambda^{11}_{\vechat y})
\hat {\mathcal M} (\hat \Phi_{\vec 0}^{12})^\dagger = \hat \Lambda_{\vechat y}^{11},
\end{split}\end{equation}
which shows that the origin of the JW string has been moved. 
Similarly, if we consider the original identifications 
$i \hat \gamma_{-\vechat  y}^{2} \hat \gamma_{\vec 0}^{1} \dot\longleftrightarrow \hat P \hat \Lambda^{22}_{- \vechat  y} \hat \Lambda^{11}_{\vec 0}$ and 
$i \hat \gamma_{\vec 0}^{2} \hat \gamma_{\vechat y}^{1} \dot\longleftrightarrow \hat \Lambda^{22}_{\vec 0} \hat \Lambda^{11}_{\vechat y}$
, we have
\begin{equation}\begin{split}\label{eq:}
\hat {\mathcal M} (\hat \Phi_{\vec 0}^{12})
(\hat P \hat \Lambda^{22}_{-\vechat y} \hat \Lambda^{11}_{\vec 0 } )
\hat {\mathcal M} (\hat \Phi_{\vec 0}^{12})^\dagger = &\hat \Lambda^{22}_{-\vechat y} \hat \Lambda^{11}_{\vec 0 };\\
\hat {\mathcal M} (\hat \Phi_{\vec 0}^{12})
( \hat \Lambda^{22}_{\vec 0} \hat \Lambda^{11}_{\vechat y} )
\hat {\mathcal M} (\hat \Phi_{\vec 0}^{12})^\dagger =& \hat P \hat \Lambda^{22}_{\vec 0} \hat \Lambda^{11}_{\vechat y},
\end{split}\end{equation}
which implies in the transformed prescription we now identify $i \hat \gamma_{-\vechat  y}^{2} \hat \gamma_{\vec 0}^{1} \dot\longleftrightarrow  \hat \Lambda^{22}_{- \vechat  y} \hat \Lambda^{11}_{\vec 0}$ and  $i \hat \gamma_{\vec 0}^{2} \hat \gamma_{\vechat y}^{1} \dot\longleftrightarrow \hat P\hat \Lambda^{22}_{\vec 0} \hat \Lambda^{11}_{\vechat y}$, i.e., the ``decorated link'' has been moved.

In closing this section, we reiterate that the considerations above assumed we place the square lattice on a torus. Had we instead considered a manifold with some open boundaries, it would be natural to fix the origin of the JW string to a point on the boundary. One can then apply a similar analysis to derive the constraints defining the physical fermionic Hilbert space. In particular, in the presence of a boundary there will generally be some ``unused'' $\hat \Phi$ operators at the edges, which are local degeneracies that can be lifted (one can think of these as decoupled ancilla quibts). Alternatively, one might also consider defining the lattice on more interesting manifolds like the real projective plane, but the subtleties and consistency conditions for defining the transformation on more general manifolds is out of the scope of the present paper; we refer the readers to Refs.\ \onlinecite{CHEN2018234, PhysRevB.100.245127,  PhysRevResearch.2.033527, PhysRevD.102.114502, Clifford} for these interesting discussions.

\section{Symmetries \label{sec:sym}}
In the previous sections we have discussed one prescription for performing a JW  transformation for spinful fermions (e.g., ordinary electrons) on the square lattice, but it may not be apparent why the (perhaps complicated) procedure is necessary. In this section, we fulfill one of our major promises, namely, we discuss how symmetries remain manifest on the bosonic side.

\subsection{Transformation of $\hat \Lambda^{\alpha \beta}$}
As we have seen from Sec.\ \ref{sec:square}, a general term in the bosonized Hamiltonian is composed of three types of operators: $\hat \Theta$, $\hat \Lambda$ and $\hat \Phi$. Heuristically, we can interpret the $\hat \Theta$ operators as the on-site terms, $\hat \Lambda$ as ``connectors'' needed for fermion hopping, and $\hat \Phi$ as the JW string. In the following, we will first show that the two indices in $\hat \Lambda^{\alpha \beta}$ transforms independently as two ${\rm SO}(4)$ vectors.
In this subsection we restrict our attention to operators defined on a single site, and so to reduce cluttering we will suppress the site index.

As a starting point, let us recall the commutation relations between the fermion bilinears and Majorana fermions. With $\alpha \neq \beta$ and $\gamma \neq \delta$, we have
\begin{equation}\begin{split}\label{eq:gamma_SO}
\left[ \hat \theta^{\alpha \beta}_0, \hat \gamma^\gamma \right] 
=&  i  (\delta^{\beta \gamma} \hat \gamma^{\alpha} -  \delta^{\alpha \gamma} \hat \gamma^{\beta} ),
\end{split}\end{equation}
where we used the notation $\hat \theta^{\alpha \beta}_{0} \equiv \frac{i}{2} \hat \gamma^\alpha \hat \gamma^\beta$ defined in Eq. \eqref{eq:theta_0}.
Let $A^{\alpha \beta}$ be a real, $4\times 4$ anti-symmetric matrix. Then $\sum_{\alpha ,\beta =1}^4 A^{\alpha \beta} \hat \theta^{\alpha \beta}_0$ is Hermitian, and we can consider the unitary operator $\hat U_0(A) \equiv e^{- i \sum_{\alpha, \beta} A^{\alpha \beta}  \hat \theta^{\alpha \beta}_0 }$. As discussed in Sec.\ \ref{sec:dot}, $\hat U_0(A)$ could be viewed as an element in ${\rm Spin}(4)$. From Eq.\ \eqref{eq:gamma_SO}, however, we see that
\begin{equation}\begin{split}\label{eq:}
\hat U_0(A) \hat \gamma^\alpha \hat U_0(A)^\dagger = \sum_{\tilde \alpha = 1}^4\left( e^{-2A}\right)^{\alpha \tilde \alpha} \hat \gamma^{\tilde \alpha},
\end{split}\end{equation}
where $e^{-2 A} \in {\rm SO}(4)$, i.e., $\hat \gamma^{\alpha}$ transforms as an ${\rm SO}(4)$ vector under conjugation by a unitary operator generated by the bilinears. 

We now consider the corresponding relations between $\hat \Theta^{\alpha \beta}$ and $\hat \Lambda^{\gamma \delta}$. From definitions (Eqs.\ \eqref{eq:Lambda_rule} and \eqref{eq:Lambda_Phi}), depending on the values of $\alpha, \beta$ and $\gamma$ we have either $[\hat \Theta^{\alpha \beta},\hat \Lambda^{\gamma \delta}] \dot = 0$ or $\{\hat \Theta^{\alpha \beta},\hat \Lambda^{\gamma \delta}\} \dot = 0$. Note that the relation is independent of $\delta$. 
Furthermore, when they anti-commute, say when $\beta = \gamma$, we have
$[\hat \Theta^{\alpha \beta},\hat \Lambda^{\beta \delta}] 
\dot =2 \hat \Theta^{\alpha \beta}\hat \Lambda^{\beta \delta} = 2 \sqrt{\hat \Gamma} \hat \Lambda^{\alpha \delta}$. 
Combining these observations with the relation $\hat \Theta^{\alpha \beta} \dot = 2 \hat \theta^{\alpha \beta}$ in Eq. \eqref{eq:Gamma_odd}, one finds
\begin{equation}\begin{split}\label{eq:}
\left[ \hat \theta^{\alpha \beta},\hat \Lambda^{\gamma \delta} \right]  \dot =
i \left( \delta^{\beta \gamma} \hat \Lambda^{\alpha \delta} - 
\delta^{\alpha \gamma} \hat \Lambda^{\beta \delta}\right),
\end{split}\end{equation}
which is in correspondence with Eq.\ \eqref{eq:gamma_SO}. This allows us to conclude that, when restricted to the $\Gamma_-$ subspace, the first index in $\hat \Lambda^{\alpha \beta}$ transforms as an ${\rm SO}(4)$ vector under conjugation by $\hat U(A) \equiv e^{- i \sum_{\alpha, \beta} A^{\alpha \beta}  \hat \theta^{\alpha \beta} }$, i.e., 
\begin{equation}\begin{split}\label{eq:}
\hat U(A) \hat \Lambda^{\alpha \beta} \hat U(A)^\dagger \dot = \sum_{\tilde \alpha= 1}^4\left( e^{-2A}\right)^{\alpha  \tilde \alpha} \hat \Lambda^{\tilde \alpha \beta}.
\end{split}\end{equation}

Similarly, we consider how $\hat \Lambda^{\alpha \beta}$ transforms under conjugation by 
\begin{equation}\begin{split}\label{eq:}
\hat V(A) \equiv  e^{- i \sum_{\alpha, \beta} A^{\alpha \beta}  \hat \phi^{\alpha \beta} },
\end{split}\end{equation}
where $\hat \phi^{\alpha \beta}$ are the parton bilinears associated with $\hat \Phi^{\alpha \beta}$ (Appendix \ref{app:RoP}).
To this end, we first notice that 
\begin{equation}\begin{split}\label{eq:}
\hat \Lambda^{\alpha \beta} = \sqrt{\hat \Gamma} \hat \Lambda^{\alpha \alpha} \hat \Phi^{\alpha \beta},
\end{split}\end{equation}
which implies $\hat \Lambda^{\alpha \beta} \hat \Phi^{\beta \gamma} = \sqrt{\hat \Gamma}^\dagger \hat \Lambda^{\alpha \gamma}$ when combined with Eq.\ \eqref{eq:Phi_prod}. The same analysis as in the preceding paragraphs gives
\begin{equation}\begin{split}\label{eq:}
[\hat \phi^{\alpha \beta}, \hat \Lambda^{\gamma \delta}] \dot = i \delta^{\alpha \delta} \hat \Lambda^{\gamma \beta} - i \delta^{\beta \delta} \hat \Lambda^{\gamma \alpha},
\end{split}\end{equation}
and correspondingly 
\begin{equation}\begin{split}\label{eq:Lambda_right}
\hat V(A) \hat \Lambda^{\alpha \beta} \hat V(A)^\dagger \dot = \sum_{\tilde \beta= 1}^4\hat \Lambda^{ \alpha \tilde \beta} \left( e^{-2A}\right)^{\tilde\beta  \beta} .
\end{split}\end{equation}
This completes our characterization on the transformation properties of $\hat \Lambda^{\alpha \beta}$.

Lastly, we note that since $[\hat \Theta^{\alpha \beta}, \hat \Phi^{\gamma \delta}] = 0$, we have
\begin{equation}\begin{split}\label{eq:}
\hat U(A) \hat \Phi^{\alpha \beta} \hat U(A)^\dagger \dot = \hat \Phi^{\alpha \beta}; ~~~
\hat V(A) \hat \Theta^{\alpha \beta} \hat V(A)^\dagger \dot = \hat \Theta^{\alpha \beta}.
\end{split}\end{equation}

\subsection{Internal unitary}
Let us consider an internal unitary symmetry in the original fermion problem. For instance, a global spin rotation about an axis $\vechat n$ by angle $\varphi$ is given by $\hat U_0(\varphi, \vechat n)  \equiv e^{- i  \varphi  \sum_{\vec r}\vechat n \cdot \hat{\vec \sigma}_{0; \vec r}}$, where the generators are defined in Eq.\ \eqref{eq:fermion_bare} (recall we have absorbed the $1/2$ factor from the spin-half nature of the electrons into the definition of $ \hat{\vec \sigma}_0$). Alternatively, one may also consider unitary symmetries generated by $\hat{\vec \chi}_0$, which do not preserve particle number, as in the discussion of $\eta$-pairing states \cite{PhysRevLett.63.2144}. From our discussion in Sec.\ \ref{sec:dot}, on the bosonic side the symmetry operators can be obtained by a mere replacement of the generators. For instance, spin rotation is now given by $\hat U_0 \dot \mapsto \hat U$, where
\begin{equation}\begin{split}\label{eq:}
\hat U(\varphi, \vechat n)  \equiv e^{- i \varphi \sum_{\vec r} \vechat n \cdot \hat{\vec \sigma}_{\vec r}},
\end{split}\end{equation}
where $\hat{\vec \sigma}_{\vec r}$ is defined in terms of the partons in Eq.\ \eqref{eq:sigma_chi}. 
More generally, we can consider any on-site unitary $ \hat U(A)$ parameterize by an anti-symmetric real matrix $A$, as defined in the previous subsection (but with a slight abuse of notation as we now sum over the generators on all sites).

Our construction guarantees that all the on-site terms transform as they should on the bosonic side. The discussion in the preceding subsection shows that the inter-site terms, which invoke the $\hat \Lambda$ and $\hat \Phi$ operators, also transform naturally. To be more explicit, let us consider the correspondence
\begin{equation}\begin{split}\label{eq:}
i \hat \gamma^{\alpha}_{\vec r} \hat \gamma^{\beta}_{\vec r'}
\dot \longleftrightarrow 
\pm \hat \Lambda^{\alpha *}_{\vec r} \hat \Phi \hat \Phi \cdots \hat \Phi \hat \Lambda^{\beta *}_{\vec r'},
\end{split}\end{equation}
where the overall sign, as well as the omitted indices, could be determined by the prescription provided in Sec.\ \ref{sec:square} \footnote{
For simplicity, we have omitted here the possibility that the global fermion parity $\hat P$ appears on the right-hand side, which would be the case if the chosen straightened arrow passes through the ``decorated link'' we fixed. In any case, since $\hat P$ is invariant under conjugation by $\hat U(A)$ our discussion is unaffected even if it is present.}. Importantly, the sign and the indices are determined solely by the choice of the straightened arrow connect $\vec r'$ to $\vec r'$, and are independent of the values of $\alpha, \beta$. As such, under an on-site unitary transformation, we have
\begin{equation}\begin{split}\label{eq:}
&\hat U_0(A) \left(i \hat \gamma^{\alpha}_{\vec r} \hat \gamma^{\beta}_{\vec r'} \right) \hat U_0(A)^\dagger \\
= & \sum_{\tilde \alpha, \tilde \beta =1}^4 
\left( e^{-2A}\right)^{\alpha \tilde \alpha} 
\left( e^{-2A}\right)^{\beta \tilde \beta} 
(i \hat \gamma^{\tilde \alpha}_{\vec r} \hat \gamma^{\tilde \beta}_{\vec r'})\\
\dot \longleftrightarrow & \sum_{\tilde \alpha, \tilde \beta =1}^4 
\left( e^{-2A}\right)^{\alpha \tilde \alpha} 
\left( e^{-2A}\right)^{\beta \tilde \beta} 
\left( 
\pm \hat \Lambda^{\tilde \alpha *}_{\vec r} \hat \Phi \hat \Phi \cdots \hat \Phi \hat \Lambda^{\tilde\beta *}_{\vec r'} 
\right)\\
\dot =& 
\hat U(A) \left( \pm \hat \Lambda^{\alpha *}_{\vec r} \hat \Phi \hat \Phi \cdots \hat \Phi \hat \Lambda^{\beta *}_{\vec r'}\right) \hat U(A)^\dagger,
\end{split}\end{equation}
verifying the claim.

\subsection{Internal anti-unitary: particle-hole and time-reversal}
Next, we show that internal anti-unitary symmetries are also manifest in the bosonized description. To achieve this goal it suffices to consider a single ``reference'' anti-unitary transformation, since any other anti-unitary symmetry can be decomposed into the product of the reference one and another unitary transformation.
To this end let us consider the anti-unitary particle-hole transformation 
\begin{equation}\begin{split}\label{eq:}
\hat \Xi_0: ~ i \mapsto -i; ~~~
\hat f^{\sigma}_{\vec r} \mapsto \hat f^{\sigma \dagger}_{\vec r};~~~
\hat f^{\sigma \dagger}_{\vec r} \mapsto \hat f^{\sigma}_{\vec r},
\end{split}\end{equation}
for $\sigma = \uparrow, \downarrow$. This transformation is particularly simple in the Majorana basis, for it maps $\hat \Xi_0: \hat \gamma^\alpha_{\vec r} \mapsto \hat \gamma^{\alpha}_{\vec r}$, and as such it flips the sign of any Hermitian fermion bilinear:
\begin{equation}\begin{split}\label{eq:}
\hat \Xi_0: i \hat \gamma^\alpha_{\vec r} \hat \gamma^\beta_{\vec r'} \mapsto - i
 \hat \gamma^\alpha_{\vec r} \hat \gamma^\beta_{\vec r'}.
\end{split}\end{equation}
$\hat \Xi_0$ maps the vacuum to the fully filled state, does not conserve the particle number, and is rarely a good symmetry in a physical system unless the system is at half-filling and all the single-particle terms $\sim i \hat \gamma^\alpha_{\vec r} \hat \gamma^\beta_{\vec r'}$ vanish identically. These criteria, however, are fulfilled in the half-filled lowest-Landau level problem and the particle-hole symmetry $\hat \Xi_0$ has played a prominent role there. Generally, it is known that for a system defined by $N$ complex fermion modes, we have $\hat \Xi_0^2 = (-1)^{N (N-1)/2}$ \cite{PhysRevX.5.031027}. In our problem, since we always consider two complex fermions per site, we have $N=2V$ and so $\hat \Xi_0^2 = (-1)^{V}$, where $V$ is the total number of sites. Furthermore, since the total number of complex fermion modes is even, we also conclude $\hat \Xi_0: \hat P \mapsto \hat P$.

On the parton side, it is natural to also consider a similar anti-unitary transformation which acts by 
\begin{equation}\begin{split}\label{eq:K_def}
\hat K:
\hat c_{\vec r} \leftrightarrow \hat c_{\vec r}^\dagger; ~~~
\hat h_{\vec r} \leftrightarrow \hat h_{\vec r}^\dagger; ~~~
\hat u_{\vec r} \leftrightarrow \hat u_{\vec r}^\dagger; ~~~
\hat d_{\vec r} \leftrightarrow \hat d_{\vec r}^\dagger.
\end{split}\end{equation}
From the preceding discussions, we can immediately conclude $\hat K: \hat \theta_{\vec r}^{\alpha \beta} \leftrightarrow - \hat \theta_{\vec r}^{\alpha \beta}$, and so from the ``root-of-parity'' form in Eq.\ \eqref{eq:h_map} we have
\begin{equation}\begin{split}\label{eq:}
\hat K: \hat \Theta^{\alpha \beta}_{\vec r}
= \sqrt{\hat \Gamma} e^{-i \pi  \hat \theta^{\alpha \beta}_{\vec r}} 
 \mapsto \sqrt{\hat \Gamma}^\dagger e^{i \pi (- \hat \theta^{\alpha \beta}_{\vec r})} \dot = -  \hat \Theta^{\alpha \beta}_{\vec r},
\end{split}\end{equation}
agreeing with the action of $\hat \Xi_0$ on $i \hat \gamma^{\alpha }_{\vec r} \hat \gamma^{\beta}_{\vec r}$. Problem arises, however, when we consider inter-site hopping terms. Since both the $\hat \Lambda$ and $\hat \Phi$ operators can be expressed in the root-of-parity form (Appendix \ref{app:RoP}), we have, for instance,
\begin{equation}\begin{split}\label{eq:}
\hat K: \hat \Lambda^{44}_{\vec r} \hat \Lambda^{33}_{\vec r + \vechat x}
~\dot \mapsto ~\hat \Lambda^{44}_{\vec r} \hat \Lambda^{33}_{\vec r + \vechat x},
\end{split}\end{equation}
which is inconsistent with the transformation of $\hat \Xi_0: i \hat \gamma^{4}_{\vec r} \hat \gamma^{3}_{\vec r + \vechat x} \mapsto -i \hat \gamma^{4}_{\vec r} \hat \gamma^{3}_{\vec r + \vechat x} $.

We now claim the problem can be remedied by defining instead 
\begin{equation}\begin{split}\label{eq:}
\hat \Xi \equiv e^{- i \pi \sum_{\vec r} \hat \phi_{\vec r}^{13}} \hat K ~\dot \propto~ \left( \prod_{\vec r} \hat \Phi^{13}_{\vec r}\right) \hat K.
\end{split}\end{equation}
First, since $\hat \Phi$ commutes with $\hat \Theta$, we retain the property $\hat \Xi: \hat \Theta^{\alpha \beta}_{\vec r} \mapsto - \hat \Theta^{\alpha \beta}_{\vec r}$.
Next, we notice that any nearest-neighbor link either takes the form $\hat \Lambda^{11}\hat \Lambda^{22}$ or $\hat \Lambda^{33}\hat \Lambda^{44}$, and precisely one of the two $\hat \Lambda$ operators will be negated upon conjugating by $\hat \Phi^{13}$. This therefore gives the required sign change 
\begin{equation}\begin{split}\label{eq:}
\hat \Xi: \hat \Lambda^{44}_{\vec r} \hat \Lambda^{33}_{\vec r + \vechat x} \mapsto& - 
\hat \Lambda^{44}_{\vec r} \hat \Lambda^{33}_{\vec r + \vechat x} ;\\
\hat \Lambda^{22}_{\vec r} \hat \Lambda^{11}_{\vec r + \vechat y} \mapsto& - 
\hat \Lambda^{22}_{\vec r} \hat \Lambda^{11}_{\vec r + \vechat y}.
\end{split}\end{equation}
Lastly, the global fermion parity $\hat P = \prod_{\vec r} \Theta^{12}_{\vec r}\Theta^{34}_{\vec r}$ is also invariant under $\hat \Xi$. This demonstrates all the generators, and hence all elements, of the fermion-even algebra transforms as expected under $\hat \Xi$ 
\footnote{
In fact, we can claim something slightly stronger: with the prescription in Sec.\ \ref{sec:global} we have the identification $\hat \gamma_{\vec 0}^1 \dot \longleftrightarrow \hat \Lambda^{11}_{\vec 0}$,  and $\Lambda^{11}_{\vec 0}$ also transforms as it should under our definition of $\hat \Xi$.}

We have seen that, in order to recover the correct transformation for all the generators of the operator algebra, we need to append an on-site unitary $e^{- i \pi \hat \phi^{13}_{\vec r}}$ to the complex conjugation $\hat K$ defined in Eq.\ \eqref{eq:K_def}. It is instructive to compute how the partons transform under $\hat \Xi$, which can be conveniently done by first decomposing the complex partons into Majorana fermions, and then interpreting the action of  $e^{- i \pi \hat \phi^{13}_{\vec r}}$ as a root of parity (Appendix \ref{app:RoP}).
The upshot is 
\begin{equation}\begin{split}\label{eq:Xi_Dirac}
\hat \Xi: \hat c_{\vec r} \mapsto - \hat h_{\vec r};& ~~~ \hat h_{\vec r} \mapsto \hat c_{\vec r};\\
\hat u_{\vec r} \mapsto - \hat d_{\vec r};& ~~~ \hat d_{\vec r} \mapsto \hat u_{\vec r},
\end{split}\end{equation}
which takes the famous ``Dirac form'' as is proposed in the half-filled lowest Landau level context \cite{PhysRevX.5.031027}. In addition, one can compute $\hat \Xi^2$ explicitly as
\begin{equation}\begin{split}\label{eq:}
\hat \Xi^2 = e^{-i 2 \pi \sum_{\vec r} \hat \phi^{13}} = \prod_{\vec r} \hat \Gamma_{\vec r} \dot= (-1)^V,
\end{split}\end{equation}
which agrees with the relation on the fermionic side $\hat \Xi_0^2 = (-1)^V$. Furthermore, before restricting to the physical $\Gamma_-$ subspace we have $\hat \Xi^2 = \prod_{\vec r} \hat \Gamma_{\vec r}$, which is the simply the total parton parity as is anticipated from the Dirac form in Eq.\ \eqref{eq:Xi_Dirac}.

For completeness, we also consider explicitly the time-reversal symmetry on the original fermions. Consider the anti-unitary operator
\begin{equation}\begin{split}\label{eq:}
\hat {\mathcal T_0}: 
\hat f^{\uparrow} \mapsto f^{\downarrow};~~~
\hat f^{\downarrow} \mapsto -f^{\uparrow}.
\end{split}\end{equation}
We can decompose it as the product of $\hat \Xi_0$ and a unitary transformation:
\begin{equation}\begin{split}\label{eq:}
\hat {\mathcal T_0} = e^{-i \pi \sum_{\vec r}\hat \chi_{0;\vec r}^2}\,\hat \Xi_0,
\end{split}\end{equation}
where the charge rotation generated by $\hat \chi_{0;\vec r}^2$ can be rationalized as undoing the charge conjugation implemented by $\hat \Xi_0$. 
We can then infer its bosonized representation 
\begin{equation}\begin{split}\label{eq:}
\hat {\mathcal T} =& e^{-i \pi \sum_{\vec r}\hat \chi_{\vec r}^2}\,\hat \Xi
= \prod_{\vec r} e^{-i \pi \left (\frac{\hat \theta^{24}_{\vec r}-\hat \theta^{13}_{\vec r}}{2} + \hat \phi^{13}_{\vec r} \right)}  \hat K,
\end{split}\end{equation}
and $\hat {\mathcal T}^2 = \hat P$, as it should.

\subsection{Spatial}
Finally, we address how spatial symmetries are represented in the bosonized description, assuming the system is placed on an even-by-even torus of size $L_x\times L_y$ and  with the prescription in Eq.\ \eqref{eq:P_insert}. We will address the generators of the space group, translations $T_{x,y}$, four-fold rotation $C_4$ about the origin, and a mirror $M_x$ about the $y=0$ line \footnote{One can also generate $T_y$ using $C_4$ and $T_x$, but we keep it here as a sanity check.}.
First, consider a translation by one unit along the $\vechat y$ direction, which acts on the fermions by 
\begin{equation}\begin{split}\label{eq:}
\hat T_{\vechat y;0} \hat \gamma^\alpha_{\vec r}\hat T_{\vechat y;0}^{-1} = 
 \hat \gamma^\alpha_{\vec r + \vechat y}.
\end{split}\end{equation}
On the bosonic side, we could as well consider a ``bare'' translation which performs the site permutation $\hat {T}_{\vechat y}^\flat: \vec r \mapsto \vec r + \vechat y$ (on a torus, we map $(x,L_y-1) \mapsto (x,0)$). Yet, this action does not generally leave the bosonized Hamiltonian invariant even if the fermionic one is translation-invariant. As an example, consider the Hamiltonian
\begin{equation}\begin{split}\label{eq:}
\sum_{\vec r} (i \hat \gamma^{2}_{\vec r - \vechat y} \hat \gamma^{1}_{\vec r})
\dot \longleftrightarrow
\sum_{\vec r \neq \vec 0} \hat \Lambda^{22}_{\vec r - \vechat y} \hat \Lambda^{11}_{\vec r } + \hat P
 \hat \Lambda^{22}_{-\vechat y} \hat \Lambda^{11}_{\vec 0}.
\end{split}\end{equation}
While the left-hand side is invariant under conjugation by $\hat T_{\vec y; 0}$, under the bare translation $\hat { T}_{\vechat y}^\flat$ the right-hand side is modified by 
\begin{equation}\begin{split}\label{eq:}
\delta \hat H = 
( \hat \Lambda^{22}_{-\vechat y} \hat \Lambda^{11}_{\vec 0}
+\hat P \hat \Lambda^{22}_{\vec 0} \hat \Lambda^{11}_{\vechat y}
) - (\hat P \hat \Lambda^{22}_{-\vechat y} \hat \Lambda^{11}_{\vec 0}
+ \hat \Lambda^{22}_{\vec 0} \hat \Lambda^{11}_{\vechat y}).
\end{split}\end{equation}
Such mismatch could be understood as arising from the movement of the decorated link: the bare translation on the bosonic side also moves the decoration from the link between $-\vechat y$ and $\vec 0$ to that between $\vec 0 $ and $\vechat y$. As discussed in Sec.\ \ref{sec:global}, such movement can be undone by $\hat {\mathcal M}(\hat \Phi^{12}_{\vec 0})$, and therefore we identify
\begin{equation}\begin{split}\label{eq:}
\hat T_{\vechat y} = \hat {\mathcal M}(\hat \Phi^{12}_{\vec 0}) \hat { T}_{\vechat y}^\flat.
\end{split}\end{equation}
Similarly, translation in the $\vechat x$ direction is given by 
\begin{equation}\begin{split}\label{eq:}
\hat T_{\vechat x} = \hat {\mathcal M}(i  \hat \Phi^{13}_{\vechat x} \hat \Phi^{41}_{\vec 0}) \hat { T}_{\vechat x}^\flat.
\end{split}\end{equation}
To appreciate this, let us consider the action of $\hat T_{\vec x}$ on $\hat \gamma^{1}_{\vec 0} \dot \longleftrightarrow\hat \Lambda^{11}_{\vec 0}$. One can evaluate $\hat T_{\vechat x} \hat \Lambda^{11}_{\vec 0} \hat T_{\vechat x}^{-1}$ readily using Eq.\ \eqref{eq:Phi_change}, and the result agrees with 
\begin{equation}\begin{split}\label{eq:}
\hat \gamma_{\vechat x}^{1} =& (-i)^3
(i \hat \gamma_{\vechat x}^{1}  \hat \gamma_{\vechat x}^{3} )
(i \hat \gamma_{\vechat x}^{3}  \hat \gamma_{\vec 0}^{4} )
(i  \hat \gamma_{\vec 0}^{4}  \hat \gamma_{\vec 0}^{1})
 \hat \gamma_{\vec 0}^{1}\\
\dot \mapsto & (-i)^3 \hat \Theta^{13}_{\vechat x} (- \hat \Lambda^{33}_{\vechat x} \hat \Lambda^{44}_{\vec 0} ) (\hat \Theta^{41}_{\vec 0}) \hat \Lambda^{11}_{\vec 0}
\dot=  -i \hat \Lambda^{11}_{\vechat x} \hat \Phi^{13}_{\vechat x} \hat \Phi^{41}_{\vec 0}.
\end{split}\end{equation}

Next, we consider the counter-clockwise $C_4$ rotation about the origin. Here, we suppose the rotation is ``spinless,'' in that it acts simply as a permutation of the sites,
\begin{equation}\begin{split}\label{eq:}
\hat C_{4;0} \hat \gamma_{\vec r}^\alpha \hat C_{4;0}^{-1} = \hat \gamma_{C_4\vec r}^\alpha,
\end{split}\end{equation}
and does not act on the internal index $\alpha$ \footnote{
In general, a spatial rotation symmetry should also act on the spin index. In our notation, one can construct such an operator by combing the ``spinless'' $C_4$ together with the corresponding internal unitary action $e^{- i \frac{\pi}{4} \hat \sigma^3 }$. Alternatively, in a system with full ${\rm SU}(2)$ spin rotation symmetry the ``spinless'' rotation considered here can also be a symmetry of the system.
}.
While the action of the bare rotation $\hat C_{4}^\flat$ is consistent with the JW transformation of the on-site operators, viz., $\hat C_{4}^\flat \hat \Theta^{\alpha \beta}_{\vec r}(\hat C_{4}^{\flat})^{-1} = \hat \Theta^{\alpha \beta}_{C_4 \vec r}$, problems again arise when we consider the inter-site processes. For instance, consider the identification $i \hat \gamma^{4}_{\vec r} \hat \gamma^{3}_{\vec r+\vechat x} \dot \longleftrightarrow \hat \Lambda^{44}_{\vec r} \hat \Lambda^{33}_{\vec r+\vechat x}$. Under the bare rotation we find
\begin{equation}\begin{split}\label{eq:}
\hat C_{4}^\flat (\hat \Lambda^{44}_{\vec r} \hat \Lambda^{33}_{\vec r+\vechat x})
(\hat C_{4}^{\flat})^{-1} 
= \hat \Lambda^{44}_{C_4 \vec r} \hat \Lambda^{33}_{C_4 \vec r+\vechat y},
\end{split}\end{equation}
which should be contrasted with 
\begin{equation}\begin{split}\label{eq:}
\hat C_{4;0} (i \hat \gamma^{4}_{\vec r} \hat \gamma^{3}_{\vec r+\vechat x})
\hat C_{4;0}^{-1} 
= 
i \hat \gamma^{4}_{C_4 \vec r} \hat \gamma^{3}_{C_4 \vec r+\vechat y}
\dot \longleftrightarrow 
\hat \Lambda^{42}_{C_4 \vec r} \hat \Lambda^{31}_{C_4 \vec r+\vechat y}.
\end{split}\end{equation}
We see that, in order for the $C_4$ rotation to be compatible with the JW transformation, we have to combine $\hat C_4^\flat$ with a suitable rotation in the second index of the $\hat \Lambda$ operators. 
By inspecting the action of $C_4$ on the various nearest-neighbor links, we see that the following transformation will be compatible with the JW transformation:
\begin{equation}\begin{split}\label{eq:Lambda_C4}
\left(
\begin{array}{cccc}
\hat \Lambda^{\alpha 1}_{\vec r} & 
\hat \Lambda^{\alpha 2}_{\vec r} & 
\hat \Lambda^{\alpha 3}_{\vec r} & 
\hat \Lambda^{\alpha 4}_{\vec r} 
\end{array}
\right)
\overset{C_4}
\mapsto 
\left(
\begin{array}{cccc}
- \hat \Lambda^{\alpha 4}_{C_4\vec r} & 
\hat \Lambda^{\alpha 3}_{C_4\vec r} & 
\hat \Lambda^{\alpha 1}_{C_4\vec r} & 
\hat \Lambda^{\alpha 2}_{C_4\vec r} 
\end{array}
\right).
\end{split}\end{equation}
As discussed in Eq.\ \eqref{eq:Lambda_right}, such rotations are generated by $\hat \phi$ with a suitable real anti-symmetric matrix $A$, which can be solved readily. Furthermore, similar to the discussions for $\hat T_x$ and $\hat T_y$, the moving of the decorated link can be accounted for by analyzing the action of $C_4$ on  $\hat \lambda_{\vec 0}^1 \dot \longleftrightarrow \hat \Lambda^{11}_{\vec 0}$. Altogether, we find
\begin{equation}\begin{split}\label{eq:}
\hat C_4 =& \hat {\mathcal M}(i \hat \Phi^{41}_{\vec 0}) \hat V_{C_4} \hat C_4^\flat;\\
\hat V_{C_4} =& e^{  -i \frac{\pi}{4} \sum_{\vec r}\left( \hat \phi_{\vec r}^{12}  + \hat \phi_{\vec r}^{34}-\sqrt{2} (\hat \phi_{\vec r}^{13}+ \hat \phi_{\vec r}^{24} +\hat \phi_{\vec r}^{14} - \hat \phi_{\vec r}^{23}  )\right)}.
\end{split}\end{equation}

Lastly, we consider the mirror symmetry $M_x$ which flips the $x$ axis, i.e., $M_x(x,y) = (-x,y)$. Repeating the same analysis as we have done for $C_4$, we require
\begin{equation}\begin{split}\label{eq:Lambda_Mx}
\left(
\begin{array}{cccc}
\hat \Lambda^{\alpha 1}_{\vec r} & 
\hat \Lambda^{\alpha 2}_{\vec r} & 
\hat \Lambda^{\alpha 3}_{\vec r} & 
\hat \Lambda^{\alpha 4}_{\vec r} 
\end{array}
\right)
\overset{M_x}
\mapsto 
\left(
\begin{array}{cccc}
\hat \Lambda^{\alpha 1}_{M_x \vec r} & 
\hat \Lambda^{\alpha 2}_{M_x \vec r} & 
\hat \Lambda^{\alpha 4}_{M_x \vec r} & 
-\hat \Lambda^{\alpha 3}_{M_x \vec r} 
\end{array}
\right),
\end{split}\end{equation}
which can be achieved with
\begin{equation}\begin{split}\label{eq:}
\hat M_x =  \hat V_{M_x} \hat M_x^\flat;~~~
\hat V_{M_x} =  e^{  -i \frac{\pi}{2} \sum_{\vec r}  \hat \phi^{34}_{\vec r} }.
\end{split}\end{equation}
Notice that $M_x$ does not move the decorated link and so $\hat M_x$ does not invoke the $\hat {\mathcal M}$ operator.

While we have identified the space group generators in the bosonized representation, the appearance of the $\hat {\mathcal M}$ operators above naturally leads one to wonder if the symmetries are now represented projectively. To this end, let us first compute 
\begin{equation}\begin{split}\label{eq:}
\hat T_{\vechat y}^{L_y} = &
\hat {\mathcal M}(\hat \Phi^{12}_{\vec 0} \hat \Phi^{12}_{\vechat y} \cdots \hat \Phi^{12}_{(L_y-1)\vechat y})  (\hat { T}_{\vechat y}^\flat)^{L_y} =  
\hat {\mathcal M}(\hat {\mathcal V}_{\vec 0}).
\end{split}\end{equation}
Since $\hat {\mathcal V}_{\vec 0} \overset{c}{=} - \hat P$, we have
\begin{equation}\begin{split}\label{eq:}
\hat {\mathcal M}(\hat {\mathcal V}_{\vec 0}) \overset{c}{=} \hat P_+ + \hat P_- (- \hat P) = 1,
\end{split}\end{equation}
i.e., $\hat T_{\vechat y}^{L_y} \overset{c}{=} 1$. Similarly, one can also check that
\begin{equation}\begin{split}\label{eq:}
\hat T_{\vechat x}^{L_x} = &\hat {\mathcal M}(- \hat {\mathcal H}_{\vec 0}) \overset{c}{=} 1.
\end{split}\end{equation}
Next, we consider the point group symmetries. Consider
\begin{equation}\begin{split}\label{eq:}
\hat C_4^4=& \left[ \hat {\mathcal M}(i \hat \Phi^{41}_{\vec 0}) \hat V_{C_4} \hat C_4^\flat\right]^4\\
=& \hat {\mathcal M} \left( 
\hat \Phi^{41}_{\vec 0}
(\hat V_{C_4} \hat \Phi^{41}_{\vec 0} \hat V_{C_4}^\dagger)
(\hat V_{C_4}^2 \hat \Phi^{41}_{\vec 0} \hat V_{C_4}^{\dagger 2})
(\hat V_{C_4}^3 \hat \Phi^{41}_{\vec 0} \hat V_{C_4}^{\dagger 3})
\right) \hat V_{C_4}^4.
\end{split}\end{equation}
We first note that
\begin{equation}\begin{split}\label{eq:}
(\hat V_{C_4})^4 = \prod_{\vec r} ( \hat \Gamma_{\vec r} \hat \Phi_{\vec r}^{12} \hat \Phi_{\vec r}^{34} ) \dot = \prod_{\vec r} \hat \Theta^{12} \hat \Theta^{34} = \hat P
\end{split}\end{equation}
This follows from observing that $ [\hat \phi_{\vec r}^{12}  + \hat \phi_{\vec r}^{34}, \sqrt{2} (\hat \phi_{\vec r}^{13} + \hat \phi_{\vec r}^{24} +\hat \phi_{\vec r}^{14} - \hat \phi_{\vec r}^{23} ] = 0$ \footnote{In case this may appear out of the blue, we comment that the two parts are actually generators for different ${\rm SU}(2)$ factors when we view the unitary as an element of ${\rm Spin}(4)$, c.f. the corresponding definitions for $\hat \theta$ in Eq.\ \eqref{eq:Spin_trans}.}, and the eigenvalues of $\sqrt{2} (\hat \phi_{\vec r}^{13}+ \hat \phi_{\vec r}^{24}  +\hat \phi_{\vec r}^{14} - \hat \phi_{\vec r}^{23} $ are $0$ and $\pm 2$. These combined give $ \hat V_{C_4}^4 =  \prod_{\vec r} e^{- i \pi (\hat \phi_{\vec r}^{12}  + \hat \phi_{\vec r}^{34})} $. The second equality follows from the definition $\hat \Phi^{\alpha \beta} = \hat \Gamma_{\vec r} \hat \Lambda^{\alpha \alpha}_{\vec r} \hat \Theta^{\alpha \beta}_{\vec r} \hat \Lambda^{\beta \beta}_{\vec r}$ together with Eq.\ \eqref{eq:Lambda_fuse}, which implies $\hat \Gamma_{\vec r} \hat \Lambda^{11}_{\vec r} \hat \Lambda^{22}_{\vec r} \hat \Lambda^{33}_{\vec r}\hat \Lambda^{44}_{\vec r} \dot= 1$. To evaluate the contribution from the $\hat {\mathcal M}$ operator, we write $\hat \Phi^{14}_{\vec 0} \dot - i \hat \Lambda^{11}_{\vec 0}\hat \Lambda^{14}_{\vec 0}$, and so using Eq.\ \eqref{eq:Lambda_C4} we have 
\begin{equation}\begin{split}\label{eq:}
\hat V_{C_4}
\hat \Phi^{14}_{\vec 0}  \hat V_{C_4}^\dagger \dot = - i (-\hat \Lambda^{14}_{\vec 0}\hat \Lambda^{12}_{\vec 0}).
\end{split}\end{equation}
We can then conclude
\begin{equation}\begin{split}\label{eq:}
& \hat {\mathcal M} \left( 
\hat \Phi^{41}_{\vec 0}
(\hat V_{C_4} \hat \Phi^{41}_{\vec 0} \hat V_{C_4}^\dagger)
(\hat V_{C_4}^2 \hat \Phi^{41}_{\vec 0} \hat V_{C_4}^{\dagger 2})
(\hat V_{C_4}^3 \hat \Phi^{41}_{\vec 0} \hat V_{C_4}^{\dagger 3})
\right)\\
\dot = &
\hat {\mathcal M} \left( 
(\hat \Lambda^{11}_{\vec 0}\hat \Lambda^{14}_{\vec 0})
(-\hat \Lambda^{14}_{\vec 0}\hat \Lambda^{12}_{\vec 0})
(-\hat \Lambda^{12}_{\vec 0}\hat \Lambda^{13}_{\vec 0})
(-\hat \Lambda^{13}_{\vec 0}\hat \Lambda^{11}_{\vec 0})
\right)\\
\dot = & \hat {\mathcal M}(-1),
\end{split}\end{equation}
which equals to $\hat P$. This shows $\hat C_4^4 = 1$.
Similarly, for $\hat M_x$ we find
\begin{equation}\begin{split}\label{eq:}
\hat M_x^2 =  \hat V_{M_x}^2 =  e^{  -i \pi\sum_{\vec r}  \hat \phi^{34}_{\vec r} }
\dot = (-i)^{L_x L_y} \prod_{n=0}^{L_y} \hat {\mathcal H}_{n \vechat y}
\overset{c}{=} 1,
\end{split}\end{equation}
where we have used the assumption that both $L_x$ and $L_y$ are even, as well as $\hat {\mathcal H}_{\vec r} \overset{c}{=} -1$.

Lastly, we verify the relations between the space group generators are preserved. Consider
\begin{equation}\begin{split}\label{eq:}
&\hat T_x \hat T_y\hat T_x^{-1} \hat T_y^{-1}\\
= &
\hat {\mathcal M}(i  \hat \Phi^{13}_{\vechat x} \hat \Phi^{41}_{\vec 0})
\hat {\mathcal M}(\hat \Phi^{12}_{\vechat x })
\hat {\mathcal M}(-i  \hat \Phi^{13}_{\vechat x+\vechat y} \hat \Phi^{41}_{\vechat y}) 
\hat {\mathcal M}(\hat \Phi^{12}_{\vec 0 })\\
\dot=& \hat {\mathcal M}(
-\hat \Phi^{24}_{\vec 0}
\hat \Phi^{32}_{\vechat x} 
\hat \Phi^{13}_{\vechat x+\vechat y} 
\hat \Phi^{41}_{\vechat y} 
),
\end{split}\end{equation}
which equals to $1$ using the plaquette constraint Eq.\ \eqref{eq:plaquette}.
By the same token, we can check
\begin{equation}\begin{split}\label{eq:}
\hat C_4 \hat T_x\hat C_4^{-1}
= &
\hat {\mathcal M}(i \hat \Phi^{41}_{\vec 0}) \hat V_{C_4} 
\hat {\mathcal M}(i  \hat \Phi^{13}_{\vechat y} \hat \Phi^{41}_{\vec 0}) 
 \hat V_{C_4} ^\dagger 
\hat {\mathcal M}(-i \hat \Phi^{41}_{\vechat y})
 \hat { T}_{\vechat y}^\flat\\
 = &
\hat {\mathcal M}(\hat \Phi^{12}_{\vec 0} )
 \hat { T}_{\vechat y}^\flat;
\end{split}\end{equation}
\begin{equation}\begin{split}\label{eq:}
\hat M_x \hat T_x\hat M_x^{-1}
= &(\hat { T}_{\vechat x}^\flat)^{-1} 
\hat V_{M_x} 
 \hat {\mathcal M}(i  \hat \Phi^{13}_{\vec 0} \hat \Phi^{41}_{\vechat  x})  \hat V_{M_x} ^\dagger \\
\dot = &(\hat { T}_{\vechat x}^\flat)^{-1} 
 \hat {\mathcal M}(-i  \hat \Phi^{13}_{\vechat x} \hat \Phi^{41}_{\vec 0}), 
\end{split}\end{equation}
i.e., the relations $\hat C_4 \hat T_x\hat C_4^{-1}=\hat T_y$ and $ \hat M_x \hat T_x \hat M_x^{-1} = T_x^{-1}$ are preserved. One can also verify $\hat M_x \hat T_y \hat M_x = \hat T_y$ readily.

\section{Qubit representation \label{sec:qubit}}
In our discussion so far, the JW transformed model is written using the fermionic partons subjected to the $\Gamma_-$ condition, which renders the system bosonic. Given the $\Gamma_-$ site Hilbert space is eight-dimensional, it is also natural to ask how we can express the bosonized model using three qubits per site. This can be achieved readily by first specifying how the eight basis states are reconciled in the two descriptions (partons vs. qubits), and then rewriting the $\Theta$, $\Lambda$ and $\Phi$ operators in the qubit notation. As we will see, the qubit notation is advantageous in that it helps connect the current construction to well-established models of topological orders like the Wen's plaquette model \cite{PhysRevLett.90.016803}. Furthermore, we will show in the next section that the qubit notation allows one to readily pass to, for instance, models of spinless fermions or the $t$-$J$ model with restricted site Hilbert spaces.
For most parts of the current section, we suppress the site index to reduce cluttering.

Let $| 0 \rangle$ be the parton vaccum on a site. The $\Gamma_-$ subspace is spanned by the one-  and three-parton states like $\hat c^\dagger | 0 \rangle$ and $\hat c^\dagger \hat u^\dagger \hat d^\dagger | 0 \rangle$. To convert into the qubit notations, we adopt the following convention:
\begin{equation}\begin{split}\label{eq:qubit_def}
\begin{array}{ll}
|\uparrow \uparrow \uparrow \rangle \equiv \hat c^\dagger | 0 \rangle; & 
|\uparrow \uparrow \downarrow \rangle \equiv \hat c^\dagger \hat u^\dagger \hat d^\dagger | 0 \rangle;\\
|\downarrow \uparrow \uparrow \rangle \equiv \hat h^\dagger | 0 \rangle; & 
|\downarrow \uparrow \downarrow \rangle \equiv \hat h^\dagger \hat u^\dagger \hat d^\dagger | 0 \rangle;\\
| \uparrow \downarrow \uparrow \rangle \equiv -i\, \hat u^\dagger | 0 \rangle; & 
|\uparrow  \downarrow \downarrow \rangle \equiv -i\,  \hat u^\dagger \hat c^\dagger \hat h^\dagger | 0 \rangle;\\
|\downarrow \downarrow \uparrow \rangle \equiv -i \,\hat d^\dagger | 0 \rangle; & 
|\downarrow  \downarrow \downarrow \rangle \equiv -i\, \hat d^\dagger \hat c^\dagger \hat h^\dagger | 0 \rangle.
\end{array}
\end{split}\end{equation}
We will label the qubits from left to right by integers $1$ to $3$, and rewrite the operators in terms of the Pauli operators $\hat X^{(j)}$, $\hat Y^{(j)}$, and $\hat Z^{(j)}$ for $j=1,2,3$. For instance, the state $|\downarrow \uparrow \uparrow \rangle \equiv \hat h^\dagger | 0 \rangle$ is the simultaneous eigenstate of $\hat Z^{(1)}$, $\hat Z^{(2)}$, and $\hat Z^{(3)}$ with eigenvalues $-1$, $+1$, and $+1$ respectively.
In this notation, the on-site terms $\hat \Theta^{\alpha \beta}$ can be rewritten as
\begin{equation}\begin{split}\label{eq:}
\begin{array}{lll}
\hat \Theta^{12} \dot = -\hat Z^{(1)};&
\hat \Theta^{13} \dot = -\hat Y^{(1)}; &
\hat \Theta^{14} \dot = \hat  X^{(1)}\hat Z^{(2)};\\
\hat \Theta^{34} \dot = -\hat  Z^{(1)} \hat Z^{(2)}; & 
\hat \Theta^{24} \dot = \hat Y^{(1)} \hat Z^{(2)}; & 
\hat \Theta^{23} \dot = \hat X^{(1)}.
\end{array}
\end{split}\end{equation}
Notice that the ``third qubit'' is never utilized. This echoes with the observation in Sec.\ \ref{sec:dot} that the single-dot description invokes an additional auxiliary qubit which is locally consumed when we construct the JW transformation on the lattice. More explicitly, following the same procedure one finds
\begin{equation}\begin{split}\label{eq:Phi_qubit}
\begin{array}{lll}
\hat \Phi^{12} \dot = \hat Z^{(3)}; &
\hat \Phi^{13} \dot = \hat Y^{(3)}; &
\hat \Phi^{14} \dot = \hat Z^{(2)}\hat X^{(3)};\\
\hat \Phi^{34} \dot = -\hat Z^{(2)}\hat Z^{(3)}; &
\hat \Phi^{24} \dot = \hat Z^{(2)}\hat Y^{(3)} &
\hat \Phi^{23} \dot = -\hat X^{(3)},
\end{array}
\end{split}\end{equation}
which act trivially on the first qubit. From the expressions above, we can see that $\hat \Theta^{\alpha \beta}$ and $\hat \Phi^{\alpha \beta}$ either act trivially or as $\hat Z^{(2)}$ on the second qubit. This can be rationalized by noticing that the on-site parity operator $\hat \Theta^{12} \hat \Theta^{34} = \hat Z^{(2)}$, which commutes with all of $\hat \Theta^{\alpha \beta}$ and $\hat \Phi^{\alpha \beta}$. In contrast, since the inter-site hopping processes are captured by the $\hat \Lambda^{\alpha \beta}$ operators, they should act like $\hat X^{(2)}$ and $\hat Y^{(2)}$. Indeed, one finds
\begin{widetext}
\begin{equation}\begin{split}\label{eq:Lambda_qubit}
\hat \Lambda^{\alpha \beta} \dot =
\left(
\begin{array}{cccc}
\hat X^{(1)}\hat X^{(2)}\hat X^{(3)} & \hat X^{(1)}\hat X^{(2)}\hat Y^{(3)} & -\hat X^{(1)}\hat X^{(2)}\hat Z^{(3)} & \hat X^{(1)}\hat Y^{(2)}\\
\hat Y^{(1)}\hat X^{(2)}\hat X^{(3)} & \hat Y^{(1)}\hat X^{(2)}\hat Y^{(3)} & -\hat Y^{(1)}\hat X^{(2)}\hat Z^{(3)} & \hat Y^{(1)}\hat Y^{(2)}\\
-\hat Z^{(1)}\hat X^{(2)}\hat X^{(3)} & -\hat Z^{(1)}\hat X^{(2)}\hat Y^{(3)} & \hat Z^{(1)}\hat X^{(2)}\hat Z^{(3)} & -\hat Z^{(1)}\hat Y^{(2)}\\
-\hat Y^{(2)}\hat X^{(3)} & -\hat Y^{(2)}\hat Y^{(3)} & \hat Y^{(2)}\hat Z^{(3)} & \hat X^{(2)}\\
\end{array}
\right)_{\alpha \beta}.
\end{split}\end{equation}
\end{widetext}

To unpack the qubit notation a bit, it is helpful to consider explicitly some physical operators. 
For instance, the single-site number and spin operators are now expressed as
\begin{equation}\begin{split}\label{eq:qubit_ex}
\hat n \dot =& 1 + \hat Z^{(1)} \hat P_+;\\
(\hat S^{x},\hat S^{y},\hat S^{z} ) \dot = & \frac{1}{2} 
(\hat X^{(1)},\hat Y^{(1)},\hat Z^{(1)}) \hat P_-,
\end{split}\end{equation}
where $\hat P_{\pm} \equiv (1 \pm \hat Z^{(2)})/2$ are projectors \footnote{Not to be confused with $\hat \Gamma_\pm$: in the physical Hilbert space we always have $\hat \Gamma \dot \mapsto -1$, but here $\hat P$ denotes the on-site fermion parity, which is dynamical and can take both $\pm 1$ values.}.
This can be interpreted as follows: the second qubit keeps track of the fermion parity on the site, and the first qubit encodes the two states belonging to each of the fixed-parity sectors. More explicitly, in Table \ref{tab:qubit_nSz} we provide the correspondence between the qubits and the physical quantum numbers, which can be readily inferred from the definition in Eq.\ \eqref{eq:qubit_def}. 
\begin{center}
\begin{table}
\caption{Physical quantum numbers associated with the first two qubits $\hat Z^{(1)}$ and $\hat Z^{(2)}$.
\label{tab:qubit_nSz}}
\begin{tabular}{cc|c}
\hline\hline
$\hat Z^{(1)}$ & $\hat Z^{(2)}$ & $(\hat n, \hat S^{z})$\\
\hline
$+1$ & $+1$ & $(2,0)$\\
$-1$ & $+1$ & $(0,0)$\\
$+1$ & $-1$ & $(1, 1/2)$\\
$-1$ & $-1$ & $(1, - 1/2)$\\
\hline \hline
\end{tabular}
\end{table}
\end{center}

Lastly, as shown in Eq.\ \eqref{eq:Phi_qubit} the third qubit is invoked when we consider the $\hat \Phi$ operators. As an example, the on-site unitary attached to the $C_4$ rotation is given by 
\begin{equation}\begin{split}\label{eq:}
\hat V_{C_4} \dot= \hat P_- \exp\left(- i \frac{\pi}{4} \hat Z^{(3)} \right) + \hat P_+ \exp \left(i \frac{\pi}{2} \frac{\hat X^{(3)} + \hat Y^{(3)}}{\sqrt{2}} \right)
\end{split}\end{equation}
The meaning of the last qubit is even more apparent when we consider the plaquette constraints in Eq.\ \eqref{eq:plaquette}:
\begin{equation}\begin{split}\label{eq:plaquette_qubit}
\hat Y^{(3)}_{\vec r }  \hat X^{(3)}_{\vec r + \vechat x} \hat Y^{(3)}_{\vec r + \vechat x+\vechat y} \hat X^{(3)}_{\vec r + \vechat y} \overset{c}{=} \hat Z_{\vec r}^{(2)} \hat Z^{(2)}_{\vec r + \vechat y},
\end{split}\end{equation}
where we have restored the site subscript.
The $YXYX$ term on the left-hand side can be identified with the Wen's plaquette model \cite{PhysRevLett.90.016803}, which is known to be equivalent to Kitaev's toric code \cite{KITAEV20032} and describes the $\mathbb Z_2$ topological order. The right-hand side, however, indicates that we do not restrict ourselves to the ground state of the Wen's plaquette model, since some plaquettes will take value $-1$ depending on the state of the second qubit, i.e., the physical on-site fermion parity. Instead, Eq.\ \eqref{eq:plaquette_qubit} can be interpreted as a flux-attachment prescription \cite{CHEN2018234, PhysRevB.100.245127,  PhysRevResearch.2.033527, PhysRevD.102.114502, Clifford}: the presence of an electron on site $\vec r$ is encoded by $\hat Z^{(2)}_{\vec r} = -1$, which translates into the flipping of the $YXYX$ terms on two adjacent plaquettes. Within the Wen's plaquette model, flipping a pair of adjacent plaquettes creates a pair of  $e$ and $m$ anyons, and this can be interpreted as the creation of a fermionic bound state. 

As a concluding remark, we observe that the explicit emergence of the Wen's plaquette model provides a convenient way to identify the bosonic state corresponding to the fermionic vacuum: when placed on an even-by-even torus, the Wen's plaquette model has a four-fold topological degeneracy \cite{PhysRevLett.90.016803}. The Wilson loop conditions Eq.\ \eqref{eq:VH_new} specify a unique state in the ground-state manifold. This state is identified with the fermionic vacuum, and any other states in the fermionic Hilbert space can be obtained by applying the JW transformed operators.

\section{Explicit square lattice models\label{sec:models}}
Our discussion so far has focused on the JW transformation of the general operators, without reference to any specific Hamiltonian. In this section, let us put the  transformation to use by bosonizing some familiar models defined on the square lattice.

\begin{widetext}
\subsection{Spinful Hubbard model}
Let us take the spinful Hubbard model as a canonical example. We parameterize the Hamiltonian as
\begin{equation}\begin{split}\label{eq:}
\hat H_{\rm H} = & -t  \sum_{\vec r}  \sum_{\substack{\delta \vec r = \vechat x, \vechat y \\ s = \uparrow, \downarrow } }(\hat f^{s \dagger}_{\vec r} \hat f^{s}_{\vec r + \delta \vec r} + {\rm h.c.} ) 
 + U \sum_{\vec r} (\hat n_{\vec r} - 1)^2 - \mu \sum_{\vec r} (\hat n_{\vec r}-1),
\end{split}\end{equation}
where ${\rm h.c.}$ stands for hermitian conjugate.
Since the on-site terms have been addressed in Eqs.\ \eqref{eq:n_map} and \eqref{eq:U_map} already, we focus on the electron hopping term.
First, we rewrite
\begin{equation}\begin{split}\label{eq:hop_rewrite}
 \sum_{s = \uparrow, \downarrow } (\hat f^{s \dagger}_{\vec r} \hat f^{s}_{\vec r'} + {\rm h.c.} )
=\frac{1}{2} \left( 
i \hat \gamma_{\vec r}^2 \hat \gamma_{\vec r'}^1 - i \hat \gamma_{\vec r}^1 \hat \gamma_{\vec r'}^2 
+i \hat \gamma_{\vec r}^4 \hat \gamma_{\vec r'}^3 - i \hat \gamma_{\vec r}^3 \hat \gamma_{\vec r'}^4
\right),
\end{split}\end{equation}
and we can perform JW transformation for each fo the bilinears following the prescription in Sec.\ \ref{sec:square}. This leads to 
\begin{equation}\begin{split}\label{eq:}
\hat H_{\rm H} \dot \mapsto &-\frac{t}{2} \sum_{\vec r} \left( 
\hat \Lambda_{\vec r}^{24}\hat \Lambda_{\vec r+\vechat x}^{13}
-\hat \Lambda_{\vec r}^{14}\hat \Lambda_{\vec r+\vechat x}^{23}
+ \hat \Lambda_{\vec r}^{44}\hat \Lambda_{\vec r+\vechat x}^{33}
-\hat \Lambda_{\vec r}^{34}\hat \Lambda_{\vec r+\vechat x}^{43}
\right) 
- \frac{t}{2}  \sideset{}{'}\sum_{\vec r}\left( 
\hat \Lambda_{\vec r}^{22}\hat \Lambda_{\vec r+\vechat y}^{11}
-\hat \Lambda_{\vec r}^{12}\hat \Lambda_{\vec r+\vechat y}^{21}
+ \hat \Lambda_{\vec r}^{42}\hat \Lambda_{\vec r+\vechat y}^{31}
-\hat \Lambda_{\vec r}^{32}\hat \Lambda_{\vec r+\vechat y}^{41}
\right)\\
&+ U \sum_{\vec r} \frac{1 + \hat \Theta^{12}_{\vec r} \hat \Theta^{34}_{\vec r}  }{2} + \mu \sum_{\vec r}   \frac{\hat \Theta^{12}_{\vec r} + \hat \Theta^{34}_{\vec r}}{2},
\end{split}\end{equation}
where the primed sum indicates we have to attach the physical, global fermion parity $\hat P$ for the terms with $\vec r = - \vechat y$. Furthermore, the model has to be restricted to the $\hat \Gamma_-$ subspace and supplemented with the plaquette constraints Eq.\ \eqref{eq:plaquette} (with the $\hat P$-dependent modification for the plaquettes indicated in Fig.\ \ref{fig:square_odd}). 

Using the explicit expressions of the $\hat \Lambda$ and $\hat \Theta$ operators in terms of the fermionic partons (Appendix \ref{app:RoP}), we can also rewrite the JW transformed Hamiltonian as
\begin{equation}\begin{split}\label{eq:Hubbard_partons}
\hat H_{\rm H}~ \dot \mapsto~& t \sum_{\vec r} \left( 
(c^\dagger_{\vec r} h^\dagger_{\vec r+\vechat x} +h^\dagger_{\vec r} c^\dagger_{\vec r+\vechat x} )
    (-u_{\vec r} d_{\vec r+\vechat x} + d_{\vec r} u_{\vec r+\vechat x} )
+ (c^\dagger_{\vec r} c_{\vec r+\vechat x}  -h^\dagger_{\vec r} h_{\vec r+\vechat x} )
    (u_{\vec r} u^\dagger_{\vec r+\vechat x} +d_{\vec r} d^\dagger_{\vec r+\vechat x} )
+ {\rm h.c.}
\right) \\
&
+t  \sideset{}{'}\sum_{\vec r} \left( 
(c^\dagger_{\vec r} h^\dagger_{\vec r+\vechat y} +h^\dagger_{\vec r} c^\dagger_{\vec r+\vechat y} )
    (-u^\dagger_{\vec r} d^\dagger_{\vec r+\vechat y} + d^\dagger_{\vec r} u^\dagger_{\vec r+\vechat y} )
+ (c^\dagger_{\vec r} c_{\vec r+\vechat y}  -h^\dagger_{\vec r} h_{\vec r+\vechat y} )
    (u^\dagger_{\vec r} u_{\vec r+\vechat y} +d^\dagger_{\vec r} d_{\vec r+\vechat y} )
+ {\rm h.c.}
\right) \\
&+ \frac{U}{2} \sum_{\vec r} 
\left[ \left( \hat c^\dagger_{\vec r} c_{\vec r}  - \hat h^\dagger_{\vec r}   \hat h_{\vec r} \right)^2 - 
\left( \hat u^\dagger_{\vec r}  u_{\vec r}  - \hat d^\dagger_{\vec r}   \hat d_{\vec r}  \right)^2 + 1 \right]
- \mu \sum_{\vec r}  (\hat c^\dagger_{\vec r} \hat c_{\vec r} - \hat h^\dagger_{\vec r} \hat h_{\vec r} ),
\end{split}\end{equation}
which is subjected to the $\hat \Gamma_-$ condition on top of the plaquette constraints. Alternatively, the $\hat \Gamma_-$ condition can be enforced explicitly by using  the qubit notation in Sec.\ \ref{sec:qubit}, which gives
\begin{equation}\begin{split}\label{eq:Hubbard_qubit}
\hat H_{\rm H} ~\dot \mapsto ~&-\frac{t}{2} \sum_{\vec r} \left( 
-\hat Y_{\vec r}^{(1)}\hat Y_{\vec r}^{(2)}
\hat X_{\vec r+\vechat x}^{(1)}\hat X_{\vec r+\vechat x}^{(2)}
+\hat X_{\vec r}^{(1)}\hat Y_{\vec r}^{(2)}
\hat Y_{\vec r+\vechat x}^{(1)}\hat X_{\vec r+\vechat x}^{(2)}
+ \hat X_{\vec r}^{(2)}
\hat Z_{\vec r+\vechat x}^{(1)}\hat X_{\vec r+\vechat x}^{(2)}
+\hat Z_{\vec r}^{(1)}\hat Y_{\vec r}^{(2)}
\hat Y_{\vec r+\vechat x}^{(2)}
\right) \hat Z_{\vec r+\vechat x}^{(3)} \\
&- \frac{t}{2}  \sideset{}{'}\sum_{\vec r}\left( 
\hat Y_{\vec r}^{(1)}\hat X_{\vec r}^{(2)}
\hat X_{\vec r+\vechat y}^{(1)}\hat X_{\vec r+\vechat y}^{(2)}
- \hat X_{\vec r}^{(1)}\hat X_{\vec r}^{(2)} 
\hat Y_{\vec r+\vechat y}^{(1)}\hat X_{\vec r+\vechat y}^{(2)} 
+ \hat Y_{\vec r}^{(2)} 
\hat Z_{\vec r+\vechat y}^{(1)}\hat X_{\vec r+\vechat y}^{(2)} 
-\hat Z_{\vec r}^{(1)}\hat X_{\vec r}^{(2)}
\hat Y_{\vec r+\vechat y}^{(2)}
\right) \hat Y_{\vec r}^{(3)} \hat X_{\vec r+\vechat y}^{(3)}\\
&+  \sum_{\vec r}  (U - \mu \hat Z_{\vec r}^{(1)} )\frac{1 + \hat Z_{\vec r}^{(2)}}{2}
.
\end{split}\end{equation}
\end{widetext}
~\\

\subsection{Spinless Hubbard model}
Next, let us consider the spinless Hubbard model on the square lattice, given by
\begin{align}\begin{split}\label{eq:}
\hat H_{\tilde{\rm H}} = & -t \sum_{\vec r}\sum_{\delta \vec r = \vechat x, \vechat y }  (\hat f^{ \dagger}_{\vec r} \hat f_{\vec r + \delta \vec r} + {\rm h.c.} ) 
- \mu \sum_{\vec r} \hat n_{\vec r}\\
& ~~
 + U\sum_{\vec r}\sum_{\delta \vec r = \vechat x, \vechat y }   \hat n_{\vec r} \hat n_{\vec r + \delta \vec r}.
\end{split}\end{align}
The JW transformation of the spinless model can be readily inferred from the spinful one we just discussed. This can be achieved by identifying the single fermion species with (say) the fermions with spin-$\downarrow$, and then assume there is no spin-$\uparrow$ fermion in the system. As one can see from Table \ref{tab:qubit_nSz}, this amounts to declaring the first qubit $\hat Z^{(1)}$ is frozen at $-1$, and the number operator is now given simply by $\hat n_{\vec r} = (1-\hat Z^{(2)}_{\vec r})/2$.
This reduces the spinful model in Eq.\ \eqref{eq:Hubbard_qubit} to 
\begin{equation}\begin{split}\label{eq:spinless_Hubbard_qubit}
\hat H_{\tilde{\rm H}} ~\dot \mapsto ~& \frac{t}{2} \sum_{\vec r} \left( 
 \hat X_{\vec r}^{(2)} \hat X_{\vec r+\vechat x}^{(2)}
+ \hat Y_{\vec r}^{(2)}
\hat Y_{\vec r+\vechat x}^{(2)}
\right) \hat Z_{\vec r+\vechat x}^{(3)} \\
&+ \frac{t}{2}  \sideset{}{'}\sum_{\vec r}\left( 
 \hat Y_{\vec r}^{(2)} \hat X_{\vec r+\vechat y}^{(2)} 
-\hat X_{\vec r}^{(2)}\hat Y_{\vec r+\vechat y}^{(2)}
\right) \hat Y_{\vec r}^{(3)} \hat X_{\vec r+\vechat y}^{(3)}\\
& ~+ \left (\frac{\mu}{2} - U \right) \sum_{\vec r}  \hat Z^{(2)}_{\vec r}
~~+ \frac{U}{4} \sum_{\substack{\vec r\\ \delta \vec r = \vechat x, \vechat y}}
\hat Z^{(2)}_{\vec r} \hat Z^{(2)}_{\vec r + \delta \vec r} 
,
\end{split}\end{equation}
where we have modified the chemical potential and interaction terms accordingly, and dropped a constant. The plaquette constraint Eq.\ \eqref{eq:plaquette_qubit} carries through. Note that, only the second and third qubits are involved in the JW transformation of $\hat H_{\tilde{\rm H}}$, and so the bosonic problem is really defined on a system with two qubits per site, i.e., the site Hilbert space is only four-dimensional. Furthermore, the particle ${\rm U}(1)$ symmetry is manifest as $e^{- i \theta \sum_{\vec r} \hat Z^{(2)}_{\vec r}}$, and, modulo the coupling with the third qubit which endows the model with a fermionic interpretation, the model can be viewed as an XY model defined on the second qubit with the chemical potential playing the role of an out-of-plane Zeeman field and density-density interaction corresponding to Ising interactions.

\subsection{$t$-$J$ model}
As our last example, let us consider the hole-doped $t$-$J$ model \cite{RevModPhys.78.17},
\begin{equation}\begin{split}\label{eq:}
\hat H_{t\text{-}J}=&\sum_{\vec r}\sum_{\delta \vec r = \vechat x, \vechat y}  \hat {\mathcal P} \left[
-t \sum_{s=\uparrow,\downarrow}
\left(
\hat f^{s \dagger }_{\vec r}\hat f^{s}_{\vec r+\delta \vec r} + {\rm h.c.}
 \right)
\right.\\
& ~~~~~~~~~~~~~~~~~~~~~~~~~\left.
+ J \left( 
\hat {\vec S}_{\vec r}\cdot \hat {\vec S}_{\vec r+\delta \vec r}
- \frac{1}{4}\hat n_{\vec r} n_{\vec r+\delta \vec r}
\right)
\right]\hat {\mathcal P},
\end{split}\end{equation}
where $\hat P$ denotes a projection eliminating the double-occupancy states.
The JW transformation for the pre-projection hopping terms is the same as that for the spinful Hubbard model, which is given in Eq.\ \eqref{eq:Hubbard_qubit} in the qubit notation. The Heisenberg interaction can also be transformed readily using Eq.\ \eqref{eq:qubit_ex}, giving
\begin{equation}\begin{split}\label{eq:}
\hat {\vec S}_{\vec r}\cdot \hat {\vec S}_{\vec r'}
~\dot \mapsto ~
\hat {\vec S}^{(1)}_{\vec r}\cdot \hat {\vec S}^{(1)}_{\vec r'}
\frac{(1-\hat Z_{\vec r}^{(2)})(1-\hat Z_{\vec r'}^{(2)})}{4},
\end{split}\end{equation}
where $\hat {\vec S}^{(1)}_{\vec r}$ denotes $(\hat X^{(1)}_{\vec r},\hat Y^{(1)}_{\vec r},\hat Z^{(1)}_{\vec r})/2$. Similarly, the density-density interaction can be written as
\begin{equation}\begin{split}\label{eq:}
\hat {n}_{\vec r} \hat {n}_{\vec r'}
\dot \mapsto 
\left( 1 + \hat Z^{(1)}_{\vec r} \frac{1+\hat Z^{(2)}_{\vec r}}{2}\right)
\left( 1 + \hat Z^{(1)}_{\vec r'} \frac{1+\hat Z^{(2)}_{\vec r'}}{2}\right).
\end{split}\end{equation}

In the qubit notation, the restriction to the three states $\{ |0\rangle, \hat f^{\uparrow\dagger}|0\rangle, f^{\downarrow\dagger}|0\rangle \}$ on each site can be conveniently achieved by first regrouping the first two qubits into a $4\times4$ matrix (per site), and then restricting it to a $3\times 3$ matrix by deleting the entries involving the double occupancy states, i.e., by deleting the first row and column corresponding to the state with $\hat Z^{(1)}_{\vec r} =1$ and $\hat Z^{(2)}_{\vec r} =1$. The resulting matrix can then be interpreted as an operator acting on a qutrit, and we may parameterize the qutrit operators using the Gell-Mann matrices $\hat \Sigma^{i}$ for $i=1,\dots, 8$.
For instance,
\begin{equation}\begin{split}\label{eq:}
\hat X_{\vec r}^{(1)}\hat X_{\vec r}^{(2)}=
\left(
\begin{array}{cccc}
0 & 0 & 0 & 1\\
0 & 0 & 1 & 0\\
0 & 1 & 0 & 0\\
1 & 0 & 0 & 0
\end{array}
\right)
\overset{\mathcal P}{\rightarrow}
\left(
\begin{array}{ccc}
 0 & 1 & 0\\
 1 & 0 & 0\\
 0 & 0 & 0
\end{array}
\right) = \hat \Sigma_{\vec r}^4.
\end{split}\end{equation}
Note that, in order to simplify the subsequent expressions, we opt to adopt a slightly different convention for the Gell-Mann matrices in which the second and third indices are exchanged. The convention, as well as the results of the restriction, are tabulated in Appendix \ref{app:Gell-Mann}.
Altogether, we can write down the JW transformation of the $t$-$J$ model (suppressing the superscript for the third qubit, which is the only qubit remaining):
\begin{widetext}
\begin{equation}\begin{split}\label{eq:tJ_qubit}
\hat H_{t\text{-}J} ~\dot \mapsto ~& +\frac{t}{2} \sum_{\vec r} \left( 
\hat \Sigma^{4}_{\vec r}\hat \Sigma^{4}_{\vec r+\vechat x}
+\hat \Sigma^{5}_{\vec r}\hat \Sigma^{5}_{\vec r+\vechat x}
+\hat \Sigma^{6}_{\vec r}\hat \Sigma^{6}_{\vec r+\vechat x}
+\hat \Sigma^{7}_{\vec r}\hat \Sigma^{7}_{\vec r+\vechat x}
\right) \hat Z_{\vec r+\vechat x}  \\
&~~+\frac{t}{2}  \sideset{}{'}\sum_{\vec r}\left( 
\hat \Sigma^{4}_{\vec r}\hat \Sigma^{5}_{\vec r+\vechat y}
-\hat \Sigma^{5}_{\vec r}\hat \Sigma^{4}_{\vec r+\vechat y}
+\hat \Sigma^{6}_{\vec r}\hat \Sigma^{7}_{\vec r+\vechat y}
- \hat \Sigma^{7}_{\vec r}\hat \Sigma^{6}_{\vec r+\vechat y}
\right) \hat Y_{\vec r} \hat X_{\vec r+\vechat y} \\
&~~~~+ \frac{J}{4} \sum_{\vec r}\sum_{\delta \vec r=\vechat x, \vechat y}\left(
\hat \Sigma^{1}_{\vec r} \hat \Sigma^{1}_{\vec r+\delta \vec r}
+ \hat \Sigma^{2}_{\vec r} \hat \Sigma^{2}_{\vec r+\delta \vec r}
+\hat \Sigma^{3}_{\vec r} \hat \Sigma^{3}_{\vec r+\delta \vec r}
-\hat n_{\vec r}  \hat n_{\vec r+\delta \vec r} 
\right)
,
\end{split}\end{equation}
with the number operator $\hat n_{\vec r} = (2 + \sqrt{3}\hat \Sigma^{8}_{\vec r})/3$ in terms of the Gell-Mann matrices, and the constraint in Eq.\ \eqref{eq:plaquette_qubit} becoming
\begin{equation}\begin{split}\label{eq:plaquette_tJ}
\hat Y_{\vec r }  \hat X_{\vec r + \vechat x} \hat Y_{\vec r + \vechat x+\vechat y} \hat X_{\vec r + \vechat y} \overset{c}{=} 
\left( 1- 2 \hat n_{\vec r} \right)
\left(1- 2 \hat n_{\vec r+\vechat y} \right)
.
\end{split}\end{equation}

\end{widetext}

\section{Other lattices \label{sec:others}}
In Secs.\ \ref{sec:square} and \ref{sec:global} we have focused exclusively on the JW transformation for the square lattice. In contrast, the single-site discussion in Sec.\ \ref{sec:dot} and, associated with that, the discussions on internal symmetries in Sec.\ \ref{sec:sym} have no apparent dependence on how the sites are connected except for the requirement that each site is resolved in to a tetrahedron with four vertices (corresponding to the four physical Majorana operators). It is, therefore, natural to expect our construction to apply more generally to any four-coordinated lattices. 
As a demonstration, in the following we will sketch how our construction can be applied to two other physically interesting cases, namely, the 2D kagome and the 3D diamond lattices. 
Notably, the following discussion will be less thorough than the parallel discussion on the square lattice. In particular, we will not address the global constraints as detailed in Sec.\ \ref{sec:global} and, relatedly, the concrete implementation of the spatial symmetries. Here, we simply note that the analysis for the square lattice can be similarly applied to these other lattices.

\subsection{Kagome}
\begin{figure}[h]
\begin{center}
{\includegraphics[width=0.4 \textwidth]{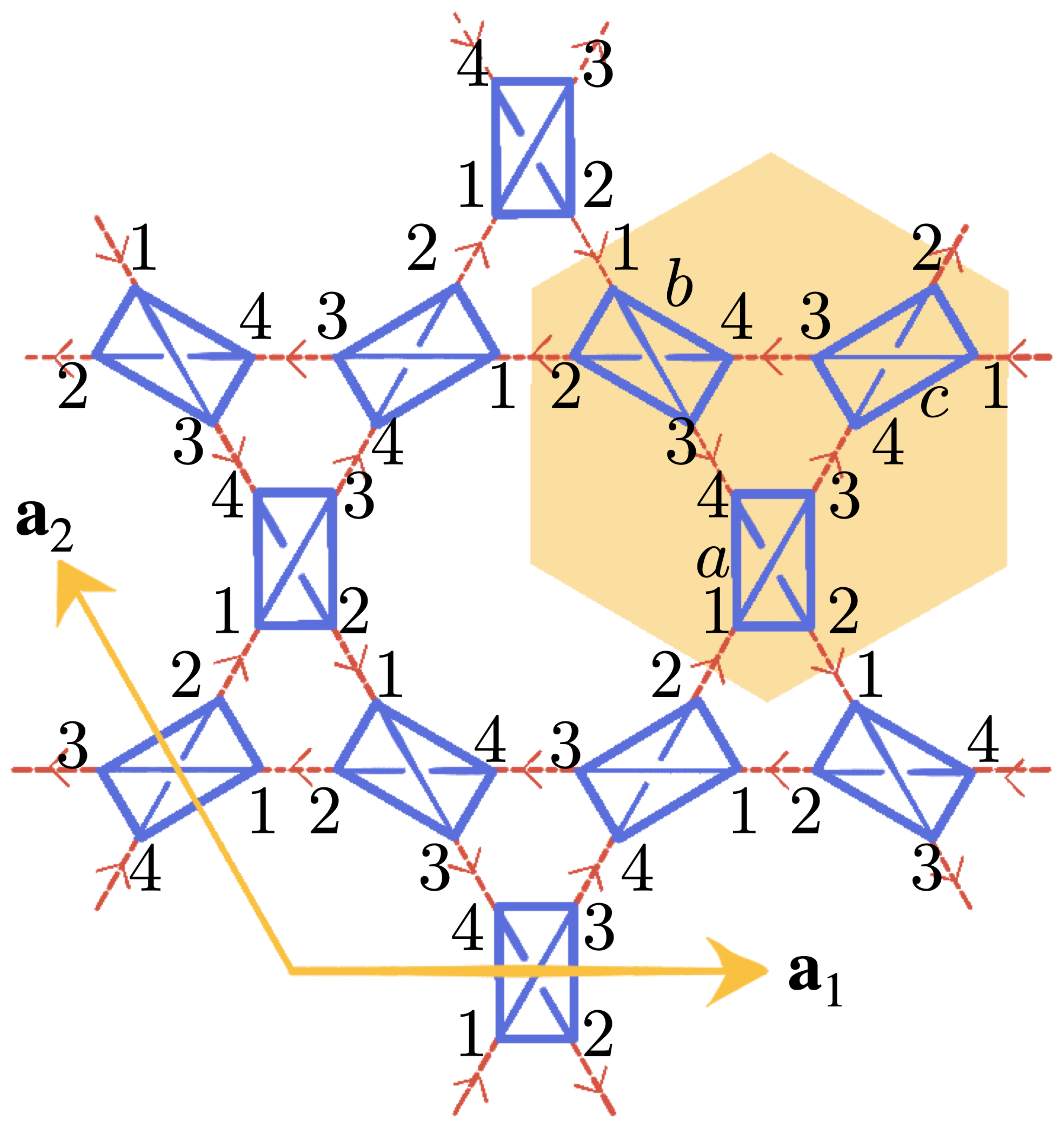}} 
\caption{Conventions for performing a Jordan-Wigner transformation on the kagome lattice. The primitive lattice vectors $\vec a_{1,2}$ are shown alongside a primitive unit cell (shaded). The three inequivalent sites in each unit cell are labeled $a$, $b$, and $c$.
\label{fig:kagome}
 }
\end{center}
\end{figure}

Let us consider a single-orbital kagome lattice with spinful electrons. The lattice consists of three sites per unit cell, which we label by $a$, $b$ and $c$. Similar to before, we resolve each site to a tetrahedron to carry out the JW transformation. A possible way to label the vertices, which correspond to the physical Majorana indices, is shown in Fig.\ \ref{fig:kagome} alongside with the unit cell and primitive lattice vectors $\vec a_{1,2}$. In addition, we also specify an orientation for each of the links connecting two nearest-neighbor sites, which determines the sign in the identification as in Eq.\ \eqref{eq:link}. 

The JW transformation is essentially done after drawing Fig.\ \ref{fig:kagome}.
To be more explicit, let us consider the plaquette constraints. As one can see from Fig.\ \ref{fig:kagome}, there are three inequivalent closed loops per unit cell $\vec r$, and each of them generate a locally independent, mutually commuting constraint:
\begin{equation}\begin{split}\label{eq:}
\hat \Phi^{34}_{\vec r^a}  \hat \Phi^{34}_{\vec r^b} \hat \Phi^{34}_{\vec r^c}
\overset{c}{=} & -1\\
\hat \Phi^{21}_{\vec r^c} 
\hat \Phi^{21}_{\vec r^b+\vec a_1}
\hat \Phi^{21}_{\vec r^a+\vec a_1 + \vec a_2}
\overset{c}{=} & -1\\
\hat \Phi^{14}_{\vec r^b}
\hat \Phi^{32}_{\vec r^c}
\hat \Phi^{14}_{\vec r^a + \vec a_1+ \vec a_2} 
\hat \Phi^{32}_{\vec r^b + \vec a_1+ \vec a_2} 
\hat \Phi^{14}_{\vec r^c + \vec a_2} 
\hat \Phi^{32}_{\vec r^a + \vec a_2} 
\overset{c}{=} & +1,
\end{split}\end{equation}
where $\vec r^{a,b,c}$ denote the coordinates of the respective sites in the unit cell at $\vec r$.
These constraints can also be rewritten in the qubit notation following Eq.\ \eqref{eq:Phi_qubit}, yielding
\begin{equation}\begin{split}\label{eq:kagome_constraint}
\prod_{ \bigtriangledown }
\hat Z^{(3)}
\overset{c}{=} & 
\prod_{ \bigtriangledown }
\hat Z^{(2)}
\\
\prod_{  \triangle }
\hat Z^{(3)}
\overset{c}{=} & 1\\
\prod_{\displaystyle \hexagon }
\hat X^{(3)}
\overset{c}{=} & 
\prod_{\displaystyle \mathrlap{\,\triangleright}{\color{gray}\hexagon}} \hat Z^{(2)},\\
\end{split}\end{equation}
where we have used $\bigtriangledown$, $\triangle$ and $\hexagon$ respectively to denote the loops in the corresponding shape, and $\mathrlap{\,\triangleright}{\color{gray}\hexagon}$ to denote the three sites of the outer hexagon connected by the inscribed triangle.

Having addressed the plaquette constraints, the procedure for JW transforming a fermionic Hamiltonian proceeds in the same way as on the square lattice. 
As a concrete example, we consider a nearest-neighbor tight-binding model with spin-rotation invariance
\begin{equation}\begin{split}\label{eq:}
\hat H_{\kappa} = - t  \sum_{\substack{\langle \vec x \vec x'\rangle \\ s = \uparrow, \downarrow } } \hat f^{s \dagger}_{\vec x} \hat f^{s}_{\vec x'} + {\rm h.c.},
\end{split}\end{equation}
where $\langle \vec x \vec x'\rangle$ denotes nearest-neighbor bonds. 
The hopping term $\hat f^{s \dagger}_{\vec x} \hat f^{s}_{\vec x'} + {\rm h.c.}$ can be rewritten in terms of the Majorana fermions $\hat \gamma^\alpha$ as in Eq.\ \eqref{eq:hop_rewrite}, and each of the bilinear $i \hat \gamma^\alpha_{\vec x} \hat \gamma^\beta_{\vec x'} $ can be recast into $\pm \hat \Lambda^{\alpha *}_{\vec x} \hat \Lambda^{\beta*}_{\vec x'}$ in the same way as discussed in Sec.\ \ref{sec:square}. To simplify the expressions, we first introduce two sets of bonds $\mathcal B^1 \equiv \{ (\vec r^a, \vec r^b) , (\vec r^b, \vec r^c),(\vec r^c, \vec r^a) \}$ and 
$\mathcal B^2 \equiv \{ (\vec r^a, \vec r^c-\vec a_1-\vec a_2) , (\vec r^b, \vec r^a+\vec a_2),(\vec r^c, \vec r^b+\vec a_1) \}$, which are respectively intra- and inter-unit cells.
As shown in Fig.\ \ref{fig:kagome}, the JW transformation utilizes different vertices for these two sets, and one finds
\begin{widetext}
\begin{equation}\begin{split}\label{eq:kagome_JW}
\hat H_{\kappa}~\dot\mapsto~
 \frac{t}{2}
\left [
\sum_{
(\vec x, \vec x') \in \mathcal B^1
} 
\left ( 
\hat \Lambda^{14}_{\vec x} \hat \Lambda^{23}_{\vec x'}
-\hat \Lambda^{24}_{\vec x} \hat \Lambda^{13}_{\vec x'}
+\hat \Lambda^{34}_{\vec x} \hat \Lambda^{43}_{\vec x'}
-\hat \Lambda^{44}_{\vec x} \hat \Lambda^{33}_{\vec x'}
\right)
+  \sum_{
(\vec x, \vec x') \in \mathcal B^2
} 
\left ( 
\hat \Lambda^{11}_{\vec x} \hat \Lambda^{22}_{\vec x'}
-\hat \Lambda^{21}_{\vec x} \hat \Lambda^{12}_{\vec x'}
+\hat \Lambda^{31}_{\vec x} \hat \Lambda^{42}_{\vec x'}
-\hat \Lambda^{41}_{\vec x} \hat \Lambda^{32}_{\vec x'}
\right)
\right],
\end{split}\end{equation}
\end{widetext}
which has to be supplemented by the plaquette constraints in Eq.\ \eqref{eq:kagome_constraint}.
The same expression in the qubit notation can be obtained through Eq.\ \eqref{eq:Lambda_qubit}. For simplicity, let us provide instead the explicit form of the JW transformed Hamiltonian in the spinless limit, obtained again by freezing $\hat Z_{\vec x}^{(1)}$ at $-1$
\begin{equation}\begin{split}\label{eq:spinless_kagome_JW}
\hat H_{\tilde \kappa}~\dot\mapsto~&
 \frac{t}{2}
\left [
\sum_{
(\vec x, \vec x') \in \mathcal B^1
} 
\left ( 
\hat X_{\vec x}^{(2)}
\hat X_{\vec x'}^{(2)}
+\hat Y_{\vec x}^{(2)}
\hat Y_{\vec x'}^{(2)} 
\right)\hat Z_{\vec x'}^{(3)} 
\right.\\
&~~~~\left.
+  \sum_{
(\vec x, \vec x') \in \mathcal B^2
} 
\left ( 
\hat Y_{\vec x}^{(2)} \hat X_{\vec x'}^{(2)}
-\hat X_{\vec x}^{(2)}\hat Y_{\vec x'}^{(2)}
\right)
\hat X_{\vec x}^{(3)}\hat Y_{\vec x'}^{(3)}
\right].
\end{split}\end{equation}
Note the similarity with the spinless model on the square lattice, Eq.\ \ref{eq:spinless_Hubbard_qubit}. In particular, as before we have effectively only two qubits per site, which are indicated by the superscripts $(2)$ and $(3)$. Together with the constraints in Eq.\ \eqref{eq:kagome_constraint}, this Hamiltonian can be interpreted as an exactly solved spin liquid hosting dispersionless emergent fermions.

\subsection{Diamond}
\begin{figure}[h]
\begin{center}
{\includegraphics[width=0.48 \textwidth]{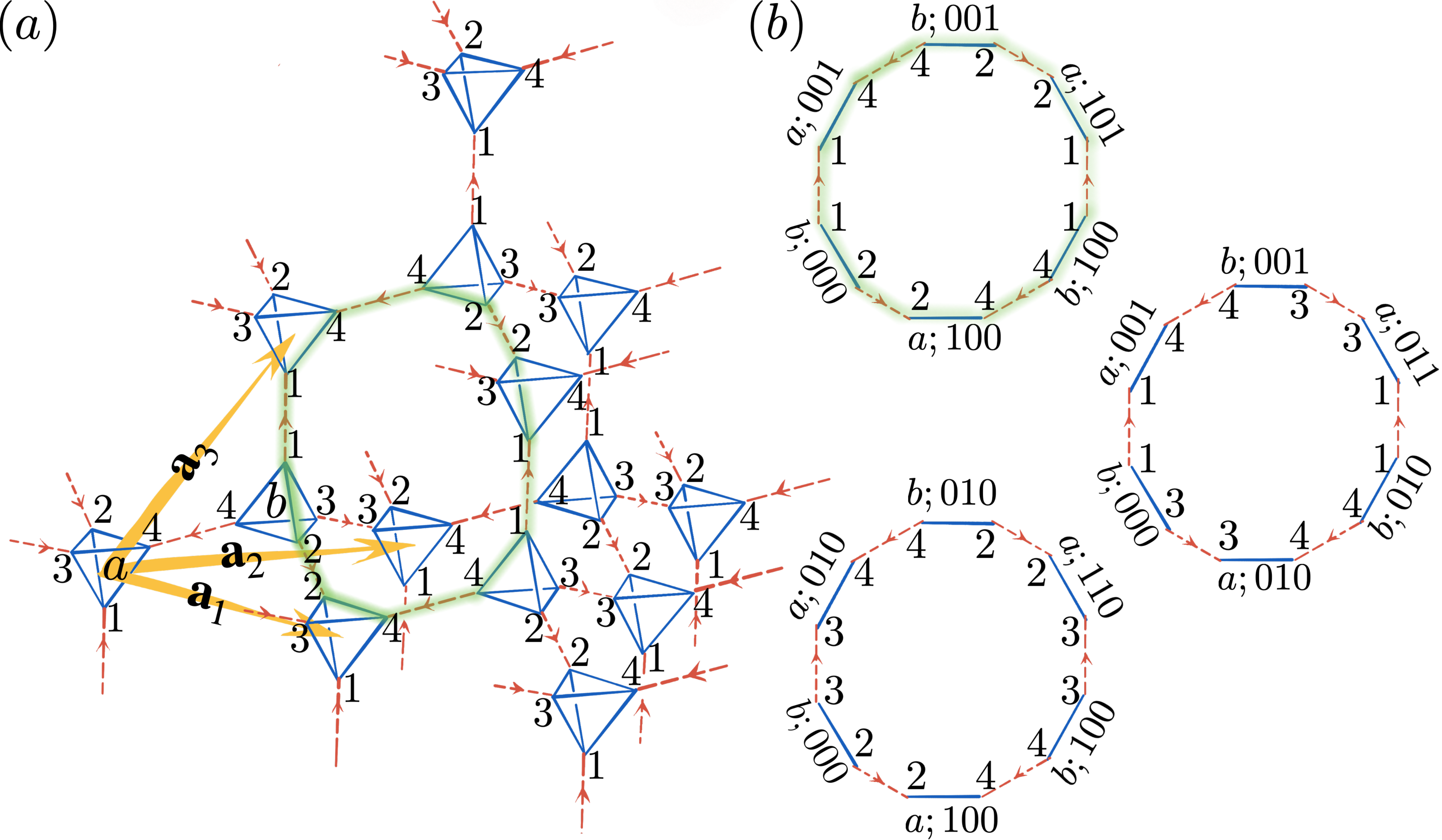}} 
\caption{Jordan-Wigner transformation on the diamond lattice. (a) The primitive lattice vectors are labeled as $\vec a_{1,2,3}$ and the two inequivalent sites in each primitive unit cell are labeled respectively as $a$ and $b$. We resolve each of the sites into a tetrahedron, and the two sublattices can be differentiated by the orientation of the tetrahedrons. One of the elementary plaquettes is highlighted in red. (b) Up to lattice translations, there are four distinct plaquettes. Three of them are shown, and the last one can be obtained by composing them. Here, $a$ and $b$ label the two inequivalent site sin each unit cell, and the notation $lmn$ denote the unit cell at $l\vec a_1+m\vec a_2+n\vec a_3$.
\label{fig:diamond}
 }
\end{center}
\end{figure}

As our last example we consider the three-dimensional diamond lattice, which is consisted of two sublattices in a face-centered cubic system. Since each site is still four-coordinated, our formalism applies directly once we pick a convention for resolving a site into a tetrahedron with labeled vertices and pick an orientation for each of the nearest-neighbor links. Our convention is shown in Fig.\ \ref{fig:diamond}. We orientate the links following an ``all-in-all-out'' pattern, i.e., the arrow always point from a site in sublattice $b$ to one in $a$. The vertices of the tetrahedron are labeled such that two vertices joined by a link are assigned the same index. We choose the primitive lattice vectors to be $\vec a_1 = (1,1,0)/2$, $\vec a_2 = (0,1,1)/2$, and $\vec a_3 = (1,0,1)/2$. In addition, we use the short-hand notation $a;lmn$ and $b;lmn$ to denote the $a$ and $b$ sites in the unit cell at $l\vec a_1+m\vec a_2 + n \vec a_3$, respectively.

Similar to above, our first step in performing the JW transformation is to identify local operator constraints coming from closed loops in the lattice. The smallest closed loop on the diamond lattice consists of six sites and can be viewed as a slightly distorted hexagon. One such loop is highlighted in green in Fig.\ \ref{fig:diamond}(a). There are four such loops per primitive unit cell, but they are not independent. In Fig.\ \ref{fig:diamond}(b) we show three loops which are related by a three-fold rotation symmetry about the axis connecting the $a$ site to the $b$ site in the home unit cell.
These loops share some common edges and when they are composed we get the fourth loop in that unit cell.  
They give rise to the plaquette constraints
\begin{equation}\begin{split}\label{eq:}
\hat \Phi^{21}_{b;000} \hat \Phi^{14}_{a;001} \hat \Phi^{42}_{b;001} \hat \Phi^{21}_{a;101}  \hat \Phi^{14}_{b;100}  \hat \Phi^{42}_{a;100}  \overset{c}{=}& 1;\\
\hat \Phi^{31}_{b;000} \hat \Phi^{14}_{a;001} \hat \Phi^{43}_{b;001} \hat \Phi^{31}_{a;011} \hat \Phi^{14}_{b;010} \hat \Phi^{43}_{a;010}   \overset{c}{=}& 1;\\
\hat \Phi^{23}_{b;000}\hat \Phi^{34}_{a;010}\hat \Phi^{42}_{b;010}\hat \Phi^{23}_{a;110}\hat \Phi^{34}_{b;100}\hat \Phi^{42}_{a;100}   \overset{c}{=}& 1.
\end{split}\end{equation}
When recast into the qubit notation, we find
\begin{equation}\begin{split}\label{eq:}
&\hat Z^{(3)}_{b;000} 
\hat X^{(3)}_{a;001} 
\hat Y^{(3)}_{b;001} 
\hat Z^{(3)}_{a;101}  
\hat X^{(3)}_{b;100}  
\hat Y^{(3)}_{a;100} \\
& ~~~~~~~~~~~~~~~~~~~~~~~~~~~~~~~~~~~~~~
 \overset{c}{=}\hat Z^{(2)}_{a;001} \hat Z^{(2)}_{b;001} \hat Z^{(2)}_{b;100} \hat Z^{(2)}_{a;100};\\
&\hat Y^{(3)}_{b;000} \hat X^{(3)}_{a;001} \hat Z^{(3)}_{b;001} \hat Y^{(3)}_{a;011} \hat X^{(3)}_{b;010} \hat Z^{(3)}_{a;010}   \\
& ~~~~~~~~~~~~~~~~~~~~~~~~~~~~~~~~~~~~~~\overset{c}{=}
\hat Z^{(2)}_{a;001}\hat Z^{(2)}_{b;001}\hat Z^{(2)}_{b;010} \hat Z^{(2)}_{a;010};\\
&\hat X^{(3)}_{b;000}\hat Z^{(3)}_{a;010} \hat Y^{(3)}_{b;010} \hat X^{(3)}_{a;110}\hat Z^{(3)}_{b;100} \hat Y^{(3)}_{a;100}   \\
& ~~~~~~~~~~~~~~~~~~~~~~~~~~~~~~~~~~~~~~\overset{c}{=}
\hat Z^{(2)}_{a;010} \hat Z^{(2)}_{b;010}  \hat Z^{(2)}_{b;100} \hat Z^{(2)}_{a;100}.
\end{split}\end{equation}

Taken at face value, we have identified four elementary loops per primitive unit cell, and each of them gives rise to a local operator constraint which reduces the dimension of the bosonic Hilbert space by two. If these four constraints per unit cell were independent, then we have locally two plaquette constraints per site and that would be over-constraining. (Recall that on both the square and the kagome lattice we have, locally, one plaquette constraint per site.) This dilemma is resolved by noticing that the constraints are not independent. In fact, as we have already pointed out, imposing three of the constraints automatically guarantee the fourth one (which we have omitted from Fig.\ \ref{fig:diamond}). More subtly, even the three constraints we wrote down are not independent when we consider all their translation copies. The detailed constraint count is discussed in Appendix \ref{app:count}. Here we only note that when we put the system on a three-torus of length $L$ along each of the three primitive lattice vectors, there are only $2(L^3-1)$ independent plaquette constraints (there are also $3$ Wilson loops). This gives a local count of $1$ plaquette constraint per site, and therefore the bosonic system is not over-constrained.

Having addressed the plaquette constraints, the JW transformation for any fermionic model defined on the diamond lattice can be carried out following the same prescription as those for the square lattice. As a concrete example, we consider once again the spinless nearest-neighbor tight-binding model $\hat H_{\tilde{\rm D}} = - t \sum_{\langle \vec x \vec x'\rangle} \hat f^\dagger_{\vec x} f_{\vec x'} + {\rm h.c.}$.
The JW transformed Hamiltonian in the qubit notation is again obtained by identifying the spinless fermion with the $f^{\downarrow}$ and so we freeze $\hat Z^{(1)}\mapsto -1$. This gives
\begin{equation}\begin{split}\label{eq:}
\hat H_{\tilde{\rm D}} ~\dot \mapsto ~
\frac{t}{2} \sum_{\substack{\vec r \\ \vec \delta = \vec 0, \vec a_1,\vec a_2,\vec a_3}}
\left( 
\hat Y_{\vec r^a +\vec \delta}^{(2)}\hat X_{\vec r^b}^{(2)}-\hat X_{\vec r^a + \vec \delta}^{(2)}\hat Y_{\vec r^b}^{(2)}
\right) \hat Q^{\vec \delta}_{\vec r^a+\vec \delta} \hat Q^{\vec \delta}_{\vec r^b},
\end{split}\end{equation}
where $\vec r^a$ and $\vec r^b$ respectively denote the location of the $a$ and $b$ sites in the unit cell at $\vec r$, and to simplify the expression we define 
\begin{equation}\begin{split}\label{eq:}
\hat Q^{\vec 0}_{\vec x} = \hat 1^{(3)}_{\vec x};~~
\hat Q^{\vec a_1}_{\vec x} = \hat Y^{(3)}_{\vec x};~~
\hat Q^{\vec a_2}_{\vec x} = \hat Z^{(3)}_{\vec x};~\&~ 
\hat Q^{\vec a_3}_{\vec x} = \hat X^{(3)}_{\vec x}.
\end{split}\end{equation}
Notice that the JW Hamiltonian takes essentially the same structure as in the previous  spinless examples: we have XY-type nearest-neighbor couplings among the second qubits, which are further decorated by direction-dependent coupling among the third qubits. Lastly, we remark that in the isotropic limit we are considering here the fermionic problem features Dirac cones at the momentum $(k_1,k_2,k_3) = (\pi, \pi,0)$ and its symmetry-related points \cite{PhysRevLett.98.106803}. Correspondingly, the bosonic problem can be viewed as an exactly solved 3D model with Dirac emergent fermions.

\section{Discussion \label{sec:discussion}}
In this work, we discuss how one could perform Jordan-Wigner transformation in higher dimensions through the introduction of fractionalized partons like spinons and chargeons/holons. Our construction is specifically tailored for spin-$1/2$ fermions, like the physical electrons, defined on a four-coordinated lattice. A merit of the present proposal is that all the relevant symmetries of the original fermionic problem are kept manifest on the bosonic side. The bosonized form of some physically interesting Hamiltonians, like the Hubbard and $t$-$J$ models, are provided. Although the bulk of the ideas are illustrated through the case of the 2D square lattice, we also sketch how the transformation can be performed on the 2D kagome and 3D diamond lattices.

Although our construction relies on the introduction of familiar partons like fermionic spinons and chargeons/holons, in the present work we have restricted our attention to the derivation of the exact JW transformation. The extent to which the usage of parton could aid our description of interesting quantum phases and phase transitions has not been pursued. There are numerous schemes to fractionalize an electron, and the relevance of any of the schemes to an actual phase of matter is a delicate question which depend not only on the Hamiltonian but also on the electron filling. In contrast, our scheme is Hamiltonian- and filling-independent. This leads to interesting interplay between the filling and the physical interpretation of the partons.
For instance, recall the on-site particle number is given by $\hat n_{\vec r} =\hat c^\dagger_{\vec r} \hat c_{\vec r} - \hat h^\dagger_{\vec r} \hat h_{\vec r} +1$. In the vacuum state, we have $ \hat c^\dagger_{\vec r} \hat c_{\vec r}\mapsto 0$ and $ \hat h^\dagger_{\vec r} \hat h_{\vec r}\mapsto 1$, whereas the spinons are subjected to the on-site parity constraint $\hat u^\dagger_{\vec r}\hat u_{\vec r} - \hat d^\dagger_{\vec r}\hat d_{\vec r}  \dot = 0$ and the plaquette constraints. 
The parity constraint could be rewritten as  $\hat u^\dagger_{\vec r}\hat u_{\vec r}  + \hat d_{\vec r} \hat d^\dagger_{\vec r}  \dot =1$, and one might interpret the spinons as the Abrikosov fermions describing the background spin liquid state on the bosonic side. In contrast, if we go to half filling of $\hat n_{\vec r} \mapsto 1$, the site Hilbert space is now described by $\hat c^\dagger_{\vec r}\hat c_{\vec r} - \hat h^\dagger_{\vec r}\hat h_{\vec r}  \mapsto 0$, and simultaneously we have $\hat u^\dagger_{\vec r}\hat u_{\vec r}  + \hat d^\dagger_{\vec r} \hat d_{\vec r}  \dot =1$. In this case, the chargeons/holons become the Abrikosov fermions associated with the background spin liquid, whereas the spinons can now be reconciled with the usual fermionic spinons (which may be confined) in the description of quantum magnets obtained by fixing $\hat n_{\vec r} \mapsto 1$. When physical electrons or holes are doped into the system, the role of spinons and chargeons/holons interchange on the doped sites. Along this line, the fermion hopping term in Eq.\ \ref{eq:Hubbard_partons} could also be interpreted as the coherent motion of spinons and chargeons/holons, for the two has to move in unison in order to keep track of the physical quantum numbers. In a way, with our JW transformation we see that doping electrons or holes into an ordinary N{\' e}el antiferromagnet is related to the doping of a (background) quantum spin liquid, a point of view which might be relevant to the physics of high temperature superconductors \cite{RevModPhys.78.17}.
Whether or not our specific description of an electron could deepen our understanding on strongly correlated electronic systems is an open question, but we note that there are already recent discussions on how the higher-dimensional JW transformation could help provide insight constructions for exotic phases of matter \cite{PhysRevResearch.2.023353,2020arXiv200212026S,PhysRevB.103.035145, PhysRevResearch.3.023120}.

Relatedly, other general aspects of the connection between our construction and the parton approach remain unexplored. For instance, our discussions on symmetries could likely be recast in the language of symmetry fractionalization \cite{CHEN20173}. In addition, we have found that the anti-unitary particle-hole symmetry is naturally represented as a Dirac time-reversal symmetry on the partons. It is an interesting problem to ask if this observation could be related to the same correspondence proposed in the context of half-filled lowest Landau level \cite{PhysRevX.5.031027}.

From another perspective, our introduction of partons could be viewed simply as an intermediary step for the construction of the JW transformed model. In fact, the reliance on partons could be completely concealed if we stay with the qubit representation in Sec.\ \ref{sec:qubit}. From this perspective, our reliance on the exceptional isomorphism ${\rm Spin}(4) \simeq  {\rm SU_s}(2)\times {\rm SU_c}(2)$ could be unsatisfactory, for it restricts us to discussing spin-$1/2$ fermions on a four-coordinated lattice. With hindsight, the key to keeping all symmetries manifest in our approach is the transformation properties of the $\hat \Lambda$ operators discussed in Sec.\ \ref{sec:sym}. One could construct alternative JW transformation schemes by first describing $\hat \Lambda$ as a fermion bilinear, which would then naturally realize the desired transformation properties and could be performed on more general lattices. This alternative approach is closely related to earlier proposals through the introduction of auxiliary Majorana fermions \cite{PhysRevLett.95.176407, Verstraete_2005} and the more general gauging procedure applied to the fermion parity symmetry \cite{PhysRevB.86.115109, PhysRevB.90.035451, PhysRevB.100.115147}. These observations are discussed in another work \cite{Kangle}.

In closing, we note that the present work only addressed the task of obtaining a symmetric bosonic Hamiltonian encoding some fermionic problem of interest. No attempt has been made to study, for instance, the phase diagram of the system of interest through its bosonic counterpart. While the bosonic description could be amenable to a large array of simulation techniques, we leave this investigation as an important future direction.

~\\
\noindent {\it Note added---} Ref.\ \onlinecite{bochniak2021bosonization} appeared while the manuscript is being finalized; part of its foci is to discuss how the approach in Refs.\ \onlinecite{Wosiek:1981mn, cmp/1103942539, PhysRevD.102.114502, Clifford} can be generalized to bosonize models with multiple fermion species per site, including in particular the problem of spin-$1/2$ electrons considered in the present manuscript.

\begin{acknowledgements}
I thank Zhen Bi, Kangle Li, Zi Yang Meng, T. Senthil, Ashvin Vishwanath, and Liujun Zou for helpful discussions.
The initial stage of this work is supported by a Pappalardo Fellowship at MIT.
\end{acknowledgements}

\appendix
\section{1D recap\label{app:1D}}
In this appendix, we first review the Jordan-Wigner transformation in one dimension, with a goal of highlighting the similarities with our construction in higher dimensions.

Consider a 1D chain of $L$ sites with one complex, spinless fermion mode per site. Equivalently, we consider two Majorana operators $\hat \gamma_{x}$ and $\hat \gamma'_{x}$ for $x \in [0,L-1]$. On the bosonic side, we consider a spin-half degree of freedom per site with the usual Pauli operators $\hat X_x$, $\hat Y_x$ and $\hat Z_x$. The Jordan-Wigner transformation is given by the identification
\begin{equation}\begin{split}\label{eq:1DJW}
\hat \gamma_x ~\longleftrightarrow~ \left(\prod_{x'=0}^{x-1} \hat Z_{x'}\right)\hat X_x;~~~
\hat \gamma'_x ~\longleftrightarrow~ \left(\prod_{x'=0}^{x-1} \hat Z_{x'}\right)\hat Y_x.
\end{split}\end{equation}
In particular, one has the on-site parity $i \hat \gamma_x \hat \gamma'_x ~\longleftrightarrow~ -\hat Z_x$.

In Eq.\ \eqref{eq:1DJW}, we have not specified the boundary condition imposed on the chain: while our prescription is natural for an open chain, subtleties arise when we consider a periodic system for which the site $x=L$ is identified with $x=0$. In particular, we notice that that the global fermion parity is $\hat P = (-1)^L \prod_{x=0}^{L-1} \hat Z_x$, and in our bosonization prescription one finds
\begin{equation}\begin{split}\label{eq:}
i \hat \gamma'_{L-1} \hat \gamma_0 \longleftrightarrow (-1)^{L-1} \hat P \hat X_{L-1} \hat X_0.
\end{split}\end{equation}
On a closed chain the left-hand side is a local operator, but on the right-hand side the presence of $\hat P$ again renders it non-local. In other words, the 1D Jordan-Wigner transformation on the circle is also only graded-local. In our lingo, we can also say that the last dashed arrow connecting the $L-1$ site to the $0$ site actually represents $(-1)^{L-1} \hat P  \left(i \hat \gamma'_{L-1} \hat \gamma_0 \right)$. This leads to subtleties in interpreting the correspondence of {\it states} between the fermionic and bosonic descriptions, and associated with that also the interpretation under the Kramers-Wannier duality. We note that such issues have been discussed in a number of excellent references, say Ref.\ \onlinecite{radicevic2019spin}, and so we will not repeat the discussion here.

As a small digression, we also sketch below how the classic transformation in 1D can also be understood in our framework. Since we have one complex fermion per site, we are interested in the representation of ${\rm Spin}(2) \simeq {\rm SO}(2)$. In our prescription we let the sole generator of ${\rm SO}(2)$ on a site $x$ be $\hat c^\dagger_x \hat c_x -  \hat h^\dagger_x \hat h$, where $\hat c$ and $\hat h$ are complex fermionic partons. To obtain a faithful representation of ${\rm Spin}(2)$ we again demand that the relation $e^{- i \pi (i \hat \gamma_x \hat \gamma'_x)} = -1$ is respected, and so we demand $e^{- i \pi ( \hat c_x^\dagger \hat c_x - \hat h_x^\dagger \hat h_x)} \dot= -1$. This again selects the $\Gamma_-$ subspace, and in this subspace we have 
\begin{equation}\begin{split}\label{eq:}
i \hat \gamma_x \hat \gamma'_x ~\dot \longleftrightarrow ~
\hat \Theta^{12}_x \equiv&
\sqrt{\hat \Gamma_x} e^{- i \frac{\pi}{2}(\hat c_x^\dagger \hat c_x - \hat h_x^\dagger \hat h_x)}\\
\dot =& \hat c_x^\dagger \hat c_x - \hat h_x^\dagger \hat h_x \dot =- \hat Z_x,
\end{split}\end{equation}
where the negative sign in $- \hat Z_x$ is customary. 
Next, we simply assert that we could choose the $\hat \Lambda^{\alpha \alpha}_x$  to be $ \hat \Lambda^{11}_x  \dot = \hat X_x$ and  $ \hat \Lambda^{22}_x  \dot = -\hat X_x$, which clearly commute. This choice amounts to the identification $
i \hat \gamma_{x}' \hat \gamma_{x+1} \dot \longleftrightarrow- \hat X_{x} \hat X_{x+1}$. The Jordan-Wigner string is then given by $\hat \Phi^{12}_{x} \equiv \hat \Gamma_x \hat \Lambda^{11}_x \hat \Theta^{12}_x \hat \Lambda^{22}_x \dot = \hat Z_x$, which matches with that in Eq.\ \eqref{eq:1DJW}. The mentioned subtlety on graded locality follows from the same considerations as we discussed in the main text for the 2D square lattice.

\section{Roots of parity \label{app:RoP}}
In this appendix we discuss some properties of a special class of operators which we dub the ``roots of parity.'' Strictly speaking, the properties outlined here are not essential to the development of the central ideas in the main text. However, as we will see the analysis of these roots of parity provide technical tricks that substantially
aid our main construction.

\subsection{Generalities}
Let us consider a system defined by $2m$ Majorana fermions $\hat \eta^{\alpha}$, $\alpha =1 ,\dots, 2m$. For the moment, we will keep $m$ a general integer. We let $\hat \Gamma$ denote the fermion parity operator 
\begin{equation}\begin{split}\label{eq:}
\hat \Gamma =  (i \hat \eta^1 \hat \eta^2) (i \hat \eta^1 \hat \eta^2)\dots  (i \hat \eta^{2m-1} \hat \eta^{2m}),
\end{split}\end{equation}
which satisfies $\hat \Gamma^2 = 1$ as it should. As discussed in the main text, we can also write the parity operator as 
\begin{equation}\begin{split}\label{eq:Gamma_Spin}
\hat \Gamma =  i^m e^{ \frac{\pi}{2}  \hat \eta^1 \hat \eta^2}\dots e^{ \frac{\pi}{2}  \hat \eta^{2m-1} \hat \eta^{2m}},
\end{split}\end{equation}
which emphasizes that, up to the phase $i^m$, we can view $\hat \Gamma$ as an element of ${\rm Spin}(2m)$. For simplicity, in the following we let $\hat \Gamma' = (-i)^m \hat \Gamma$, which can be viewed as an element of ${\rm Spin}(2m)$ directly.

Given ${\rm Spin}(2m)$ is a double cover of ${\rm SO}(2m)$, we can determine an element of ${\rm Spin}(2m)$ up to an overall sign of $\pm 1$ by studying its image in ${\rm SO}(2m)$. In our context, given any $\hat g \in  {\rm Spin}(2m)$ generated by the fermion bilinears $i \hat \eta^{\alpha} \hat \eta^{\beta}$, we can consider its action 
\begin{equation}\begin{split}\label{eq:}
\hat g \hat \eta^{\alpha} \hat g^\dagger = \sum_{\beta} R_{\hat g}^{\alpha \beta} \hat \eta^{\beta},
\end{split}\end{equation}
where $R_g$ is an element of ${\rm SO}(2m)$. Since $\hat \Gamma' \hat \eta^{\alpha} \hat \Gamma^{'\dagger} = - \hat \eta^{\alpha}$, we see that it corresponds to the element $R_{\hat \Gamma'} = - \openone_{2m\times 2m}$. Interpreting its action classically, $R_{\hat \Gamma'} $ is a global $\pi$-rotation which flips all of the $2m$ axes. 
(Note that the angle is $\pi/2$ in the exponent in Eq.\ \eqref{eq:Gamma_Spin}, although its conjugation action corresponds to a $\pi$ rotation. This is reminiscent of the behavior of half-integral spins as representation of ${\rm SU}(2)$, which double covers ${\rm SO}(3)$.)

More concretely, we could first group the $2m$ axes into $m$ sets of orthogonal planes, and then perform a $\pi$ rotation within each of the planes to arrive at $R_{\hat \Gamma'}$.  Note that $R_{\hat \Gamma'}$ is independent of how we group the axes into planes, and similarly we do not need to specify the sense of rotations involved. Yet, given any fixed choice of planes and sense of rotation, we can define a ``root'' of $\hat \Gamma'$ through rotating only by $\pi/2$ instead of $\pi$. More concretely, let $\zeta$ be a permutation of the $2m$ labels $1,2,\dots, 2m$, and for brevity we let $\zeta_\alpha = \zeta(\alpha)$, i.e.,  the index obtained by applying the permutation $\zeta$ to $\alpha$. 
The permutation $\zeta$ specifies $m$ orthogonal planes, with the $k$-th plane spanned by the $\zeta_{2k-1}$ and $\zeta_{2k}$ axes. The sense of rotation is also fixed by the ordering of $\zeta_{2k-1}$ vs $\zeta_{2k}$. Note that, permuting the $m$ orthogonal planes will lead to the same rotation, and so distinct permutations $\zeta \neq \zeta'$ could correspond to the same action.

Now, we define
\begin{equation}\begin{split}\label{eq:}
\hat \zeta \equiv 
e^{\frac{\pi}{4} \hat \eta^{\zeta_1} \eta^{\zeta_2}} e^{\frac{\pi}{4} \hat \eta^{\zeta_3} \eta^{ \zeta_4}}
\dots e^{\frac{\pi}{4} \hat \eta^{\zeta_{2m-1}} \eta^{ \zeta_{2m}}},
\end{split}\end{equation}
which is an element in ${\rm Spin}(2m)$ satisfying $R_{\hat \zeta^2} = R_{\hat \Gamma'}$. This implies $\hat \zeta^2 = \pm \hat \Gamma'$, and, in fact, we could determine this more explicitly through a direct computation:
\begin{equation}\begin{split}\label{eq:zeta_Gamma}
\hat \zeta ^2 = \eta^{\zeta_1} \eta^{\zeta_2} \dots \eta^{\zeta_{2m}} = {\rm sgn}(\zeta) \hat \Gamma',
\end{split}\end{equation}
where ${\rm sgn}(\zeta) $ denotes the sign of $\zeta$ as a permutation. We will refer to $\hat \zeta$ as a ``root of parity.''

As specified, the geometric action of $R_{\hat\zeta}$ can be readily summarized: it is a $\pi/2$ clockwise rotation in each of the $\zeta_{2k-1}$-$\zeta_{2k}$ planes for $k=1,\dots, m$. Correspondingly, it maps the Majorana operators as
\begin{equation}\begin{split}\label{eq:eta_map}
\hat \zeta \hat \eta^{\zeta_{2k-1}}\hat \zeta^\dagger  = -\hat  \eta^{\zeta_{2k}};~~~
\hat \zeta \hat \eta^{\zeta_{2k}}\hat \zeta^\dagger  = \hat \eta^{\zeta_{2k-1}}.
\end{split}\end{equation}
The action of $\hat \zeta$ can be represented graphically by a collection of arrows connecting dots corresponding to the Majorana operators. For instance, the diagram
\feynmandiagram [horizontal=a to b] {
  a[dot, label=above:$\zeta_{2k-1}$]-- [anti fermion] b[dot, label=above:$\zeta_{2k}$],
};
represents Eq.\ \eqref{eq:eta_map} with the sign encoded by the direction of the arrow.

\subsection{Case of $m=4$}
We now specialize to the case when the Majorana operators correspond to the spinons and chargon/holons we introduced in the main text. There, the enlarged site Hilbert space is defined by $m=4$ complex fermions, and we see from Eq.\ \eqref{eq:Gamma_Spin} that $\hat \Gamma = \hat \Gamma'$. A special property of the $m=4$ problem is that, when restricted to the $\Gamma_-$ sector, the roots of parity with ${\rm sgn}(\zeta)=1$ can be identified with the generating fermion bilinear. This can be seen by first expanding 
\begin{equation}\begin{split}\label{eq:}
\hat \zeta =&
\exp\left(\frac{\pi}{4} \sum_{k=1}^4 \hat \eta^{\zeta_{2k-1}} \hat \eta^{\zeta_{2k}} \right)\\
= & \frac{1}{4} \prod_{k=1}^4 \left(1 + \hat \eta^{\zeta_{2k-1}} \hat \eta^{\zeta_{2k}}  \right)\\
=& \frac{1}{4} \left(
1 
+ \hat \eta^{\zeta_{1}} \hat \eta^{\zeta_{2}} 
+ \hat \eta^{\zeta_{3}} \hat \eta^{\zeta_{4}} 
+ \hat \eta^{\zeta_{5}} \hat \eta^{\zeta_{6}} 
+ \hat \eta^{\zeta_{7}} \hat \eta^{\zeta_{8}} \right.\\
&~~~~~~~~~\left.
+ \hat \eta^{\zeta_{1}} \hat \eta^{\zeta_{2}}  \hat \eta^{\zeta_{3}} \hat \eta^{\zeta_{4}} 
+ \hat \eta^{\zeta_{5}} \hat \eta^{\zeta_{6}}  \hat \eta^{\zeta_{7}} \hat \eta^{\zeta_{8}} 
+ \cdots
 \right),
\end{split}\end{equation}
and then noticing that the terms can be grouped in pairs as 
\begin{equation}\begin{split}\label{eq:}
1 + \hat \eta^{\zeta_{1}} \hat \eta^{\zeta_{2}}  \cdots \hat \eta^{\zeta_{8}} 
 = &  1+  \hat \Gamma\\
\hat \eta^{\zeta_{1}} \hat \eta^{\zeta_{2}}  + \hat \eta^{\zeta_{3}} \hat \eta^{\zeta_{4}}  \cdots \hat \eta^{\zeta_{8}} 
 = &  (1 - \hat \Gamma)  \hat \eta^{\zeta_{1}} \hat \eta^{\zeta_{2}} \\
\hat \eta^{\zeta_{1}} \hat \eta^{\zeta_{2}}  \hat \eta^{\zeta_{3}} \hat \eta^{\zeta_{4}}  
+ \hat \eta^{\zeta_{5}} \hat \eta^{\zeta_{6}}  \hat \eta^{\zeta_{7}} \hat \eta^{\zeta_{8}}  
 = &  (1 + \hat \Gamma) \hat \eta^{\zeta_{1}} \hat \eta^{\zeta_{2}} \hat \eta^{\zeta_{3}} \hat \eta^{\zeta_{4}}  \\ 
 & \vdots
\end{split}\end{equation}
where we have used Eq.\ \eqref{eq:zeta_Gamma} with ${\rm sgn}(\zeta)=1$.  As such, one can conclude 
\begin{equation}\begin{split}\label{eq:}
\hat \zeta =&
\frac{1 - \hat \Gamma}{2} \left( \frac{
\hat \eta^{\zeta_{1}} \hat \eta^{\zeta_{2}} 
+ \hat \eta^{\zeta_{3}} \hat \eta^{\zeta_{4}} 
+ \hat \eta^{\zeta_{5}} \hat \eta^{\zeta_{6}} 
+ \hat \eta^{\zeta_{7}} \hat \eta^{\zeta_{8}}
}{2} \right)\\
&+ \frac{1 + \hat \Gamma}{2}   \left( \frac{1 + \hat \eta^{\zeta_{1}} \hat \eta^{\zeta_{2}}  \hat \eta^{\zeta_{3}} \hat \eta^{\zeta_{4}}  + \dots}{2} \right),
\end{split}\end{equation}
where $\frac{1 \pm \hat \Gamma}{2}$ are projectors into the $\Gamma_{\pm}$ subspaces, and we have not documented all the terms active in the $\Gamma_+$ subspace since that will not be our focus. In general, the grouping involves terms with $2k$ and $2(m-k)$ Majorana operators, and one sees that for $m>4$ the action of $\hat \zeta$ will no longer agree with a fermion bilinear in the $\Gamma_-$ subspace.

\begin{figure}[h]
\begin{center}
{\includegraphics[width=0.45 \textwidth]{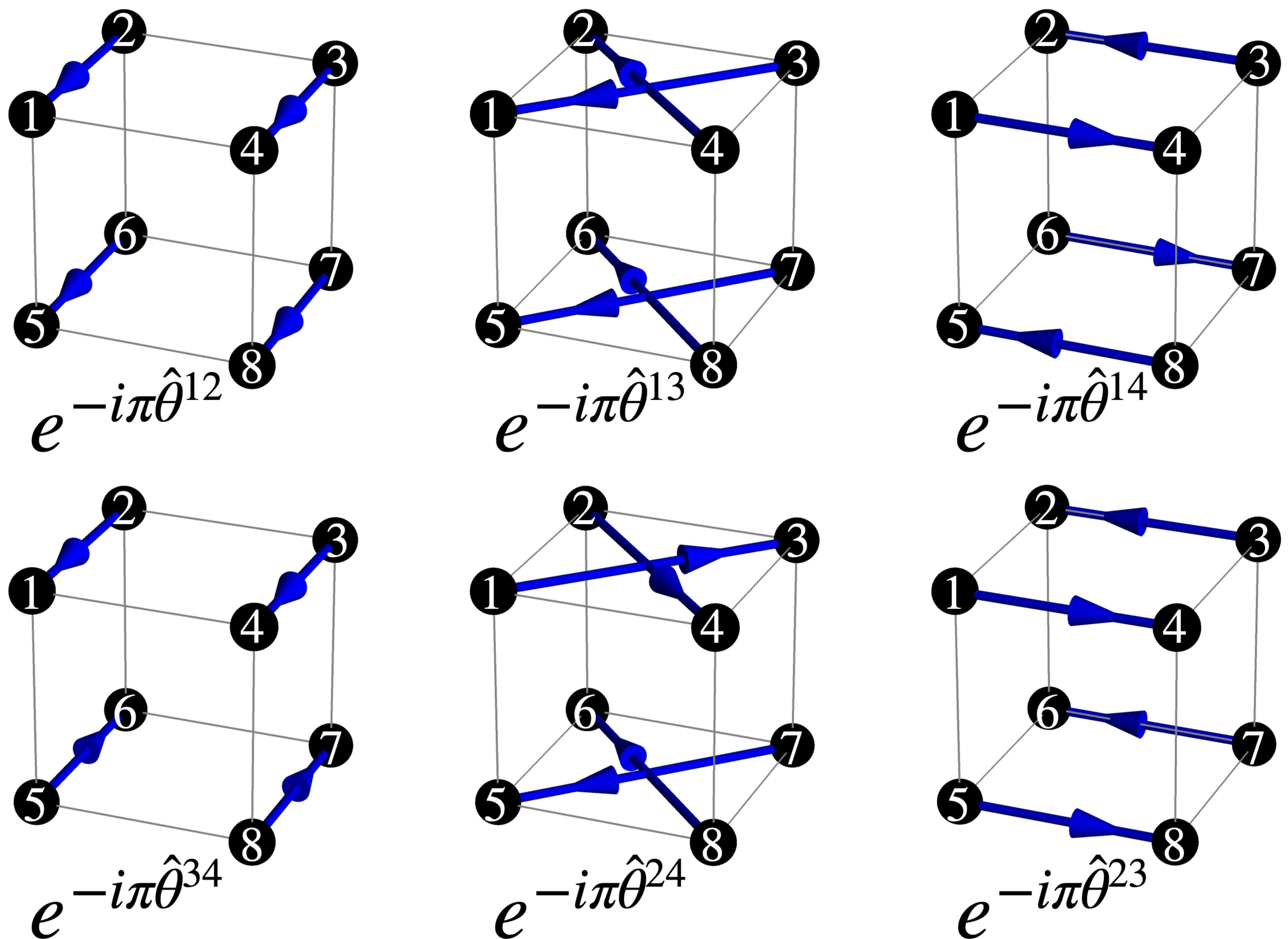}} 
\caption{Graphical notation for the roots of parity corresponding to $e^{- i \pi \hat \theta^{\alpha \beta}}$.
\label{fig:rop_theta}
 }
\end{center}
\end{figure}

Next, let us pick a convention and express the complex fermionic partons explicitly in terms of $\hat \eta^\alpha$. Let
\begin{equation}\begin{split}\label{eq:}
\begin{array}{ll}
\hat c = \frac{1}{2}(\hat \eta^{1} - i \hat \eta^{2}); &
\hat h = \frac{1}{2}(\hat \eta^{3} - i \hat \eta^{4});\\
\hat u = \frac{1}{2}(\hat \eta^{5} - i \hat \eta^{6}); &
\hat d = \frac{1}{2}(\hat \eta^{7} - i \hat \eta^{8}).
\end{array}
\end{split}\end{equation}
From this, we could express the generators $\hat \theta^{\alpha \beta}$ as
\begin{equation}\begin{split}\label{eq:theta_eta}
\hat \theta^{12} =& \frac{i}{4} \left( 
\hat \eta^1 \hat \eta^2+  \hat \eta^4  \hat \eta^3 + \hat \eta^5 \hat \eta^6 + \hat \eta^8  \hat \eta^7  \right);\\
\hat \theta^{34} =&  \frac{i}{4} \left(
\hat \eta^1 \hat \eta^2 +   \hat \eta^4 \hat \eta^3 + \hat \eta^6 \hat \eta^5 +  \hat \eta^7 \hat \eta^8
\right);\\
\hat \theta^{13} =&  \frac{i}{4} \left(
\hat \eta^1 \hat \eta^3  +  \hat \eta^2 \hat \eta^4+ \hat \eta^5 \hat \eta^7 + \hat \eta^6 \hat \eta^8 
 \right);\\
\hat \theta^{24} =&\frac{i}{4} \left(
\hat \eta^3 \hat \eta^1 + \hat \eta^4 \hat \eta^2 + \hat \eta^5 \hat \eta^7+  \hat \eta^6 \hat \eta^8
 \right);\\
\hat \theta^{14} =&  \frac{i}{4} \left(
\hat \eta^2 \hat \eta^3 + \hat \eta^4 \hat \eta^1 +  \hat \eta^7\hat \eta^6 + \hat \eta^5 \hat \eta^8 
 \right);\\
\hat \theta^{23} =& \frac{i}{4} \left(
\hat \eta^2 \hat \eta^3 + \hat \eta^4 \hat \eta^1 + \hat \eta^6 \hat \eta^7 + \hat \eta^8 \hat \eta^5
 \right),
\end{split}\end{equation}
and so $e^{- i \pi \hat \theta^{\alpha \beta}}$ is indeed a root of parity. One can also verify that all the permutations involved are even, and so $\hat \Theta^{\alpha \beta}$  agrees with the generating fermion bilinear in the $\Gamma_-$ subspace, giving
\begin{equation}\begin{split}\label{eq:GammaMinus}
\left. \sqrt{\hat \Gamma} e^{- i \pi \hat \theta^{\alpha \beta}} \right|_{\hat \Gamma\mapsto -1} = 
\left. 2\hat \theta^{\alpha \beta} \right|_{\hat  \Gamma\mapsto -1} .
\end{split}\end{equation}
Furthermore, as discussed around Eq.\ \eqref{eq:eta_map} one can adopt a graphical notation for these roots of parity (up to the overall sign), as is done in Fig.\ \ref{fig:rop_theta}.

\begin{figure}[h]
\begin{center}
{\includegraphics[width=0.45 \textwidth]{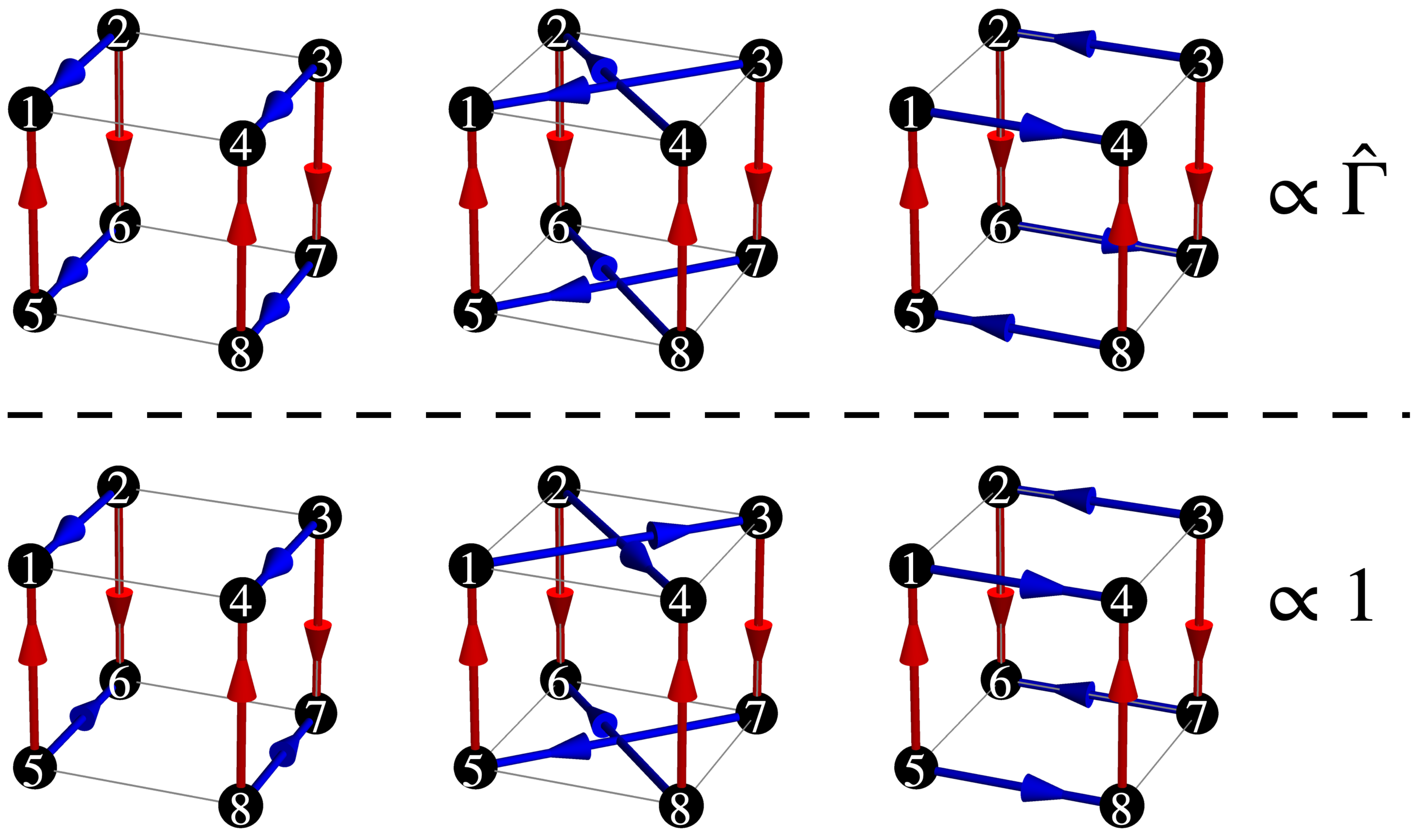}} 
\caption{Commutation between roots of parity. Let $\hat b$ and $\hat r$ be two roots of parity which are represented graphically by blue and red arrows, respectively. The action of the product $\hat b \hat r \hat b^{-1} \hat r^{-1} = \hat b \hat r \hat b \hat r $ on a Majorana operator, represented by a vertex, can be inferred by following the arrows in the red-blue-red-blue sequence, with the final sign determined by the oddness of the number of ``flipped arrows'' one encounters. From this prescription, one sees that the action of the first row in the figure corresponds to $\pm \hat \Gamma$, and that of the second row corresponds to $\pm 1$. Here, the red arrows correspond to $\hat \Lambda^{11}$, and the blue arrows correspond to $\Theta^{\alpha\beta}$ in the same order as in Fig.\ \ref{fig:rop_theta}.
\label{fig:rop_lambda}
 }
\end{center}
\end{figure}

An advantage of the graphical notation is that the effective commutation/ anti-commutation between certain roots of parity in the $\Gamma_-$ subspace can be readily inferred (Fig.\ \ref{fig:rop_lambda}).
Let us define 
\begin{equation}\begin{split}\label{eq:lambda_eta}
\hat \lambda^{11} =& \frac{i}{4} \left( 
\hat \eta^1 \hat \eta^5+  \hat \eta^6  \hat \eta^2 + \hat \eta^4 \hat \eta^8 + \hat \eta^7  \hat \eta^3  \right);\\
\hat \lambda^{22} =& \frac{i}{4} \left( 
\hat \eta^1 \hat \eta^5+  \hat \eta^6  \hat \eta^2 +  \hat \eta^8 \hat \eta^4 +   \hat \eta^3  \hat \eta^7\right);\\
\hat \lambda^{33} =& \frac{i}{4} \left( 
\hat \eta^1 \hat \eta^5+   \hat \eta^2 \hat \eta^6  + \hat \eta^8 \hat \eta^4 + \hat \eta^7  \hat \eta^3  \right);\\
\hat \lambda^{44} =& \frac{i}{4} \left( 
\hat \eta^1 \hat \eta^5+    \hat \eta^2 \hat \eta^6 + \hat \eta^4 \hat \eta^8 + \hat \eta^3  \hat \eta^7  \right).
\end{split}\end{equation}
One can check that $e^{- i \pi \hat \lambda^{\alpha \alpha}}$ are all roots of parity, and they all commute with each other. 
Similar to before, we define the Hermitian version of these roots of parity by 
\begin{equation}\begin{split}\label{eq:}
\hat \Lambda^{\alpha \alpha} \equiv \sqrt{\hat \Gamma} e^{- i \pi \hat \lambda^{\alpha \alpha}},
\end{split}\end{equation}
for $\alpha =1,\dots, 4$ and there is {\it no} summation on the repeated index.
Importantly, we defined $\lambda^{\alpha \alpha}$ such that
\begin{equation}\begin{split}\label{eq:}
\hat \Theta^{\alpha \beta} \hat \Lambda^{\gamma \gamma}
= \left\{
\begin{array}{lc}
\hat \Gamma  \hat \Lambda^{\gamma \gamma} \hat \Theta^{\alpha \beta} & \text{if}~ \alpha = \gamma ~\text{or}~\beta = \gamma\\
 \hat \Lambda^{\gamma \gamma}  \hat \Theta^{\alpha \beta} & \text{otherwise}
\end{array}
\right.
,
\end{split}\end{equation}
which provides the needed effective commutation/ anti-commutation in the $\Gamma_-$ subspace for constructing the fermion bilinears connecting different sites. For instance, the case of $\hat \Lambda^{11}$ is illustrated in Fig.\ \ref{fig:rop_lambda}.

\begin{figure}[h]
\begin{center}
{\includegraphics[width=0.45 \textwidth]{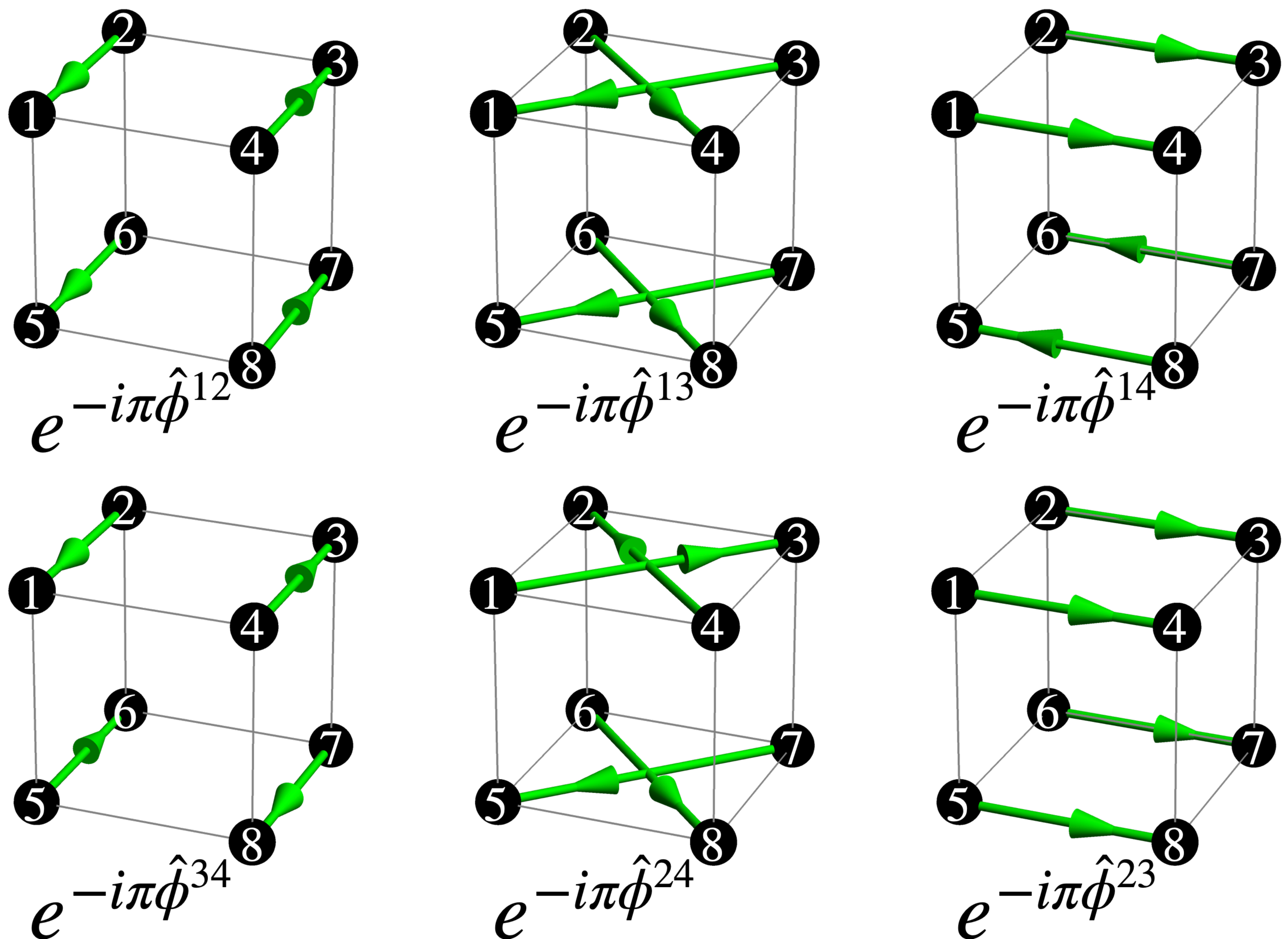}} 
\caption{Graphical notation for the roots of parity corresponding to $e^{- i \pi \hat \phi^{\alpha \beta}}$.
\label{fig:rop_phi}
 }
\end{center}
\end{figure}

\begin{figure}[h]
\begin{center}
{\includegraphics[width=0.45 \textwidth]{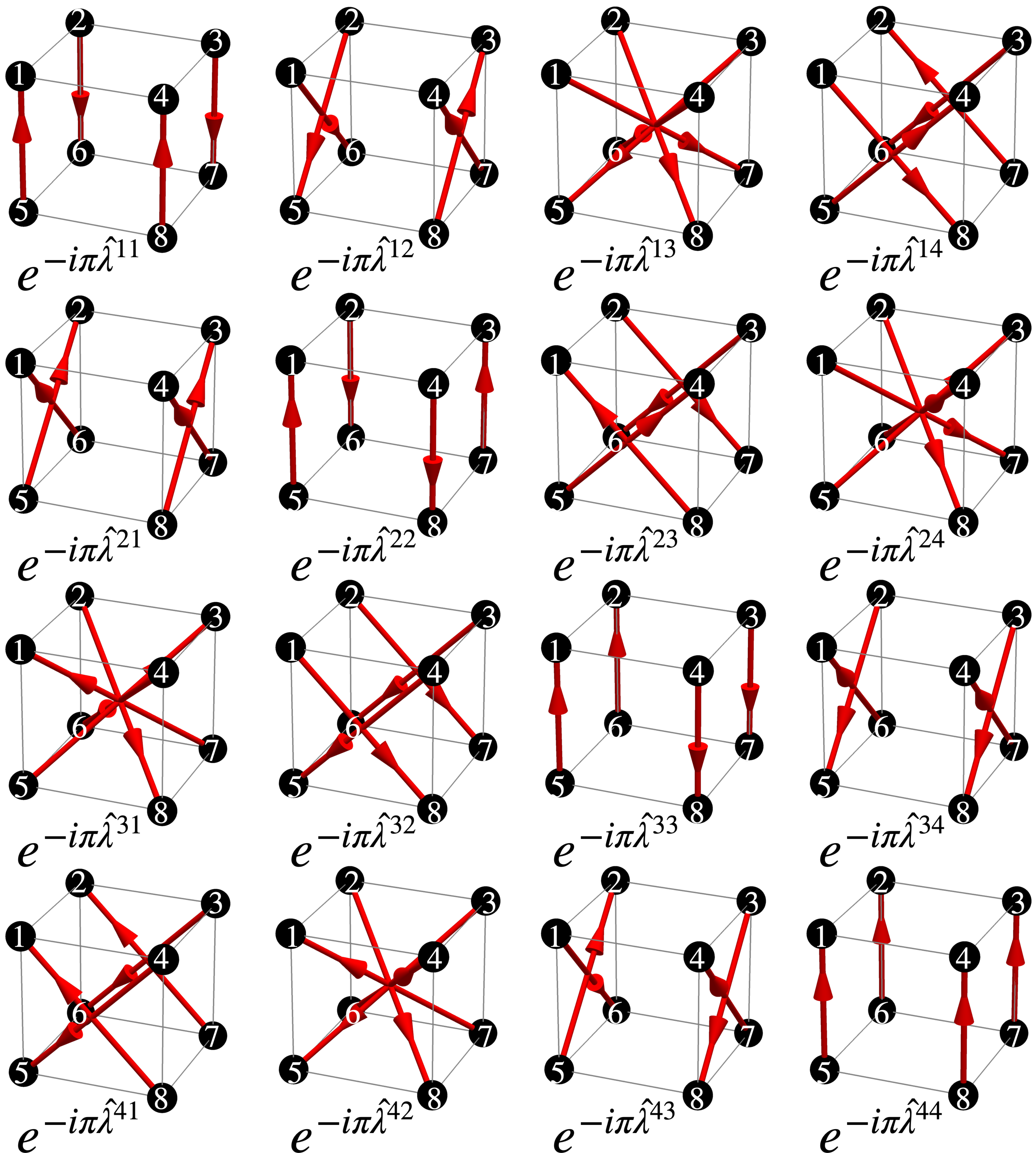}} 
\caption{Graphical notation for the roots of parity corresponding to $e^{- i \pi \hat \lambda^{\alpha \beta}}$.
\label{fig:rop_lambda_full}
 }
\end{center}
\end{figure}

In the main text, we have defined additional operators $\hat \Lambda^{\alpha \beta} = \sqrt{\hat \Gamma}^\dagger \hat \Theta^{\alpha\beta} \hat \Lambda^{\beta \beta}$ and $\hat \Phi^{\alpha \beta} = \hat \Gamma \hat \Lambda^{\alpha \alpha} \hat \Theta^{\alpha\beta} \hat \Lambda^{\beta \beta}$. The products between the operators can actually be readily analyzed using the graphical notation discussed, and one can see that all of these operators take the root-of-parity form, namely, 
\begin{equation}\begin{split}\label{eq:}
\hat \Lambda^{\alpha \beta} = \sqrt{\hat \Gamma} e^{- i \pi \hat \lambda^{\alpha \beta}};~~~
\hat \Phi^{\alpha \beta} = \sqrt{\hat \Gamma} e^{- i \pi \hat \phi^{\alpha \beta}},
\end{split}\end{equation}
for some fermion bilinears $\hat \lambda^{\alpha \beta}$ and $\hat \phi^{\alpha \beta}$ similar to those in Eqs. \eqref{eq:theta_eta} and \eqref{eq:lambda_eta}. The explicit form of these operators are provided in the graphical notation in Figs.\ \ref{fig:rop_phi} and \ref{fig:rop_lambda_full}.

Although the operators $\hat \Theta$, $\hat \Phi$, and $\hat \Lambda$ are already fully specified by Figs.\ \ref{fig:rop_theta}, \ref{fig:rop_phi}, and \ref{fig:rop_lambda_full}, for completeness we also provide their  explicit expressions in terms of the complex fermionic partons. Since Eq.\ \eqref{eq:GammaMinus} ensures these operators all agree with a fermion bilinear in the $\Gamma_-$ subspace, it will be convenient to introduce notations which exploit such simplification. 
Let us define the two-component spinors
\begin{equation}\begin{split}\label{eq:}
\hat {\bm \psi} \equiv 
\left(
\begin{array}{c}
\hat c\\
\hat h
\end{array}
\right);~~~~~
\hat {\bm \varphi} \equiv 
\left(
\begin{array}{c}
\hat u\\
\hat d
\end{array}
\right),
\end{split}\end{equation}
and we adopt the notation
\begin{equation}\begin{split}\label{eq:}
\hat {\bm \psi}^* = 
\left(
\begin{array}{c}
\hat c^\dagger\\
\hat h^\dagger
\end{array}
\right);~~
\hat {\bm \psi}^T = 
\left(
\begin{array}{cc}
\hat c & \hat h
\end{array}
\right);~~
\hat {\bm \psi}^\dagger = 
\left(
\begin{array}{cc}
\hat c^\dagger & \hat h^\dagger
\end{array}
\right).
\end{split}\end{equation}
$\hat {\bm \varphi}^*$, $\hat {\bm \varphi}^T$ and $\hat {\bm \varphi}^\dagger$ are similarly defined.
In terms of these spinors, we can express the fermion bilinears in terms of the Pauli matrices $\tau$. Explicitly,  we find
\begin{widetext}
\begin{equation}\begin{split}\label{eq:}
\begin{array}{lll}
\hat \Theta^{12}~\dot=~-\hat {\bm \psi}^\dagger\tau^3\hat {\bm \psi}-\hat {\bm \varphi}^\dagger\tau^3\hat {\bm \varphi}; & 
\hat \Theta^{13}~\dot=~ -\hat {\bm \psi}^\dagger\tau^2\hat {\bm \psi}-\hat {\bm \varphi}^\dagger\tau^2\hat {\bm \varphi};&
\hat \Theta^{14}~\dot=~\hat {\bm \psi}^\dagger\tau^1\hat {\bm \psi}-\hat {\bm \varphi}^\dagger\tau^1\hat {\bm \varphi};\\
\hat \Theta^{23}~\dot=~\hat {\bm \psi}^\dagger\tau^1\hat {\bm \psi}+\hat {\bm \varphi}^\dagger\tau^1\hat {\bm \varphi}; &
\hat \Theta^{24}~\dot=~\hat {\bm \psi}^\dagger\tau^2\hat {\bm \psi}-\hat {\bm \varphi}^\dagger\tau^2\hat {\bm \varphi}; &
\hat \Theta^{34}~\dot=~-\hat {\bm \psi}^\dagger\tau^3\hat {\bm \psi}+\hat {\bm \varphi}^\dagger\tau^3\hat {\bm \varphi}.
\end{array}
\end{split}\end{equation}

\begin{equation}\begin{split}\label{eq:}
\begin{array}{ll}
\hat \Phi^{12}~\dot=~
	-\hat {\bm \psi}^\dagger\hat {\bm \psi}-\hat {\bm \varphi}^\dagger\hat {\bm \varphi}; &
\hat \Phi^{23}~\dot=~
	-\frac{i}{2}\hat {\bm \psi}^\dagger\tau^2\hat {\bm \psi}^*-\frac{i}{2}\hat {\bm \varphi}^\dagger\tau^2\hat {\bm \varphi}^*+\frac{i}{2}\hat {\bm \psi}^T\tau^2\hat {\bm \psi}+\frac{i}{2}\hat {\bm \varphi}^T\tau^2\hat {\bm \varphi}
; \\	
\hat \Phi^{13}~\dot=~
	-\frac{1}{2}\hat {\bm \psi}^\dagger\tau^2\hat {\bm \psi}^*-\frac{1}{2}\hat {\bm \varphi}^\dagger\tau^2\hat {\bm \varphi}^*-\frac{1}{2}\hat {\bm \psi}^T\tau^2\hat {\bm \psi}-\frac{1}{2}\hat {\bm \varphi}^T\tau^2\hat {\bm \varphi}; &
\hat \Phi^{24}~\dot=~
	\frac{1}{2}\hat {\bm \psi}^\dagger\tau^2\hat {\bm \psi}^*-\frac{1}{2}\hat {\bm \varphi}^\dagger\tau^2\hat {\bm \varphi}^*+\frac{1}{2}\hat {\bm \psi}^T\tau^2\hat {\bm \psi}-\frac{1}{2}\hat {\bm \varphi}^T\tau^2\hat {\bm \varphi}; \\	
\hat \Phi^{14}~\dot=~
	-\frac{i}{2}\hat {\bm \psi}^\dagger\tau^2\hat {\bm \psi}^*+\frac{i}{2}\hat {\bm \varphi}^\dagger\tau^2\hat {\bm \varphi}^*+\frac{i}{2}\hat {\bm \psi}^T\tau^2\hat {\bm \psi}-\frac{i}{2}\hat {\bm \varphi}^T\tau^2\hat {\bm \varphi};&
\hat \Phi^{34}~\dot=~
	-\hat {\bm \psi}^\dagger\hat {\bm \psi}+\hat {\bm \varphi}^\dagger\hat {\bm \varphi}.
\end{array}
\end{split}\end{equation}

\begin{equation}\begin{split}\label{eq:}
\begin{array}{llll}
\hat \Lambda^{11}~\dot=~
	i\hat {\bm \psi}^\dagger\tau^3\hat {\bm \varphi}^*-i\hat {\bm \varphi}^T\tau^3\hat {\bm \psi}; &
\hat \Lambda^{12}~\dot=~
	-\hat {\bm \psi}^\dagger\tau^3\hat {\bm \varphi}^*-\hat {\bm \varphi}^T\tau^3\hat {\bm \psi}; &
\hat \Lambda^{13}~\dot=~
	-i\hat {\bm \psi}^\dagger\tau^1\hat {\bm \varphi}+i\hat {\bm \varphi}^\dagger\tau^1\hat {\bm \psi}; & 
\hat \Lambda^{14}~\dot=~
	\hat {\bm \psi}^\dagger\tau^1\hat {\bm \varphi}+\hat {\bm \varphi}^\dagger\tau^1\hat {\bm \psi};\\
\hat \Lambda^{21}~\dot=~
	\hat {\bm \psi}^\dagger\hat {\bm \varphi}^*+\hat {\bm \varphi}^T\hat {\bm \psi}; &
\hat \Lambda^{22}~\dot=~
	i\hat {\bm \psi}^\dagger\hat {\bm \varphi}^*-i\hat {\bm \varphi}^T\hat {\bm \psi}; & 
\hat \Lambda^{23}~\dot=~
	-i\hat {\bm \psi}^\dagger\tau^2\hat {\bm \varphi}+i\hat {\bm \varphi}^\dagger\tau^2\hat {\bm \psi}; & 
\hat \Lambda^{24}~\dot=~
	\hat {\bm \psi}^\dagger\tau^2\hat {\bm \varphi}+\hat {\bm \varphi}^\dagger\tau^2\hat {\bm \psi};\\
\hat \Lambda^{31}~\dot=~
	i\hat {\bm \psi}^\dagger\tau^1\hat {\bm \varphi}^*-i\hat {\bm \varphi}^T\tau^1\hat {\bm \psi}; &
\hat \Lambda^{32}~\dot=~
	-\hat {\bm \psi}^\dagger\tau^1\hat {\bm \varphi}^*-\hat {\bm \varphi}^T\tau^1\hat {\bm \psi}; &
\hat \Lambda^{33}~\dot=~
	i\hat {\bm \psi}^\dagger\tau^3\hat {\bm \varphi}-i\hat {\bm \varphi}^\dagger\tau^3\hat {\bm \psi}; &
\hat \Lambda^{34}~\dot=~
	-\hat {\bm \psi}^\dagger\tau^3\hat {\bm \varphi}-\hat {\bm \varphi}^\dagger\tau^3\hat {\bm \psi};\\
\hat \Lambda^{41}~\dot=~
	i\hat {\bm \psi}^\dagger\tau^2\hat {\bm \varphi}^*-i\hat {\bm \varphi}^T\tau^2\hat {\bm \psi}; &
\hat \Lambda^{42}~\dot=~
	-\hat {\bm \psi}^\dagger\tau^2\hat {\bm \varphi}^*-\hat {\bm \varphi}^T\tau^2\hat {\bm \psi}; &
\hat \Lambda^{43}~\dot=~
	\hat {\bm \psi}^\dagger\hat {\bm \varphi}+\hat {\bm \varphi}^\dagger\hat {\bm \psi}; &
\hat \Lambda^{44}~\dot=~
	i\hat {\bm \psi}^\dagger\hat {\bm \varphi}-i\hat {\bm \varphi}^\dagger\hat {\bm \psi}.
\end{array}
\end{split}\end{equation}

\section{Restriction to Gell-Mann matrices \label{app:Gell-Mann}}
We first document here the notation for the Gell-Mann matrices we used:
\begin{equation}\begin{split}\label{eq:}
\hat \Sigma^{1}=&
\left(
\begin{array}{ccc}
 0 & 0 & 1 \\
 0 & 0 & 0 \\
 1 & 0 & 0 \\
\end{array}
\right); ~
\hat \Sigma^{2}=
\left(
\begin{array}{ccc}
 0 & 0 & -i \\
 0 & 0 & 0 \\
 i & 0 & 0 \\
\end{array}
\right); ~
\hat \Sigma^{3}=
\left(
\begin{array}{ccc}
 1 & 0 & 0 \\
 0 & 0 & 0 \\
 0 & 0 & -1 \\
\end{array}
\right)\\
\hat \Sigma^{4}=&
\left(
\begin{array}{ccc}
 0 & 1 & 0 \\
 1 & 0 & 0 \\
 0 & 0 & 0 \\
\end{array}
\right); ~
\hat \Sigma^{5}=
\left(
\begin{array}{ccc}
 0 & -i & 0 \\
 i & 0 & 0 \\
 0 & 0 & 0 \\
\end{array}
\right);~~
\hat \Sigma^{6}=
\left(
\begin{array}{ccc}
 0 & 0 & 0 \\
 0 & 0 & 1 \\
 0 & 1 & 0 \\
\end{array}
\right)\\
\hat \Sigma^{7}=&
\left(
\begin{array}{ccc}
 0 & 0 & 0 \\
 0 & 0 & i \\
 0 & -i & 0 \\
\end{array}
\right);~
\hat \Sigma^{8}=
\left(
\begin{array}{ccc}
 \frac{1}{\sqrt{3}} & 0 & 0 \\
 0 & -\frac{2}{\sqrt{3}} & 0 \\
 0 & 0 & \frac{1}{\sqrt{3}} \\
\end{array}
\right).
\end{split}\end{equation}
Note that we have interchanged the second and third indices in the matrices when compared with the more common convention. This is done intentionally such that $\hat \Sigma^{1,2,3}$ will correspond to the usual Pauli matrices, and hence the physical spin operators, in the single-occupancy subspace.

The restriction of the two-qubit operators to the qutrit, achieved by deleting the first row and column, is then found to be 
\begin{equation}\begin{split}\label{eq:}
\begin{array}{lllll}
\hat X^{(1)}\overset{\mathcal P}{\rightarrow} \hat \Sigma^{1}; & 
\hat X^{(2)}\overset{\mathcal P}{\rightarrow} \hat \Sigma^6; &
\hat X^{(1)}\hat X^{(2)}\overset{\mathcal P}{\rightarrow} \hat \Sigma^4; &
\hat Y^{(1)}\hat X^{(2)}\overset{\mathcal P}{\rightarrow} \hat \Sigma^5; &
\hat Z^{(1)}\hat X^{(2)}\overset{\mathcal P}{\rightarrow} -\hat \Sigma^6 ;\\
\hat Y^{(1)}\overset{\mathcal P}{\rightarrow} \hat \Sigma^{2}; &
\hat Y^{(2)}\overset{\mathcal P}{\rightarrow} -\hat \Sigma^{7}; &
\hat X^{(1)}\hat Y^{(2)}\overset{\mathcal P}{\rightarrow} - \hat \Sigma^5; &
\hat Y^{(1)}\hat Y^{(2)}\overset{\mathcal P}{\rightarrow} \hat \Sigma^4; &
\hat Z^{(1)}\hat Y^{(2)}\overset{\mathcal P}{\rightarrow}  \hat \Sigma^7; \\
\hat Z^{(1)}\overset{\mathcal P}{\rightarrow}  -\frac{1}{3} + \hat \Sigma^{3} + \frac{1}{\sqrt{3}}\hat \Sigma^{8}; & 
\hat Z^{(2)}\overset{\mathcal P}{\rightarrow} - \frac{1}{3}  - \frac{2}{\sqrt{3}} \hat \Sigma^{8}; &
\hat X^{(1)}\hat Z^{(2)}\overset{\mathcal P}{\rightarrow} - \hat \Sigma^1; &
\hat Y^{(1)}\hat Z^{(2)}\overset{\mathcal P}{\rightarrow} -\hat \Sigma^2; &
\hat Z^{(1)}\hat Z^{(2)}\overset{\mathcal P}{\rightarrow} -\frac{1}{3} - \hat \Sigma^3 + \frac{1}{\sqrt{3}} \hat \Sigma^8.
\end{array}
\end{split}\end{equation}
\end{widetext}

\section{Loop counting \label{app:count}}
Here we discuss how to count the number of independent loops on the specific lattices we considered, exploiting their translation invariance. For our purposes, it suffices to invoke topology in a very minimal manner. Readers who are interested in a more proper introduction to the machinery in a similar context can consult, for instance, Ref.\ \onlinecite{Clifford}.

\subsection{Square}
As a warm-up, let us first revisit the square lattice. We suppose the system is placed on a $L_x \times L_y$ torus. We have one site $p_{\vec r}$ in the unit cell at $\vec r$, and to the unit cell containing $p_{\vec r}$ we can allocate two oriented nearest-neighbor links 
\begin{equation}\begin{split}\label{eq:}
b_{\vec r}^1 \equiv (p_{\vec r + \vechat x}, p_{\vec r} );~~~~
b_{\vec r}^2 \equiv (  p_{\vec r + \vechat y}, p_{\vec r}).
\end{split}\end{equation}
Here, the notation means that $b_{\vec r}^1$ is the link represented by the arrow pointing from $p_{\vec r + \vechat x}$ to $p_{\vec r}$, and $b^2_{\vec r}$ is similarly defined.
Our goal is to find all the independent generating loops such that any loop on the lattice can be viewed as composition of the generators. 

Technically, this is achieved by inspecting the kernel of the boundary map $\partial$ from the free abelian group generated by the edges (i.e., $1$-chain) to that generated by the vertices (i.e., $0$-chain).
As a linear operator, $\partial$ can be viewed simply as a $(L_xL_y) \times (2L_xL_y)$ matrix. Thanks to translation invariance we may further pass to the momentum space and write $\partial$ as the direct sum of $L_x L_y$ matrices, each $1\times 2$ dimensional. Explicitly, we have
\begin{equation}\begin{split}\label{eq:}
\partial \simeq \bigoplus_{\vec k}  \partial(\vec k); ~~~
 \partial (\vec k) \equiv
\left(
\begin{array}{cc}
1-e^{- i k_1} & 1-e^{- i k_2}
\end{array}
\right),
\end{split}\end{equation}
where $\vec k = 2 \pi(m/L_x, n/L_y)$ for integers $m,n$ takes the usual $L_x L_y$ discrete values in the Brillouin zone.
Here, the two columns correspond to the two types of links we have in each unit cell. The first represents $\partial b_{\vec r}^1 = p_{\vec r} - p_{\vec r + \vechat x}$, which upon Fourier transform becomes $1-e^{-ik_1}$. Similarly, the second column corresponds to $\partial b_{\vec r}^2= p_{\vec r} - p_{\vec r + \vechat y}$. One can expect from the rank-nullity theorem that each of the $\partial(\vec k)$ yields a null vector, which can be seen explicitly through 
\begin{equation}\begin{split}\label{eq:}
\ell_{\vec k} \equiv
\left(
\begin{array}{c}
1-e^{- i k_2} \\ 
-1+e^{- i k_1}
\end{array}
\right);~~~
\partial(\vec k)
\ell_{\vec k}
=0.
\end{split}\end{equation}
To unpack the calculation, we can Fourier transform back to the real space in the same spirit as constructing Wannier functions. The $1$ in the first entry corresponds to the link $b_{\vec 0}^1 = (p_{ \vechat x},p_{\vec 0} ) $, whereas the term $-e^{- i k_2}$ corresponds to the same link shifted along $\vechat y$ and reversed, i.e., $- b_{\vechat y}^1 = (p_{ \vechat y},p_{  \vechat x+\vechat y} ) $. The second entry can be interpreted similarly, and altogether our momentum-space solution translates into 
\begin{equation}\begin{split}\label{eq:loop_formal}
& b_{\vec 0}^1  - b_{\vec 0}^2 - b_{\vechat y}^1  + b_{\vechat x}^2\\
=&(p_ {\vechat x},p_{\vec 0} )  +  (p_{\vec 0},p_{  \vechat y} )   + (p_{ \vechat y}, p_{ \vechat x+\vechat y} ) + (p_{ \vechat x + \vechat y},p_{  \vechat x} ),
\end{split}\end{equation}
which is manifestly the closed loop surrounding a plaquette cornered at $\vec 0$.

Naively, one may expect there to be $L_x L_y$ independent loops, obtained by translating the mentioned loop by any lattice translation. Yet, one should notice that $\ell_{\vec k = \vec 0}$ vanishes, and so does $\partial(\vec k = \vec 0)$. In terms of solution counting, $\partial(\vec k = \vec 0)$ has two null vectors, and each of $\partial(\vec k \neq \vec0)$ has one, and so we have a total of $L_xL_y+1$ generators. The vanishing of  $\ell_{\vec k = \vec 0}$ implies that when we sum over all the translated copies of Eq.\ \eqref{eq:loop_formal}, we do not have any loop left, i.e., the $L_xL_y$ loops are not linearly independent. This can be verified explicitly by seeing that in the sum each link is traversed twice but in opposite directions. As such, there are only $L_xL_y-1$ independent loops defined on the plaquettes. The remaining two loops are then, unsurprisingly, the nontrivial loops on the torus, which corresponds to the two null vectors of $\partial(\vec k=\vec 0)$. 

\subsection{Kagome}
As our second example, we consider the kagome lattice. Following the labeling convention in Fig.\ \ref{fig:kagome}, the six nearest-neighbor links associated with each unit cell are
\begin{equation}\begin{split}\label{eq:}
\begin{array}{ll}
b^{1}_{\vec r} = (p_{\vec r^b}, p_{\vec r^a}); &  b^{4}_{\vec r} = (p_{\vec r^c -\vec a_1-\vec a_2},p_{\vec r^a});\\
b^{2}_{\vec r} = (p_{\vec r^c},p_{\vec r^b}); & b^{5}_{\vec r} = (p_{\vec r^a+\vec a_2},p_{\vec r^b});\\
b^{3}_{\vec r} = (p_{\vec r^a},p_{\vec r^c}); & b^{6}_{\vec r} = (p_{\vec r^b + \vec a_1},p_{\vec r^c}).
\end{array}
\end{split}\end{equation}
The $\partial(\vec k)$ is now a $3\times 6$ matrix, since we have three sites and six links per unit cell. It reads
\begin{equation}\begin{split}\label{eq:}
\partial(\vec k) = 
\left(
\begin{array}{cccccc}
1 & 0 & -1 & 1 & -e^{-i k_2} & 0 \\
-1 & 1 & 0 & 0 & 1 & -e^{-i k_1} \\
0 & -1 & 1 & -e^{i (k_1+k_2)} & 0 & 1
\end{array}
\right).
\end{split}\end{equation}
We again expect from rank-nullity theorem that there will generically be $3$ null vectors at every $\vec k$. They are given by
\begin{equation}\begin{split}\label{eq:}
\ell^1_{\vec k}=
\left(
\begin{array}{c}
1\\
1\\
1\\
0\\
0\\
0
\end{array}
\right);
\ell^2_{\vec k}=
\left(
\begin{array}{c}
0\\
0\\
0\\
1\\
e^{-k_2}\\
e^{i (k_1+k_2)}
\end{array}
\right);
\ell^3_{\vec k}=
\left(
\begin{array}{c}
-1\\
-e^{i(k_1+k_2)}\\
-e^{i k_1}\\
1\\
e^{i (k_1+k_2)}\\
e^{i k_1}
\end{array}
\right).
 \end{split}\end{equation}
For general values of $\vec k$, the three solutions are independent. However, at $\vec k=\vec 0$ we have $\ell^3_{\vec 0} =-\ell^1_{\vec 0}+\ell^2_{\vec 0}$.  Similarly, one linear relation exists among the three rows of $\delta(\vec 0)$, and so the null space of $\delta(\vec 0)$ is $4$-dimensional, with two of them being $\ell^{1,2}_{\vec 0}$ and the other two corresponding to the Wilson loops.

\subsection{Diamond}
The diamond lattice contains two sublattices, i.e., there are two inequivalent (but symmetry-related) sites in each unit cell.
Let $\vec r^a$ and $\vec r^b$ be the coordinates of the two sublattices in the unit cell at $\vec r$, and let $\vec a_{i=1,2,3}$ be the three primitive lattice vectors. We should allocate four nearest-neighbor links to each unit cell, and we pick 
\begin{equation}\begin{split}\label{eq:}
\begin{array}{ll}
b^{1}_{\vec r} = (p_{\vec r^b}, p_{\vec r^a}); & 
b^{2}_{\vec r} = (p_{\vec r^b}, p_{\vec r^a+\vec a_1}); \\
b^{3}_{\vec r} = (p_{\vec r^b}, p_{\vec r^a+\vec a_2}); & 
b^{4}_{\vec r} = (p_{\vec r^b}, p_{\vec r^a+\vec a_3}).
\end{array}
\end{split}\end{equation}
This gives
\begin{equation}\begin{split}\label{eq:}
\partial(\vec k) = 
\left(
\begin{array}{cccc}
1 & e^{- i k_1} & e^{- i k_2} & e^{- i k_3}\\
-1 & -1 & -1 & -1
\end{array}
\right),
\end{split}\end{equation}
and we assume the system is placed on a torus of length $L$ along each of the three primitive lattice vectors. Again, from rank-nullity we expect $\partial(\vec k)$ to yield two null vectors generically. However, from the discussion in the main text we see that there are four sets of closed loops with different orientations
\begin{align}\begin{split}\label{eq:}
\ell^1_{\vec k}=
\left(
\begin{array}{c}
e^{- i k_1}- e^{-i k_3} \\
 e^{-i k_3} -1 \\
0\\
1 - e^{- i k_1} 
\end{array}
\right); ~~~
\ell^2_{\vec k}=
\left(
\begin{array}{c}
e^{-i k_3} -e^{- i k_2} \\
0 \\
1-  e^{-i k_3} \\
e^{- i k_2} -1
\end{array}
\right)
\\
\ell^3_{\vec k}=
\left(
\begin{array}{c}
e^{-i k_2}- e^{- i k_1} \\
1- e^{-i k_2} \\
e^{- i k_1} -1 \\
0
\end{array}
\right)
; ~~~
\ell^4_{\vec k}=\left(
\begin{array}{c}
0\\
e^{-i k_2} - e^{- i k_3}\\
e^{- i k_3} - e^{- i k_1}\\
e^{- i k_1} -  e^{- i k_2}
\end{array}
\right)
.
\end{split}\end{align}
In order for the two perspectives to be consistent, there must be two linear relations among $\ell_{\vec k}^{i=1,2,3,4}$. Indeed, one can verify
\begin{equation}\begin{split}\label{eq:}
\ell_{\vec k}^1+\ell_{\vec k}^2+\ell_{\vec k}^3 + \ell_{\vec k}^4 =& 0;\\
(1- e^{-i k_2}) \ell_{\vec k}^1+ (1-e^{-i k_1})\ell_{\vec k}^2
+ (1- e^{-i k_3}) \ell_{\vec k}^3 =& 0.
\end{split}\end{equation}
Furthermore, although each of the $\ell_{\vec k}^{i}$ vanishes on a line in the Brillouin zone (e.g., $\ell_{\vec k}^1$ vanishes along $(0,k_2,0)$), these lines intersect only at the origin and so for any $\vec k \neq 0$ we still have two independent solutions spanned by the four $\ell_{\vec k}^i$. For the exceptional point of $\vec k = 0$, all the four vectors $\ell^i_{\vec k =0}$ vanish and the two rows of $\partial(\vec k =\vec 0)$ become degenerate, and so there are three null vectors. Altogether, we conclude there are $2(L^3-1)+3 = 2L^3+1$ generators, among which $3$ of them correspond to the nontrivial loops on the three torus.

\bibliography{references}

\begin{thebibliography}{97}%
\makeatletter
\providecommand \@ifxundefined [1]{%
 \@ifx{#1\undefined}
}%
\providecommand \@ifnum [1]{%
 \ifnum #1\expandafter \@firstoftwo
 \else \expandafter \@secondoftwo
 \fi
}%
\providecommand \@ifx [1]{%
 \ifx #1\expandafter \@firstoftwo
 \else \expandafter \@secondoftwo
 \fi
}%
\providecommand \natexlab [1]{#1}%
\providecommand \enquote  [1]{``#1''}%
\providecommand \bibnamefont  [1]{#1}%
\providecommand \bibfnamefont [1]{#1}%
\providecommand \citenamefont [1]{#1}%
\providecommand \href@noop [0]{\@secondoftwo}%
\providecommand \href [0]{\begingroup \@sanitize@url \@href}%
\providecommand \@href[1]{\@@startlink{#1}\@@href}%
\providecommand \@@href[1]{\endgroup#1\@@endlink}%
\providecommand \@sanitize@url [0]{\catcode `\\12\catcode `\$12\catcode
  `\&12\catcode `\#12\catcode `\^12\catcode `\_12\catcode `\%12\relax}%
\providecommand \@@startlink[1]{}%
\providecommand \@@endlink[0]{}%
\providecommand \url  [0]{\begingroup\@sanitize@url \@url }%
\providecommand \@url [1]{\endgroup\@href {#1}{\urlprefix }}%
\providecommand \urlprefix  [0]{URL }%
\providecommand \Eprint [0]{\href }%
\providecommand \doibase [0]{https://doi.org/}%
\providecommand \selectlanguage [0]{\@gobble}%
\providecommand \bibinfo  [0]{\@secondoftwo}%
\providecommand \bibfield  [0]{\@secondoftwo}%
\providecommand \translation [1]{[#1]}%
\providecommand \BibitemOpen [0]{}%
\providecommand \bibitemStop [0]{}%
\providecommand \bibitemNoStop [0]{.\EOS\space}%
\providecommand \EOS [0]{\spacefactor3000\relax}%
\providecommand \BibitemShut  [1]{\csname bibitem#1\endcsname}%
\let\auto@bib@innerbib\@empty
\bibitem [{\citenamefont {Jordan}\ and\ \citenamefont {Wigner}(1928)}]{JW}%
  \BibitemOpen
  \bibfield  {author} {\bibinfo {author} {\bibfnamefont {P.}~\bibnamefont
  {Jordan}}\ and\ \bibinfo {author} {\bibfnamefont {E.}~\bibnamefont
  {Wigner}},\ }\bibfield  {title} {\bibinfo {title} {{\"U}ber das paulische
  {\"a}quivalenzverbot},\ }\href {https://doi.org/10.1007/BF01331938}
  {\bibfield  {journal} {\bibinfo  {journal} {Zeitschrift f{\"u}r Physik}\
  }\textbf {\bibinfo {volume} {47}},\ \bibinfo {pages} {631} (\bibinfo {year}
  {1928})}\BibitemShut {NoStop}%
\bibitem [{\citenamefont {Huerta}\ and\ \citenamefont
  {Zanelli}(1993)}]{PhysRevLett.71.3622}%
  \BibitemOpen
  \bibfield  {author} {\bibinfo {author} {\bibfnamefont {L.}~\bibnamefont
  {Huerta}}\ and\ \bibinfo {author} {\bibfnamefont {J.}~\bibnamefont
  {Zanelli}},\ }\bibfield  {title} {\bibinfo {title} {Bose-fermi transformation
  in three-dimensional space},\ }\href
  {https://doi.org/10.1103/PhysRevLett.71.3622} {\bibfield  {journal} {\bibinfo
   {journal} {Phys. Rev. Lett.}\ }\textbf {\bibinfo {volume} {71}},\ \bibinfo
  {pages} {3622} (\bibinfo {year} {1993})}\BibitemShut {NoStop}%
\bibitem [{\citenamefont {Burgess}\ \emph {et~al.}(1994)\citenamefont
  {Burgess}, \citenamefont {L{\"u}tken},\ and\ \citenamefont
  {Quevedo}}]{BURGESS199418}%
  \BibitemOpen
  \bibfield  {author} {\bibinfo {author} {\bibfnamefont {C.}~\bibnamefont
  {Burgess}}, \bibinfo {author} {\bibfnamefont {C.}~\bibnamefont
  {L{\"u}tken}},\ and\ \bibinfo {author} {\bibfnamefont {F.}~\bibnamefont
  {Quevedo}},\ }\bibfield  {title} {\bibinfo {title} {Bosonization in higher
  dimensions},\ }\href
  {https://doi.org/https://doi.org/10.1016/0370-2693(94)00963-5} {\bibfield
  {journal} {\bibinfo  {journal} {Physics Letters B}\ }\textbf {\bibinfo
  {volume} {336}},\ \bibinfo {pages} {18} (\bibinfo {year} {1994})}\BibitemShut
  {NoStop}%
\bibitem [{\citenamefont {Kopietz}(1997)}]{Kopietz1997}%
  \BibitemOpen
  \bibfield  {author} {\bibinfo {author} {\bibfnamefont {P.}~\bibnamefont
  {Kopietz}},\ }\href@noop {} {\emph {\bibinfo {title} {Bosonization of
  Interacting Fermions in Arbitrary Dimensions}}},\ \bibinfo {series} {Lecture
  Notes in Physics Monographs}, Vol.~\bibinfo {volume} {48}\ (\bibinfo
  {publisher} {Springer-Verlag Berlin Heidelberg},\ \bibinfo {year}
  {1997})\BibitemShut {NoStop}%
\bibitem [{\citenamefont {Pasquier}\ and\ \citenamefont
  {Haldane}(1998)}]{PASQUIER1998719}%
  \BibitemOpen
  \bibfield  {author} {\bibinfo {author} {\bibfnamefont {V.}~\bibnamefont
  {Pasquier}}\ and\ \bibinfo {author} {\bibfnamefont {F.}~\bibnamefont
  {Haldane}},\ }\bibfield  {title} {\bibinfo {title} {A dipole interpretation
  of the v = 1/2 state},\ }\href
  {https://doi.org/https://doi.org/10.1016/S0550-3213(98)00069-8} {\bibfield
  {journal} {\bibinfo  {journal} {Nuclear Physics B}\ }\textbf {\bibinfo
  {volume} {516}},\ \bibinfo {pages} {719} (\bibinfo {year}
  {1998})}\BibitemShut {NoStop}%
\bibitem [{\citenamefont {Read}(1998)}]{PhysRevB.58.16262}%
  \BibitemOpen
  \bibfield  {author} {\bibinfo {author} {\bibfnamefont {N.}~\bibnamefont
  {Read}},\ }\bibfield  {title} {\bibinfo {title} {Lowest-landau-level theory
  of the quantum hall effect: The fermi-liquid-like state of bosons at filling
  factor one},\ }\href {https://doi.org/10.1103/PhysRevB.58.16262} {\bibfield
  {journal} {\bibinfo  {journal} {Phys. Rev. B}\ }\textbf {\bibinfo {volume}
  {58}},\ \bibinfo {pages} {16262} (\bibinfo {year} {1998})}\BibitemShut
  {NoStop}%
\bibitem [{\citenamefont {Gaiotto}\ and\ \citenamefont
  {Kapustin}(2016)}]{Gaiotto2016}%
  \BibitemOpen
  \bibfield  {author} {\bibinfo {author} {\bibfnamefont {D.}~\bibnamefont
  {Gaiotto}}\ and\ \bibinfo {author} {\bibfnamefont {A.}~\bibnamefont
  {Kapustin}},\ }\bibfield  {title} {\bibinfo {title} {Spin tqfts and fermionic
  phases of matter},\ }\href {https://doi.org/10.1142/S0217751X16450445}
  {\bibfield  {journal} {\bibinfo  {journal} {International Journal of Modern
  Physics A}\ }\textbf {\bibinfo {volume} {31}},\ \bibinfo {pages} {1645044}
  (\bibinfo {year} {2016})}\BibitemShut {NoStop}%
\bibitem [{\citenamefont {Seiberg}\ \emph {et~al.}(2016)\citenamefont
  {Seiberg}, \citenamefont {Senthil}, \citenamefont {Wang},\ and\ \citenamefont
  {Witten}}]{SEIBERG2016395}%
  \BibitemOpen
  \bibfield  {author} {\bibinfo {author} {\bibfnamefont {N.}~\bibnamefont
  {Seiberg}}, \bibinfo {author} {\bibfnamefont {T.}~\bibnamefont {Senthil}},
  \bibinfo {author} {\bibfnamefont {C.}~\bibnamefont {Wang}},\ and\ \bibinfo
  {author} {\bibfnamefont {E.}~\bibnamefont {Witten}},\ }\bibfield  {title}
  {\bibinfo {title} {A duality web in 2+1 dimensions and condensed matter
  physics},\ }\href {https://doi.org/https://doi.org/10.1016/j.aop.2016.08.007}
  {\bibfield  {journal} {\bibinfo  {journal} {Annals of Physics}\ }\textbf
  {\bibinfo {volume} {374}},\ \bibinfo {pages} {395} (\bibinfo {year}
  {2016})}\BibitemShut {NoStop}%
\bibitem [{\citenamefont {Karch}\ and\ \citenamefont
  {Tong}(2016)}]{PhysRevX.6.031043}%
  \BibitemOpen
  \bibfield  {author} {\bibinfo {author} {\bibfnamefont {A.}~\bibnamefont
  {Karch}}\ and\ \bibinfo {author} {\bibfnamefont {D.}~\bibnamefont {Tong}},\
  }\bibfield  {title} {\bibinfo {title} {Particle-vortex duality from 3d
  bosonization},\ }\href {https://doi.org/10.1103/PhysRevX.6.031043} {\bibfield
   {journal} {\bibinfo  {journal} {Phys. Rev. X}\ }\textbf {\bibinfo {volume}
  {6}},\ \bibinfo {pages} {031043} (\bibinfo {year} {2016})}\BibitemShut
  {NoStop}%
\bibitem [{\citenamefont {Đorđe Radi{\v c}evi{\'c}}(2016)}]{JHEP032016131}%
  \BibitemOpen
  \bibfield  {author} {\bibinfo {author} {\bibnamefont {Đorđe Radi{\v
  c}evi{\'c}}},\ }\bibfield  {title} {\bibinfo {title} {Disorder operators in
  chern-simons-fermion theories},\ }\href
  {https://doi.org/10.1007/JHEP03(2016)131} {\bibfield  {journal} {\bibinfo
  {journal} {Journal of High Energy Physics}\ }\textbf {\bibinfo {volume}
  {2016}},\ \bibinfo {pages} {131} (\bibinfo {year} {2016})}\BibitemShut
  {NoStop}%
\bibitem [{\citenamefont {Kapustin}\ and\ \citenamefont
  {Thorngren}(2017)}]{Kapustin2017}%
  \BibitemOpen
  \bibfield  {author} {\bibinfo {author} {\bibfnamefont {A.}~\bibnamefont
  {Kapustin}}\ and\ \bibinfo {author} {\bibfnamefont {R.}~\bibnamefont
  {Thorngren}},\ }\bibfield  {title} {\bibinfo {title} {Fermionic spt phases in
  higher dimensions and bosonization},\ }\href
  {https://doi.org/10.1007/JHEP10(2017)080} {\bibfield  {journal} {\bibinfo
  {journal} {Journal of High Energy Physics}\ }\textbf {\bibinfo {volume}
  {2017}},\ \bibinfo {pages} {80} (\bibinfo {year} {2017})}\BibitemShut
  {NoStop}%
\bibitem [{\citenamefont {Bhardwaj}\ \emph {et~al.}(2017)\citenamefont
  {Bhardwaj}, \citenamefont {Gaiotto},\ and\ \citenamefont
  {Kapustin}}]{Bhardwaj2017}%
  \BibitemOpen
  \bibfield  {author} {\bibinfo {author} {\bibfnamefont {L.}~\bibnamefont
  {Bhardwaj}}, \bibinfo {author} {\bibfnamefont {D.}~\bibnamefont {Gaiotto}},\
  and\ \bibinfo {author} {\bibfnamefont {A.}~\bibnamefont {Kapustin}},\
  }\bibfield  {title} {\bibinfo {title} {State sum constructions of spin-tfts
  and string net constructions of fermionic phases of matter},\ }\href
  {https://doi.org/10.1007/JHEP04(2017)096} {\bibfield  {journal} {\bibinfo
  {journal} {Journal of High Energy Physics}\ }\textbf {\bibinfo {volume}
  {2017}},\ \bibinfo {pages} {96} (\bibinfo {year} {2017})}\BibitemShut
  {NoStop}%
\bibitem [{\citenamefont {Chen}\ \emph
  {et~al.}(2018{\natexlab{a}})\citenamefont {Chen}, \citenamefont {Son},
  \citenamefont {Wang},\ and\ \citenamefont {Raghu}}]{PhysRevLett.120.016602}%
  \BibitemOpen
  \bibfield  {author} {\bibinfo {author} {\bibfnamefont {J.-Y.}\ \bibnamefont
  {Chen}}, \bibinfo {author} {\bibfnamefont {J.~H.}\ \bibnamefont {Son}},
  \bibinfo {author} {\bibfnamefont {C.}~\bibnamefont {Wang}},\ and\ \bibinfo
  {author} {\bibfnamefont {S.}~\bibnamefont {Raghu}},\ }\bibfield  {title}
  {\bibinfo {title} {Exact boson-fermion duality on a 3d euclidean lattice},\
  }\href {https://doi.org/10.1103/PhysRevLett.120.016602} {\bibfield  {journal}
  {\bibinfo  {journal} {Phys. Rev. Lett.}\ }\textbf {\bibinfo {volume} {120}},\
  \bibinfo {pages} {016602} (\bibinfo {year} {2018}{\natexlab{a}})}\BibitemShut
  {NoStop}%
\bibitem [{\citenamefont {Đorđe Radi{\v
  c}evi{\'c}}(2019)}]{radicevic2019spin}%
  \BibitemOpen
  \bibfield  {author} {\bibinfo {author} {\bibnamefont {Đorđe Radi{\v
  c}evi{\'c}}},\ }\href@noop {} {\bibinfo {title} {Spin structures and exact
  dualities in low dimensions}} (\bibinfo {year} {2019}),\ \Eprint
  {https://arxiv.org/abs/1809.07757} {arXiv:1809.07757} \BibitemShut {NoStop}%
\bibitem [{\citenamefont {Senthil}\ \emph {et~al.}(2019)\citenamefont
  {Senthil}, \citenamefont {Son}, \citenamefont {Wang},\ and\ \citenamefont
  {Xu}}]{SENTHIL20191}%
  \BibitemOpen
  \bibfield  {author} {\bibinfo {author} {\bibfnamefont {T.}~\bibnamefont
  {Senthil}}, \bibinfo {author} {\bibfnamefont {D.~T.}\ \bibnamefont {Son}},
  \bibinfo {author} {\bibfnamefont {C.}~\bibnamefont {Wang}},\ and\ \bibinfo
  {author} {\bibfnamefont {C.}~\bibnamefont {Xu}},\ }\bibfield  {title}
  {\bibinfo {title} {Duality between (2+1)d quantum critical points},\ }\href
  {https://doi.org/https://doi.org/10.1016/j.physrep.2019.09.001} {\bibfield
  {journal} {\bibinfo  {journal} {Physics Reports}\ }\textbf {\bibinfo {volume}
  {827}},\ \bibinfo {pages} {1} (\bibinfo {year} {2019})},\ \bibinfo {note}
  {duality between (2+1)d quantum critical points}\BibitemShut {NoStop}%
\bibitem [{\citenamefont {Karch}\ \emph {et~al.}(2019)\citenamefont {Karch},
  \citenamefont {Tong},\ and\ \citenamefont
  {Turner}}]{10.21468/SciPostPhys.7.1.007}%
  \BibitemOpen
  \bibfield  {author} {\bibinfo {author} {\bibfnamefont {A.}~\bibnamefont
  {Karch}}, \bibinfo {author} {\bibfnamefont {D.}~\bibnamefont {Tong}},\ and\
  \bibinfo {author} {\bibfnamefont {C.}~\bibnamefont {Turner}},\ }\bibfield
  {title} {\bibinfo {title} {{A Web of 2d Dualities: ${\bf Z}_2$ Gauge Fields
  and Arf Invariants}},\ }\href {https://doi.org/10.21468/SciPostPhys.7.1.007}
  {\bibfield  {journal} {\bibinfo  {journal} {SciPost Phys.}\ }\textbf
  {\bibinfo {volume} {7}},\ \bibinfo {pages} {7} (\bibinfo {year}
  {2019})}\BibitemShut {NoStop}%
\bibitem [{\citenamefont {Thorngren}(2020)}]{Thorngren2020}%
  \BibitemOpen
  \bibfield  {author} {\bibinfo {author} {\bibfnamefont {R.}~\bibnamefont
  {Thorngren}},\ }\bibfield  {title} {\bibinfo {title} {Anomalies and
  bosonization},\ }\href {https://doi.org/10.1007/s00220-020-03830-0}
  {\bibfield  {journal} {\bibinfo  {journal} {Communications in Mathematical
  Physics}\ }\textbf {\bibinfo {volume} {378}},\ \bibinfo {pages} {1775}
  (\bibinfo {year} {2020})}\BibitemShut {NoStop}%
\bibitem [{\citenamefont {Huang}\ and\ \citenamefont
  {Lee}(2021)}]{huang2021nonabelian}%
  \BibitemOpen
  \bibfield  {author} {\bibinfo {author} {\bibfnamefont {Y.-T.}\ \bibnamefont
  {Huang}}\ and\ \bibinfo {author} {\bibfnamefont {D.-H.}\ \bibnamefont
  {Lee}},\ }\href@noop {} {\bibinfo {title} {Non-abelian bosonization in two
  and three spatial dimensions and applications}} (\bibinfo {year} {2021}),\
  \Eprint {https://arxiv.org/abs/2104.07015} {arXiv:2104.07015} \BibitemShut
  {NoStop}%
\bibitem [{\citenamefont {Wosiek}(1982)}]{Wosiek:1981mn}%
  \BibitemOpen
  \bibfield  {author} {\bibinfo {author} {\bibfnamefont {J.}~\bibnamefont
  {Wosiek}},\ }\bibfield  {title} {\bibinfo {title} {{A LOCAL REPRESENTATION
  FOR FERMIONS ON A LATTICE}},\ }\href@noop {} {\bibfield  {journal} {\bibinfo
  {journal} {Acta Phys. Polon. B}\ }\textbf {\bibinfo {volume} {13}},\ \bibinfo
  {pages} {543} (\bibinfo {year} {1982})}\BibitemShut {NoStop}%
\bibitem [{\citenamefont {Szczerba}(1985)}]{cmp/1103942539}%
  \BibitemOpen
  \bibfield  {author} {\bibinfo {author} {\bibfnamefont {A.~M.}\ \bibnamefont
  {Szczerba}},\ }\bibfield  {title} {\bibinfo {title} {{Spins and fermions on
  arbitrary lattices}},\ }\href {https://doi.org/cmp/1103942539} {\bibfield
  {journal} {\bibinfo  {journal} {Communications in Mathematical Physics}\
  }\textbf {\bibinfo {volume} {98}},\ \bibinfo {pages} {513 } (\bibinfo {year}
  {1985})}\BibitemShut {NoStop}%
\bibitem [{\citenamefont {Fradkin}(1989)}]{PhysRevLett.63.322}%
  \BibitemOpen
  \bibfield  {author} {\bibinfo {author} {\bibfnamefont {E.}~\bibnamefont
  {Fradkin}},\ }\bibfield  {title} {\bibinfo {title} {Jordan-wigner
  transformation for quantum-spin systems in two dimensions and fractional
  statistics},\ }\href {https://doi.org/10.1103/PhysRevLett.63.322} {\bibfield
  {journal} {\bibinfo  {journal} {Phys. Rev. Lett.}\ }\textbf {\bibinfo
  {volume} {63}},\ \bibinfo {pages} {322} (\bibinfo {year} {1989})}\BibitemShut
  {NoStop}%
\bibitem [{\citenamefont {Wang}(1991)}]{PhysRevB.43.3786}%
  \BibitemOpen
  \bibfield  {author} {\bibinfo {author} {\bibfnamefont {Y.~R.}\ \bibnamefont
  {Wang}},\ }\bibfield  {title} {\bibinfo {title} {Ground state of the
  two-dimensional antiferromagnetic heisenberg model studied using an extended
  wigner-jordon transformation},\ }\href
  {https://doi.org/10.1103/PhysRevB.43.3786} {\bibfield  {journal} {\bibinfo
  {journal} {Phys. Rev. B}\ }\textbf {\bibinfo {volume} {43}},\ \bibinfo
  {pages} {3786} (\bibinfo {year} {1991})}\BibitemShut {NoStop}%
\bibitem [{\citenamefont {Weng}\ \emph {et~al.}(2000)\citenamefont {Weng},
  \citenamefont {Sheng},\ and\ \citenamefont {Ting}}]{WENG200067}%
  \BibitemOpen
  \bibfield  {author} {\bibinfo {author} {\bibfnamefont {Z.}~\bibnamefont
  {Weng}}, \bibinfo {author} {\bibfnamefont {D.}~\bibnamefont {Sheng}},\ and\
  \bibinfo {author} {\bibfnamefont {C.}~\bibnamefont {Ting}},\ }\bibfield
  {title} {\bibinfo {title} {Understanding high-tc cuprates based on the phase
  string theory of doped antiferromagnet},\ }\href
  {https://doi.org/https://doi.org/10.1016/S0921-4534(00)00391-9} {\bibfield
  {journal} {\bibinfo  {journal} {Physica C: Superconductivity}\ }\textbf
  {\bibinfo {volume} {341-348}},\ \bibinfo {pages} {67} (\bibinfo {year}
  {2000})}\BibitemShut {NoStop}%
\bibitem [{\citenamefont {Batista}\ and\ \citenamefont
  {Ortiz}(2001)}]{PhysRevLett.86.1082}%
  \BibitemOpen
  \bibfield  {author} {\bibinfo {author} {\bibfnamefont {C.~D.}\ \bibnamefont
  {Batista}}\ and\ \bibinfo {author} {\bibfnamefont {G.}~\bibnamefont
  {Ortiz}},\ }\bibfield  {title} {\bibinfo {title} {Generalized jordan-wigner
  transformations},\ }\href {https://doi.org/10.1103/PhysRevLett.86.1082}
  {\bibfield  {journal} {\bibinfo  {journal} {Phys. Rev. Lett.}\ }\textbf
  {\bibinfo {volume} {86}},\ \bibinfo {pages} {1082} (\bibinfo {year}
  {2001})}\BibitemShut {NoStop}%
\bibitem [{\citenamefont {Dobrov}(2003)}]{Dobrov_2003}%
  \BibitemOpen
  \bibfield  {author} {\bibinfo {author} {\bibfnamefont {S.~V.}\ \bibnamefont
  {Dobrov}},\ }\bibfield  {title} {\bibinfo {title} {On the spin-fermion
  connection},\ }\href {https://doi.org/10.1088/0305-4470/36/39/101} {\bibfield
   {journal} {\bibinfo  {journal} {Journal of Physics A: Mathematical and
  General}\ }\textbf {\bibinfo {volume} {36}},\ \bibinfo {pages} {L503}
  (\bibinfo {year} {2003})}\BibitemShut {NoStop}%
\bibitem [{\citenamefont {Ball}(2005)}]{PhysRevLett.95.176407}%
  \BibitemOpen
  \bibfield  {author} {\bibinfo {author} {\bibfnamefont {R.~C.}\ \bibnamefont
  {Ball}},\ }\bibfield  {title} {\bibinfo {title} {Fermions without fermion
  fields},\ }\href {https://doi.org/10.1103/PhysRevLett.95.176407} {\bibfield
  {journal} {\bibinfo  {journal} {Phys. Rev. Lett.}\ }\textbf {\bibinfo
  {volume} {95}},\ \bibinfo {pages} {176407} (\bibinfo {year}
  {2005})}\BibitemShut {NoStop}%
\bibitem [{\citenamefont {Verstraete}\ and\ \citenamefont
  {Cirac}(2005)}]{Verstraete_2005}%
  \BibitemOpen
  \bibfield  {author} {\bibinfo {author} {\bibfnamefont {F.}~\bibnamefont
  {Verstraete}}\ and\ \bibinfo {author} {\bibfnamefont {J.~I.}\ \bibnamefont
  {Cirac}},\ }\bibfield  {title} {\bibinfo {title} {Mapping local hamiltonians
  of fermions to local hamiltonians of spins},\ }\href
  {https://doi.org/10.1088/1742-5468/2005/09/p09012} {\bibfield  {journal}
  {\bibinfo  {journal} {Journal of Statistical Mechanics: Theory and
  Experiment}\ }\textbf {\bibinfo {volume} {2005}},\ \bibinfo {pages} {P09012}
  (\bibinfo {year} {2005})}\BibitemShut {NoStop}%
\bibitem [{\citenamefont {Chen}\ \emph {et~al.}(2007)\citenamefont {Chen},
  \citenamefont {Fang}, \citenamefont {Hu},\ and\ \citenamefont
  {Yao}}]{PhysRevB.75.144401}%
  \BibitemOpen
  \bibfield  {author} {\bibinfo {author} {\bibfnamefont {H.-D.}\ \bibnamefont
  {Chen}}, \bibinfo {author} {\bibfnamefont {C.}~\bibnamefont {Fang}}, \bibinfo
  {author} {\bibfnamefont {J.}~\bibnamefont {Hu}},\ and\ \bibinfo {author}
  {\bibfnamefont {H.}~\bibnamefont {Yao}},\ }\bibfield  {title} {\bibinfo
  {title} {Quantum phase transition in the quantum compass model},\ }\href
  {https://doi.org/10.1103/PhysRevB.75.144401} {\bibfield  {journal} {\bibinfo
  {journal} {Phys. Rev. B}\ }\textbf {\bibinfo {volume} {75}},\ \bibinfo
  {pages} {144401} (\bibinfo {year} {2007})}\BibitemShut {NoStop}%
\bibitem [{\citenamefont {Feng}\ \emph {et~al.}(2007)\citenamefont {Feng},
  \citenamefont {Zhang},\ and\ \citenamefont {Xiang}}]{PhysRevLett.98.087204}%
  \BibitemOpen
  \bibfield  {author} {\bibinfo {author} {\bibfnamefont {X.-Y.}\ \bibnamefont
  {Feng}}, \bibinfo {author} {\bibfnamefont {G.-M.}\ \bibnamefont {Zhang}},\
  and\ \bibinfo {author} {\bibfnamefont {T.}~\bibnamefont {Xiang}},\ }\bibfield
   {title} {\bibinfo {title} {Topological characterization of quantum phase
  transitions in a spin-$1/2$ model},\ }\href
  {https://doi.org/10.1103/PhysRevLett.98.087204} {\bibfield  {journal}
  {\bibinfo  {journal} {Phys. Rev. Lett.}\ }\textbf {\bibinfo {volume} {98}},\
  \bibinfo {pages} {087204} (\bibinfo {year} {2007})}\BibitemShut {NoStop}%
\bibitem [{\citenamefont {Chen}\ and\ \citenamefont
  {Hu}(2007)}]{PhysRevB.76.193101}%
  \BibitemOpen
  \bibfield  {author} {\bibinfo {author} {\bibfnamefont {H.-D.}\ \bibnamefont
  {Chen}}\ and\ \bibinfo {author} {\bibfnamefont {J.}~\bibnamefont {Hu}},\
  }\bibfield  {title} {\bibinfo {title} {Exact mapping between classical and
  topological orders in two-dimensional spin systems},\ }\href
  {https://doi.org/10.1103/PhysRevB.76.193101} {\bibfield  {journal} {\bibinfo
  {journal} {Phys. Rev. B}\ }\textbf {\bibinfo {volume} {76}},\ \bibinfo
  {pages} {193101} (\bibinfo {year} {2007})}\BibitemShut {NoStop}%
\bibitem [{\citenamefont {Chen}\ and\ \citenamefont
  {Nussinov}(2008)}]{Chen_2008}%
  \BibitemOpen
  \bibfield  {author} {\bibinfo {author} {\bibfnamefont {H.-D.}\ \bibnamefont
  {Chen}}\ and\ \bibinfo {author} {\bibfnamefont {Z.}~\bibnamefont
  {Nussinov}},\ }\bibfield  {title} {\bibinfo {title} {Exact results of the
  kitaev model on a hexagonal lattice: spin states, string and brane
  correlators, and anyonic excitations},\ }\href
  {https://doi.org/10.1088/1751-8113/41/7/075001} {\bibfield  {journal}
  {\bibinfo  {journal} {Journal of Physics A: Mathematical and Theoretical}\
  }\textbf {\bibinfo {volume} {41}},\ \bibinfo {pages} {075001} (\bibinfo
  {year} {2008})}\BibitemShut {NoStop}%
\bibitem [{\citenamefont {Cobanera}\ \emph {et~al.}(2011)\citenamefont
  {Cobanera}, \citenamefont {Ortiz},\ and\ \citenamefont
  {Nussinov}}]{Cobanera2011}%
  \BibitemOpen
  \bibfield  {author} {\bibinfo {author} {\bibfnamefont {E.}~\bibnamefont
  {Cobanera}}, \bibinfo {author} {\bibfnamefont {G.}~\bibnamefont {Ortiz}},\
  and\ \bibinfo {author} {\bibfnamefont {Z.}~\bibnamefont {Nussinov}},\
  }\bibfield  {title} {\bibinfo {title} {The bond-algebraic approach to
  dualities},\ }\href {https://doi.org/10.1080/00018732.2011.619814} {\bibfield
   {journal} {\bibinfo  {journal} {Advances in Physics}\ }\textbf {\bibinfo
  {volume} {60}},\ \bibinfo {pages} {679} (\bibinfo {year} {2011})}\BibitemShut
  {NoStop}%
\bibitem [{\citenamefont {Nussinov}\ \emph {et~al.}(2012)\citenamefont
  {Nussinov}, \citenamefont {Ortiz},\ and\ \citenamefont
  {Cobanera}}]{PhysRevB.86.085415}%
  \BibitemOpen
  \bibfield  {author} {\bibinfo {author} {\bibfnamefont {Z.}~\bibnamefont
  {Nussinov}}, \bibinfo {author} {\bibfnamefont {G.}~\bibnamefont {Ortiz}},\
  and\ \bibinfo {author} {\bibfnamefont {E.}~\bibnamefont {Cobanera}},\
  }\bibfield  {title} {\bibinfo {title} {Arbitrary dimensional majorana
  dualities and architectures for topological matter},\ }\href
  {https://doi.org/10.1103/PhysRevB.86.085415} {\bibfield  {journal} {\bibinfo
  {journal} {Phys. Rev. B}\ }\textbf {\bibinfo {volume} {86}},\ \bibinfo
  {pages} {085415} (\bibinfo {year} {2012})}\BibitemShut {NoStop}%
\bibitem [{\citenamefont {Zohar}\ and\ \citenamefont
  {Cirac}(2018)}]{PhysRevB.98.075119}%
  \BibitemOpen
  \bibfield  {author} {\bibinfo {author} {\bibfnamefont {E.}~\bibnamefont
  {Zohar}}\ and\ \bibinfo {author} {\bibfnamefont {J.~I.}\ \bibnamefont
  {Cirac}},\ }\bibfield  {title} {\bibinfo {title} {Eliminating fermionic
  matter fields in lattice gauge theories},\ }\href
  {https://doi.org/10.1103/PhysRevB.98.075119} {\bibfield  {journal} {\bibinfo
  {journal} {Phys. Rev. B}\ }\textbf {\bibinfo {volume} {98}},\ \bibinfo
  {pages} {075119} (\bibinfo {year} {2018})}\BibitemShut {NoStop}%
\bibitem [{\citenamefont {Minami}(2016)}]{Minami2016}%
  \BibitemOpen
  \bibfield  {author} {\bibinfo {author} {\bibfnamefont {K.}~\bibnamefont
  {Minami}},\ }\bibfield  {title} {\bibinfo {title} {Solvable hamiltonians and
  fermionization transformations obtained from operators satisfying specific
  commutation relations},\ }\href {https://doi.org/10.7566/JPSJ.85.024003}
  {\bibfield  {journal} {\bibinfo  {journal} {Journal of the Physical Society
  of Japan}\ }\textbf {\bibinfo {volume} {85}},\ \bibinfo {pages} {024003}
  (\bibinfo {year} {2016})}\BibitemShut {NoStop}%
\bibitem [{\citenamefont {Chen}\ \emph
  {et~al.}(2018{\natexlab{b}})\citenamefont {Chen}, \citenamefont {Kapustin},\
  and\ \citenamefont {Radi{\v c}evi{\'c}}}]{CHEN2018234}%
  \BibitemOpen
  \bibfield  {author} {\bibinfo {author} {\bibfnamefont {Y.-A.}\ \bibnamefont
  {Chen}}, \bibinfo {author} {\bibfnamefont {A.}~\bibnamefont {Kapustin}},\
  and\ \bibinfo {author} {\bibfnamefont {D.}~\bibnamefont {Radi{\v
  c}evi{\'c}}},\ }\bibfield  {title} {\bibinfo {title} {Exact bosonization in
  two spatial dimensions and a new class of lattice gauge theories},\ }\href
  {https://doi.org/https://doi.org/10.1016/j.aop.2018.03.024} {\bibfield
  {journal} {\bibinfo  {journal} {Annals of Physics}\ }\textbf {\bibinfo
  {volume} {393}},\ \bibinfo {pages} {234} (\bibinfo {year}
  {2018}{\natexlab{b}})}\BibitemShut {NoStop}%
\bibitem [{\citenamefont {Chen}\ and\ \citenamefont
  {Kapustin}(2019)}]{PhysRevB.100.245127}%
  \BibitemOpen
  \bibfield  {author} {\bibinfo {author} {\bibfnamefont {Y.-A.}\ \bibnamefont
  {Chen}}\ and\ \bibinfo {author} {\bibfnamefont {A.}~\bibnamefont
  {Kapustin}},\ }\bibfield  {title} {\bibinfo {title} {Bosonization in three
  spatial dimensions and a 2-form gauge theory},\ }\href
  {https://doi.org/10.1103/PhysRevB.100.245127} {\bibfield  {journal} {\bibinfo
   {journal} {Phys. Rev. B}\ }\textbf {\bibinfo {volume} {100}},\ \bibinfo
  {pages} {245127} (\bibinfo {year} {2019})}\BibitemShut {NoStop}%
\bibitem [{\citenamefont {Minami}(2019)}]{MINAMI2019465}%
  \BibitemOpen
  \bibfield  {author} {\bibinfo {author} {\bibfnamefont {K.}~\bibnamefont
  {Minami}},\ }\bibfield  {title} {\bibinfo {title} {Honeycomb lattice kitaev
  model with wen--toric-code interactions, and anyon excitations},\ }\href
  {https://doi.org/https://doi.org/10.1016/j.nuclphysb.2018.12.029} {\bibfield
  {journal} {\bibinfo  {journal} {Nuclear Physics B}\ }\textbf {\bibinfo
  {volume} {939}},\ \bibinfo {pages} {465} (\bibinfo {year}
  {2019})}\BibitemShut {NoStop}%
\bibitem [{\citenamefont {Chen}(2020)}]{PhysRevResearch.2.033527}%
  \BibitemOpen
  \bibfield  {author} {\bibinfo {author} {\bibfnamefont {Y.-A.}\ \bibnamefont
  {Chen}},\ }\bibfield  {title} {\bibinfo {title} {Exact bosonization in
  arbitrary dimensions},\ }\href
  {https://doi.org/10.1103/PhysRevResearch.2.033527} {\bibfield  {journal}
  {\bibinfo  {journal} {Phys. Rev. Research}\ }\textbf {\bibinfo {volume}
  {2}},\ \bibinfo {pages} {033527} (\bibinfo {year} {2020})}\BibitemShut
  {NoStop}%
\bibitem [{\citenamefont {Bochniak}\ \emph {et~al.}(2020)\citenamefont
  {Bochniak}, \citenamefont {Ruba}, \citenamefont {Wosiek},\ and\ \citenamefont
  {Wyrzykowski}}]{PhysRevD.102.114502}%
  \BibitemOpen
  \bibfield  {author} {\bibinfo {author} {\bibfnamefont {A.}~\bibnamefont
  {Bochniak}}, \bibinfo {author} {\bibfnamefont {B.~d.~z.}\ \bibnamefont
  {Ruba}}, \bibinfo {author} {\bibfnamefont {J.}~\bibnamefont {Wosiek}},\ and\
  \bibinfo {author} {\bibfnamefont {A.}~\bibnamefont {Wyrzykowski}},\
  }\bibfield  {title} {\bibinfo {title} {Constraints of kinematic bosonization
  in two and higher dimensions},\ }\href
  {https://doi.org/10.1103/PhysRevD.102.114502} {\bibfield  {journal} {\bibinfo
   {journal} {Phys. Rev. D}\ }\textbf {\bibinfo {volume} {102}},\ \bibinfo
  {pages} {114502} (\bibinfo {year} {2020})}\BibitemShut {NoStop}%
\bibitem [{\citenamefont {Bochniak}\ and\ \citenamefont
  {Ruba}(2020)}]{Clifford}%
  \BibitemOpen
  \bibfield  {author} {\bibinfo {author} {\bibfnamefont {A.}~\bibnamefont
  {Bochniak}}\ and\ \bibinfo {author} {\bibfnamefont {B.}~\bibnamefont
  {Ruba}},\ }\bibfield  {title} {\bibinfo {title} {Bosonization based on
  clifford algebras and its gauge theoretic interpretation},\ }\href
  {https://doi.org/10.1007/JHEP12(2020)118} {\bibfield  {journal} {\bibinfo
  {journal} {Journal of High Energy Physics}\ }\textbf {\bibinfo {volume}
  {2020}},\ \bibinfo {pages} {118} (\bibinfo {year} {2020})}\BibitemShut
  {NoStop}%
\bibitem [{\citenamefont {Ogura}\ \emph {et~al.}(2020)\citenamefont {Ogura},
  \citenamefont {Imamura}, \citenamefont {Kameyama}, \citenamefont {Minami},\
  and\ \citenamefont {Sato}}]{PhysRevB.102.245118}%
  \BibitemOpen
  \bibfield  {author} {\bibinfo {author} {\bibfnamefont {M.}~\bibnamefont
  {Ogura}}, \bibinfo {author} {\bibfnamefont {Y.}~\bibnamefont {Imamura}},
  \bibinfo {author} {\bibfnamefont {N.}~\bibnamefont {Kameyama}}, \bibinfo
  {author} {\bibfnamefont {K.}~\bibnamefont {Minami}},\ and\ \bibinfo {author}
  {\bibfnamefont {M.}~\bibnamefont {Sato}},\ }\bibfield  {title} {\bibinfo
  {title} {Geometric criterion for solvability of lattice spin systems},\
  }\href {https://doi.org/10.1103/PhysRevB.102.245118} {\bibfield  {journal}
  {\bibinfo  {journal} {Phys. Rev. B}\ }\textbf {\bibinfo {volume} {102}},\
  \bibinfo {pages} {245118} (\bibinfo {year} {2020})}\BibitemShut {NoStop}%
\bibitem [{\citenamefont {Meier}\ \emph {et~al.}(2021)\citenamefont {Meier},
  \citenamefont {Waltner}, \citenamefont {Braun},\ and\ \citenamefont
  {Guhr}}]{Meier_2021}%
  \BibitemOpen
  \bibfield  {author} {\bibinfo {author} {\bibfnamefont {F.}~\bibnamefont
  {Meier}}, \bibinfo {author} {\bibfnamefont {D.}~\bibnamefont {Waltner}},
  \bibinfo {author} {\bibfnamefont {P.}~\bibnamefont {Braun}},\ and\ \bibinfo
  {author} {\bibfnamefont {T.}~\bibnamefont {Guhr}},\ }\bibfield  {title}
  {\bibinfo {title} {A mapping between the spin and fermion algebra},\ }\href
  {https://doi.org/10.1088/1751-8121/ac13dc} {\bibfield  {journal} {\bibinfo
  {journal} {Journal of Physics A: Mathematical and Theoretical}\ }\textbf
  {\bibinfo {volume} {54}},\ \bibinfo {pages} {345201} (\bibinfo {year}
  {2021})}\BibitemShut {NoStop}%
\bibitem [{\citenamefont {Bravyi}\ and\ \citenamefont
  {Kitaev}(2002)}]{BRAVYI2002210}%
  \BibitemOpen
  \bibfield  {author} {\bibinfo {author} {\bibfnamefont {S.~B.}\ \bibnamefont
  {Bravyi}}\ and\ \bibinfo {author} {\bibfnamefont {A.~Y.}\ \bibnamefont
  {Kitaev}},\ }\bibfield  {title} {\bibinfo {title} {Fermionic quantum
  computation},\ }\href
  {https://doi.org/https://doi.org/10.1006/aphy.2002.6254} {\bibfield
  {journal} {\bibinfo  {journal} {Annals of Physics}\ }\textbf {\bibinfo
  {volume} {298}},\ \bibinfo {pages} {210} (\bibinfo {year}
  {2002})}\BibitemShut {NoStop}%
\bibitem [{\citenamefont {Seeley}\ \emph {et~al.}(2012)\citenamefont {Seeley},
  \citenamefont {Richard},\ and\ \citenamefont {Love}}]{Seeley2012}%
  \BibitemOpen
  \bibfield  {author} {\bibinfo {author} {\bibfnamefont {J.~T.}\ \bibnamefont
  {Seeley}}, \bibinfo {author} {\bibfnamefont {M.~J.}\ \bibnamefont
  {Richard}},\ and\ \bibinfo {author} {\bibfnamefont {P.~J.}\ \bibnamefont
  {Love}},\ }\bibfield  {title} {\bibinfo {title} {The bravyi-kitaev
  transformation for quantum computation of electronic structure},\ }\href
  {https://doi.org/10.1063/1.4768229} {\bibfield  {journal} {\bibinfo
  {journal} {The Journal of Chemical Physics}\ }\textbf {\bibinfo {volume}
  {137}},\ \bibinfo {pages} {224109} (\bibinfo {year} {2012})}\BibitemShut
  {NoStop}%
\bibitem [{\citenamefont {Setia}\ and\ \citenamefont
  {Whitfield}(2018)}]{setia2018bravyi}%
  \BibitemOpen
  \bibfield  {author} {\bibinfo {author} {\bibfnamefont {K.}~\bibnamefont
  {Setia}}\ and\ \bibinfo {author} {\bibfnamefont {J.~D.}\ \bibnamefont
  {Whitfield}},\ }\bibfield  {title} {\bibinfo {title} {Bravyi-kitaev superfast
  simulation of electronic structure on a quantum computer},\ }\href
  {https://aip.scitation.org/doi/full/10.1063/1.5019371} {\bibfield  {journal}
  {\bibinfo  {journal} {The Journal of chemical physics}\ }\textbf {\bibinfo
  {volume} {148}},\ \bibinfo {pages} {164104} (\bibinfo {year}
  {2018})}\BibitemShut {NoStop}%
\bibitem [{\citenamefont {Setia}\ \emph {et~al.}(2019)\citenamefont {Setia},
  \citenamefont {Bravyi}, \citenamefont {Mezzacapo},\ and\ \citenamefont
  {Whitfield}}]{setia2019superfast}%
  \BibitemOpen
  \bibfield  {author} {\bibinfo {author} {\bibfnamefont {K.}~\bibnamefont
  {Setia}}, \bibinfo {author} {\bibfnamefont {S.}~\bibnamefont {Bravyi}},
  \bibinfo {author} {\bibfnamefont {A.}~\bibnamefont {Mezzacapo}},\ and\
  \bibinfo {author} {\bibfnamefont {J.~D.}\ \bibnamefont {Whitfield}},\
  }\bibfield  {title} {\bibinfo {title} {Superfast encodings for fermionic
  quantum simulation},\ }\href
  {https://journals.aps.org/prresearch/abstract/10.1103/PhysRevResearch.1.033033}
  {\bibfield  {journal} {\bibinfo  {journal} {Physical Review Research}\
  }\textbf {\bibinfo {volume} {1}},\ \bibinfo {pages} {033033} (\bibinfo {year}
  {2019})}\BibitemShut {NoStop}%
\bibitem [{\citenamefont {Derby}\ \emph {et~al.}(2021)\citenamefont {Derby},
  \citenamefont {Klassen}, \citenamefont {Bausch},\ and\ \citenamefont
  {Cubitt}}]{PhysRevB.104.035118}%
  \BibitemOpen
  \bibfield  {author} {\bibinfo {author} {\bibfnamefont {C.}~\bibnamefont
  {Derby}}, \bibinfo {author} {\bibfnamefont {J.}~\bibnamefont {Klassen}},
  \bibinfo {author} {\bibfnamefont {J.}~\bibnamefont {Bausch}},\ and\ \bibinfo
  {author} {\bibfnamefont {T.}~\bibnamefont {Cubitt}},\ }\bibfield  {title}
  {\bibinfo {title} {Compact fermion to qubit mappings},\ }\href
  {https://doi.org/10.1103/PhysRevB.104.035118} {\bibfield  {journal} {\bibinfo
   {journal} {Phys. Rev. B}\ }\textbf {\bibinfo {volume} {104}},\ \bibinfo
  {pages} {035118} (\bibinfo {year} {2021})}\BibitemShut {NoStop}%
\bibitem [{\citenamefont {Derby}\ and\ \citenamefont
  {Klassen}(2021)}]{derby2021compact}%
  \BibitemOpen
  \bibfield  {author} {\bibinfo {author} {\bibfnamefont {C.}~\bibnamefont
  {Derby}}\ and\ \bibinfo {author} {\bibfnamefont {J.}~\bibnamefont
  {Klassen}},\ }\bibfield  {title} {\bibinfo {title} {A compact fermion to
  qubit mapping part 2: Alternative lattice geometries},\ }\href
  {https://arxiv.org/abs/2101.10735} {\bibfield  {journal} {\bibinfo  {journal}
  {arXiv preprint arXiv:2101.10735}\ } (\bibinfo {year} {2021})}\BibitemShut
  {NoStop}%
\bibitem [{\citenamefont {Wen}(2004)}]{Wen2004}%
  \BibitemOpen
  \bibfield  {author} {\bibinfo {author} {\bibfnamefont {X.-G.}\ \bibnamefont
  {Wen}},\ }\href@noop {} {\emph {\bibinfo {title} {Quantum Field Theory of
  Many-Body Systems: From the Origin of Sound to an Origin of Light and
  Electrons}}}\ (\bibinfo  {publisher} {Oxford University Press},\ \bibinfo
  {year} {2004})\BibitemShut {NoStop}%
\bibitem [{\citenamefont {Levin}\ and\ \citenamefont
  {Gu}(2012)}]{PhysRevB.86.115109}%
  \BibitemOpen
  \bibfield  {author} {\bibinfo {author} {\bibfnamefont {M.}~\bibnamefont
  {Levin}}\ and\ \bibinfo {author} {\bibfnamefont {Z.-C.}\ \bibnamefont {Gu}},\
  }\bibfield  {title} {\bibinfo {title} {Braiding statistics approach to
  symmetry-protected topological phases},\ }\href
  {https://doi.org/10.1103/PhysRevB.86.115109} {\bibfield  {journal} {\bibinfo
  {journal} {Phys. Rev. B}\ }\textbf {\bibinfo {volume} {86}},\ \bibinfo
  {pages} {115109} (\bibinfo {year} {2012})}\BibitemShut {NoStop}%
\bibitem [{\citenamefont {Swingle}(2014)}]{PhysRevB.90.035451}%
  \BibitemOpen
  \bibfield  {author} {\bibinfo {author} {\bibfnamefont {B.}~\bibnamefont
  {Swingle}},\ }\bibfield  {title} {\bibinfo {title} {Interplay between short-
  and long-range entanglement in symmetry-protected phases},\ }\href
  {https://doi.org/10.1103/PhysRevB.90.035451} {\bibfield  {journal} {\bibinfo
  {journal} {Phys. Rev. B}\ }\textbf {\bibinfo {volume} {90}},\ \bibinfo
  {pages} {035451} (\bibinfo {year} {2014})}\BibitemShut {NoStop}%
\bibitem [{\citenamefont {Barkeshli}\ \emph {et~al.}(2019)\citenamefont
  {Barkeshli}, \citenamefont {Bonderson}, \citenamefont {Cheng},\ and\
  \citenamefont {Wang}}]{PhysRevB.100.115147}%
  \BibitemOpen
  \bibfield  {author} {\bibinfo {author} {\bibfnamefont {M.}~\bibnamefont
  {Barkeshli}}, \bibinfo {author} {\bibfnamefont {P.}~\bibnamefont
  {Bonderson}}, \bibinfo {author} {\bibfnamefont {M.}~\bibnamefont {Cheng}},\
  and\ \bibinfo {author} {\bibfnamefont {Z.}~\bibnamefont {Wang}},\ }\bibfield
  {title} {\bibinfo {title} {Symmetry fractionalization, defects, and gauging
  of topological phases},\ }\href {https://doi.org/10.1103/PhysRevB.100.115147}
  {\bibfield  {journal} {\bibinfo  {journal} {Phys. Rev. B}\ }\textbf {\bibinfo
  {volume} {100}},\ \bibinfo {pages} {115147} (\bibinfo {year}
  {2019})}\BibitemShut {NoStop}%
\bibitem [{\citenamefont {Lee}\ \emph {et~al.}(2006)\citenamefont {Lee},
  \citenamefont {Nagaosa},\ and\ \citenamefont {Wen}}]{RevModPhys.78.17}%
  \BibitemOpen
  \bibfield  {author} {\bibinfo {author} {\bibfnamefont {P.~A.}\ \bibnamefont
  {Lee}}, \bibinfo {author} {\bibfnamefont {N.}~\bibnamefont {Nagaosa}},\ and\
  \bibinfo {author} {\bibfnamefont {X.-G.}\ \bibnamefont {Wen}},\ }\bibfield
  {title} {\bibinfo {title} {Doping a mott insulator: Physics of
  high-temperature superconductivity},\ }\href
  {https://doi.org/10.1103/RevModPhys.78.17} {\bibfield  {journal} {\bibinfo
  {journal} {Rev. Mod. Phys.}\ }\textbf {\bibinfo {volume} {78}},\ \bibinfo
  {pages} {17} (\bibinfo {year} {2006})}\BibitemShut {NoStop}%
\bibitem [{\citenamefont {Si}\ and\ \citenamefont {Steglich}(2010)}]{Si1161}%
  \BibitemOpen
  \bibfield  {author} {\bibinfo {author} {\bibfnamefont {Q.}~\bibnamefont
  {Si}}\ and\ \bibinfo {author} {\bibfnamefont {F.}~\bibnamefont {Steglich}},\
  }\bibfield  {title} {\bibinfo {title} {Heavy fermions and quantum phase
  transitions},\ }\href {https://doi.org/10.1126/science.1191195} {\bibfield
  {journal} {\bibinfo  {journal} {Science}\ }\textbf {\bibinfo {volume}
  {329}},\ \bibinfo {pages} {1161} (\bibinfo {year} {2010})}\BibitemShut
  {NoStop}%
\bibitem [{\citenamefont {Barnes}(1976)}]{Barnes_1976}%
  \BibitemOpen
  \bibfield  {author} {\bibinfo {author} {\bibfnamefont {S.~E.}\ \bibnamefont
  {Barnes}},\ }\bibfield  {title} {\bibinfo {title} {New method for the
  anderson model},\ }\href {https://doi.org/10.1088/0305-4608/6/7/018}
  {\bibfield  {journal} {\bibinfo  {journal} {Journal of Physics F: Metal
  Physics}\ }\textbf {\bibinfo {volume} {6}},\ \bibinfo {pages} {1375}
  (\bibinfo {year} {1976})}\BibitemShut {NoStop}%
\bibitem [{\citenamefont {Barnes}(1977)}]{Barnes_1977}%
  \BibitemOpen
  \bibfield  {author} {\bibinfo {author} {\bibfnamefont {S.~E.}\ \bibnamefont
  {Barnes}},\ }\bibfield  {title} {\bibinfo {title} {New method for the
  anderson model. {II}. the u=0 limit},\ }\href
  {https://doi.org/10.1088/0305-4608/7/12/022} {\bibfield  {journal} {\bibinfo
  {journal} {Journal of Physics F: Metal Physics}\ }\textbf {\bibinfo {volume}
  {7}},\ \bibinfo {pages} {2637} (\bibinfo {year} {1977})}\BibitemShut
  {NoStop}%
\bibitem [{\citenamefont {Coleman}(1984)}]{PhysRevB.29.3035}%
  \BibitemOpen
  \bibfield  {author} {\bibinfo {author} {\bibfnamefont {P.}~\bibnamefont
  {Coleman}},\ }\bibfield  {title} {\bibinfo {title} {New approach to the
  mixed-valence problem},\ }\href {https://doi.org/10.1103/PhysRevB.29.3035}
  {\bibfield  {journal} {\bibinfo  {journal} {Phys. Rev. B}\ }\textbf {\bibinfo
  {volume} {29}},\ \bibinfo {pages} {3035} (\bibinfo {year}
  {1984})}\BibitemShut {NoStop}%
\bibitem [{\citenamefont {Xu}\ and\ \citenamefont
  {Sachdev}(2010)}]{PhysRevLett.105.057201}%
  \BibitemOpen
  \bibfield  {author} {\bibinfo {author} {\bibfnamefont {C.}~\bibnamefont
  {Xu}}\ and\ \bibinfo {author} {\bibfnamefont {S.}~\bibnamefont {Sachdev}},\
  }\bibfield  {title} {\bibinfo {title} {Majorana liquids: The complete
  fractionalization of the electron},\ }\href
  {https://doi.org/10.1103/PhysRevLett.105.057201} {\bibfield  {journal}
  {\bibinfo  {journal} {Phys. Rev. Lett.}\ }\textbf {\bibinfo {volume} {105}},\
  \bibinfo {pages} {057201} (\bibinfo {year} {2010})}\BibitemShut {NoStop}%
\bibitem [{\citenamefont {Castelnovo}\ \emph {et~al.}(2012)\citenamefont
  {Castelnovo}, \citenamefont {Moessner},\ and\ \citenamefont
  {Sondhi}}]{Castelnovo2012}%
  \BibitemOpen
  \bibfield  {author} {\bibinfo {author} {\bibfnamefont {C.}~\bibnamefont
  {Castelnovo}}, \bibinfo {author} {\bibfnamefont {R.}~\bibnamefont
  {Moessner}},\ and\ \bibinfo {author} {\bibfnamefont {S.}~\bibnamefont
  {Sondhi}},\ }\bibfield  {title} {\bibinfo {title} {Spin ice,
  fractionalization, and topological order},\ }\href
  {https://doi.org/10.1146/annurev-conmatphys-020911-125058} {\bibfield
  {journal} {\bibinfo  {journal} {Annual Review of Condensed Matter Physics}\
  }\textbf {\bibinfo {volume} {3}},\ \bibinfo {pages} {35} (\bibinfo {year}
  {2012})}\BibitemShut {NoStop}%
\bibitem [{\citenamefont {Savary}\ and\ \citenamefont
  {Balents}(2016)}]{Savary_2016}%
  \BibitemOpen
  \bibfield  {author} {\bibinfo {author} {\bibfnamefont {L.}~\bibnamefont
  {Savary}}\ and\ \bibinfo {author} {\bibfnamefont {L.}~\bibnamefont
  {Balents}},\ }\bibfield  {title} {\bibinfo {title} {Quantum spin liquids: a
  review},\ }\href {https://doi.org/10.1088/0034-4885/80/1/016502} {\bibfield
  {journal} {\bibinfo  {journal} {Reports on Progress in Physics}\ }\textbf
  {\bibinfo {volume} {80}},\ \bibinfo {pages} {016502} (\bibinfo {year}
  {2016})}\BibitemShut {NoStop}%
\bibitem [{\citenamefont {Chen}(2017)}]{CHEN20173}%
  \BibitemOpen
  \bibfield  {author} {\bibinfo {author} {\bibfnamefont {X.}~\bibnamefont
  {Chen}},\ }\bibfield  {title} {\bibinfo {title} {Symmetry fractionalization
  in two dimensional topological phases},\ }\href
  {https://doi.org/https://doi.org/10.1016/j.revip.2017.02.002} {\bibfield
  {journal} {\bibinfo  {journal} {Reviews in Physics}\ }\textbf {\bibinfo
  {volume} {2}},\ \bibinfo {pages} {3} (\bibinfo {year} {2017})}\BibitemShut
  {NoStop}%
\bibitem [{\citenamefont {Thomson}\ and\ \citenamefont
  {Sachdev}(2018)}]{PhysRevX.8.011012}%
  \BibitemOpen
  \bibfield  {author} {\bibinfo {author} {\bibfnamefont {A.}~\bibnamefont
  {Thomson}}\ and\ \bibinfo {author} {\bibfnamefont {S.}~\bibnamefont
  {Sachdev}},\ }\bibfield  {title} {\bibinfo {title} {Fermionic spinon theory
  of square lattice spin liquids near the n\'eel state},\ }\href
  {https://doi.org/10.1103/PhysRevX.8.011012} {\bibfield  {journal} {\bibinfo
  {journal} {Phys. Rev. X}\ }\textbf {\bibinfo {volume} {8}},\ \bibinfo {pages}
  {011012} (\bibinfo {year} {2018})}\BibitemShut {NoStop}%
\bibitem [{\citenamefont {Zhang}\ and\ \citenamefont
  {Sachdev}(2020)}]{PhysRevResearch.2.023172}%
  \BibitemOpen
  \bibfield  {author} {\bibinfo {author} {\bibfnamefont {Y.-H.}\ \bibnamefont
  {Zhang}}\ and\ \bibinfo {author} {\bibfnamefont {S.}~\bibnamefont
  {Sachdev}},\ }\bibfield  {title} {\bibinfo {title} {From the pseudogap metal
  to the fermi liquid using ancilla qubits},\ }\href
  {https://doi.org/10.1103/PhysRevResearch.2.023172} {\bibfield  {journal}
  {\bibinfo  {journal} {Phys. Rev. Research}\ }\textbf {\bibinfo {volume}
  {2}},\ \bibinfo {pages} {023172} (\bibinfo {year} {2020})}\BibitemShut
  {NoStop}%
\bibitem [{Note1()}]{Note1}%
  \BibitemOpen
  \bibinfo {note} {In one dimension, the JW transformation proceeds in both
  ways, namely, one could also femrionize a bosonic system. In higher
  dimensions, certain bosonic problems may also admit a femrionic description,
  but this is not generic. Our focus here is the bosonization of generic
  femrionic systems.}\BibitemShut {Stop}%
\bibitem [{\citenamefont {Corboz}\ \emph
  {et~al.}(2010{\natexlab{a}})\citenamefont {Corboz}, \citenamefont {Evenbly},
  \citenamefont {Verstraete},\ and\ \citenamefont
  {Vidal}}]{PhysRevA.81.010303}%
  \BibitemOpen
  \bibfield  {author} {\bibinfo {author} {\bibfnamefont {P.}~\bibnamefont
  {Corboz}}, \bibinfo {author} {\bibfnamefont {G.}~\bibnamefont {Evenbly}},
  \bibinfo {author} {\bibfnamefont {F.}~\bibnamefont {Verstraete}},\ and\
  \bibinfo {author} {\bibfnamefont {G.}~\bibnamefont {Vidal}},\ }\bibfield
  {title} {\bibinfo {title} {Simulation of interacting fermions with
  entanglement renormalization},\ }\href
  {https://doi.org/10.1103/PhysRevA.81.010303} {\bibfield  {journal} {\bibinfo
  {journal} {Phys. Rev. A}\ }\textbf {\bibinfo {volume} {81}},\ \bibinfo
  {pages} {010303} (\bibinfo {year} {2010}{\natexlab{a}})}\BibitemShut
  {NoStop}%
\bibitem [{\citenamefont {Corboz}\ and\ \citenamefont
  {Vidal}(2009)}]{PhysRevB.80.165129}%
  \BibitemOpen
  \bibfield  {author} {\bibinfo {author} {\bibfnamefont {P.}~\bibnamefont
  {Corboz}}\ and\ \bibinfo {author} {\bibfnamefont {G.}~\bibnamefont {Vidal}},\
  }\bibfield  {title} {\bibinfo {title} {Fermionic multiscale entanglement
  renormalization ansatz},\ }\href {https://doi.org/10.1103/PhysRevB.80.165129}
  {\bibfield  {journal} {\bibinfo  {journal} {Phys. Rev. B}\ }\textbf {\bibinfo
  {volume} {80}},\ \bibinfo {pages} {165129} (\bibinfo {year}
  {2009})}\BibitemShut {NoStop}%
\bibitem [{\citenamefont {Corboz}\ \emph
  {et~al.}(2010{\natexlab{b}})\citenamefont {Corboz}, \citenamefont {Or\'us},
  \citenamefont {Bauer},\ and\ \citenamefont {Vidal}}]{PhysRevB.81.165104}%
  \BibitemOpen
  \bibfield  {author} {\bibinfo {author} {\bibfnamefont {P.}~\bibnamefont
  {Corboz}}, \bibinfo {author} {\bibfnamefont {R.}~\bibnamefont {Or\'us}},
  \bibinfo {author} {\bibfnamefont {B.}~\bibnamefont {Bauer}},\ and\ \bibinfo
  {author} {\bibfnamefont {G.}~\bibnamefont {Vidal}},\ }\bibfield  {title}
  {\bibinfo {title} {Simulation of strongly correlated fermions in two spatial
  dimensions with fermionic projected entangled-pair states},\ }\href
  {https://doi.org/10.1103/PhysRevB.81.165104} {\bibfield  {journal} {\bibinfo
  {journal} {Phys. Rev. B}\ }\textbf {\bibinfo {volume} {81}},\ \bibinfo
  {pages} {165104} (\bibinfo {year} {2010}{\natexlab{b}})}\BibitemShut
  {NoStop}%
\bibitem [{\citenamefont {Shukla}\ \emph {et~al.}(2020)\citenamefont {Shukla},
  \citenamefont {Ellison},\ and\ \citenamefont
  {Fidkowski}}]{PhysRevB.101.155105}%
  \BibitemOpen
  \bibfield  {author} {\bibinfo {author} {\bibfnamefont {S.~K.}\ \bibnamefont
  {Shukla}}, \bibinfo {author} {\bibfnamefont {T.~D.}\ \bibnamefont
  {Ellison}},\ and\ \bibinfo {author} {\bibfnamefont {L.}~\bibnamefont
  {Fidkowski}},\ }\bibfield  {title} {\bibinfo {title} {Tensor network approach
  to two-dimensional bosonization},\ }\href
  {https://doi.org/10.1103/PhysRevB.101.155105} {\bibfield  {journal} {\bibinfo
   {journal} {Phys. Rev. B}\ }\textbf {\bibinfo {volume} {101}},\ \bibinfo
  {pages} {155105} (\bibinfo {year} {2020})}\BibitemShut {NoStop}%
\bibitem [{\citenamefont {Schollw\"ock}(2005)}]{RevModPhys.77.259}%
  \BibitemOpen
  \bibfield  {author} {\bibinfo {author} {\bibfnamefont {U.}~\bibnamefont
  {Schollw\"ock}},\ }\bibfield  {title} {\bibinfo {title} {The density-matrix
  renormalization group},\ }\href {https://doi.org/10.1103/RevModPhys.77.259}
  {\bibfield  {journal} {\bibinfo  {journal} {Rev. Mod. Phys.}\ }\textbf
  {\bibinfo {volume} {77}},\ \bibinfo {pages} {259} (\bibinfo {year}
  {2005})}\BibitemShut {NoStop}%
\bibitem [{\citenamefont {Fishman}\ \emph {et~al.}(2020)\citenamefont
  {Fishman}, \citenamefont {White},\ and\ \citenamefont
  {Stoudenmire}}]{itensor}%
  \BibitemOpen
  \bibfield  {author} {\bibinfo {author} {\bibfnamefont {M.}~\bibnamefont
  {Fishman}}, \bibinfo {author} {\bibfnamefont {S.~R.}\ \bibnamefont {White}},\
  and\ \bibinfo {author} {\bibfnamefont {E.~M.}\ \bibnamefont {Stoudenmire}},\
  }\href@noop {} {\bibinfo {title} {The \mbox{ITensor} software library for
  tensor network calculations}} (\bibinfo {year} {2020}),\ \Eprint
  {https://arxiv.org/abs/2007.14822} {arXiv:2007.14822} \BibitemShut {NoStop}%
\bibitem [{\citenamefont {Kaul}\ \emph {et~al.}(2008)\citenamefont {Kaul},
  \citenamefont {Kim}, \citenamefont {Sachdev},\ and\ \citenamefont
  {Senthil}}]{Ribhu2008}%
  \BibitemOpen
  \bibfield  {author} {\bibinfo {author} {\bibfnamefont {R.~K.}\ \bibnamefont
  {Kaul}}, \bibinfo {author} {\bibfnamefont {Y.~B.}\ \bibnamefont {Kim}},
  \bibinfo {author} {\bibfnamefont {S.}~\bibnamefont {Sachdev}},\ and\ \bibinfo
  {author} {\bibfnamefont {T.}~\bibnamefont {Senthil}},\ }\bibfield  {title}
  {\bibinfo {title} {Algebraic charge liquids},\ }\href
  {https://doi.org/10.1038/nphys790} {\bibfield  {journal} {\bibinfo  {journal}
  {Nature Physics}\ }\textbf {\bibinfo {volume} {4}},\ \bibinfo {pages} {28}
  (\bibinfo {year} {2008})}\BibitemShut {NoStop}%
\bibitem [{\citenamefont {Son}(2015)}]{PhysRevX.5.031027}%
  \BibitemOpen
  \bibfield  {author} {\bibinfo {author} {\bibfnamefont {D.~T.}\ \bibnamefont
  {Son}},\ }\bibfield  {title} {\bibinfo {title} {Is the composite fermion a
  dirac particle?},\ }\href {https://doi.org/10.1103/PhysRevX.5.031027}
  {\bibfield  {journal} {\bibinfo  {journal} {Phys. Rev. X}\ }\textbf {\bibinfo
  {volume} {5}},\ \bibinfo {pages} {031027} (\bibinfo {year}
  {2015})}\BibitemShut {NoStop}%
\bibitem [{\citenamefont {Kitaev}(2006)}]{KITAEV20062}%
  \BibitemOpen
  \bibfield  {author} {\bibinfo {author} {\bibfnamefont {A.}~\bibnamefont
  {Kitaev}},\ }\bibfield  {title} {\bibinfo {title} {Anyons in an exactly
  solved model and beyond},\ }\href
  {https://doi.org/https://doi.org/10.1016/j.aop.2005.10.005} {\bibfield
  {journal} {\bibinfo  {journal} {Annals of Physics}\ }\textbf {\bibinfo
  {volume} {321}},\ \bibinfo {pages} {2} (\bibinfo {year} {2006})},\ \bibinfo
  {note} {january Special Issue}\BibitemShut {NoStop}%
\bibitem [{Note2()}]{Note2}%
  \BibitemOpen
  \bibinfo {note} {We implicitly assumed we write the Hamiltonian as an
  expansion over a unitary operator basis. For instance, while complex fermion
  hopping terms like $\protect \hat f^\dagger \protect \hat f' + {\protect \rm
  h.c.}$ is not unitary in general, we may expand the term using Majorana
  bilinears which are unitary.}\BibitemShut {Stop}%
\bibitem [{\citenamefont {Affleck}\ \emph {et~al.}(1988)\citenamefont
  {Affleck}, \citenamefont {Zou}, \citenamefont {Hsu},\ and\ \citenamefont
  {Anderson}}]{PhysRevB.38.745}%
  \BibitemOpen
  \bibfield  {author} {\bibinfo {author} {\bibfnamefont {I.}~\bibnamefont
  {Affleck}}, \bibinfo {author} {\bibfnamefont {Z.}~\bibnamefont {Zou}},
  \bibinfo {author} {\bibfnamefont {T.}~\bibnamefont {Hsu}},\ and\ \bibinfo
  {author} {\bibfnamefont {P.~W.}\ \bibnamefont {Anderson}},\ }\bibfield
  {title} {\bibinfo {title} {Su(2) gauge symmetry of the large-$u$ limit of the
  hubbard model},\ }\href {https://doi.org/10.1103/PhysRevB.38.745} {\bibfield
  {journal} {\bibinfo  {journal} {Phys. Rev. B}\ }\textbf {\bibinfo {volume}
  {38}},\ \bibinfo {pages} {745} (\bibinfo {year} {1988})}\BibitemShut
  {NoStop}%
\bibitem [{\citenamefont {Dagotto}\ \emph {et~al.}(1988)\citenamefont
  {Dagotto}, \citenamefont {Fradkin},\ and\ \citenamefont
  {Moreo}}]{PhysRevB.38.2926}%
  \BibitemOpen
  \bibfield  {author} {\bibinfo {author} {\bibfnamefont {E.}~\bibnamefont
  {Dagotto}}, \bibinfo {author} {\bibfnamefont {E.}~\bibnamefont {Fradkin}},\
  and\ \bibinfo {author} {\bibfnamefont {A.}~\bibnamefont {Moreo}},\ }\bibfield
   {title} {\bibinfo {title} {Su(2) gauge invariance and order parameters in
  strongly coupled electronic systems},\ }\href
  {https://doi.org/10.1103/PhysRevB.38.2926} {\bibfield  {journal} {\bibinfo
  {journal} {Phys. Rev. B}\ }\textbf {\bibinfo {volume} {38}},\ \bibinfo
  {pages} {2926} (\bibinfo {year} {1988})}\BibitemShut {NoStop}%
\bibitem [{Note3()}]{Note3}%
  \BibitemOpen
  \bibinfo {note} {This may appear to be trivially guaranteed; but we emphasize
  that $\protect \hat S^i$ represents a term in the Hamiltonian, whereas
  $\protect \hat \sigma ^i$ is interpreted as a symmetry generator. The fact
  that the two agree in the $\Gamma $-odd subspace is a special property of the
  current problem with $4$ complex partons (Appendix \ref
  {app:RoP}).}\BibitemShut {Stop}%
\bibitem [{Note4()}]{Note4}%
  \BibitemOpen
  \bibinfo {note} {As a cautionary remark, for reasons detailed in Sec.\ \ref
  {sec:global} it is not generally possible to make such an identification for
  all arrows on the lattice. The more accurate statement here is as follows: to
  each arrow we attach an operator, and the algebra of the operators is the
  same as that obtained by replacing each arrow by $i \protect \hat \gamma
  ^\alpha _{\protect \mathbf {r}} \protect \hat \gamma ^\beta _{\protect
  \mathbf {r}'}$.}\BibitemShut {Stop}%
\bibitem [{Note5()}]{Note5}%
  \BibitemOpen
  \bibinfo {note} {One could alternatively assert that if we reverse an arrow
  connecting sites $\protect \mathbf {r}$ and $\protect \mathbf {r}'$, then we
  attach the term $\protect \hat \Gamma _{\protect \mathbf {r}}$ to the
  operator. Once we restrict to the $\Gamma _-$-subspace of all the sites the
  two definitions become equivalent.}\BibitemShut {Stop}%
\bibitem [{Note6()}]{Note6}%
  \BibitemOpen
  \bibinfo {note} {We impose this mostly for simplicity; one could also relax
  our construction and treat the sign in the operator identification also as a
  variable, and then solve the constraints.}\BibitemShut {Stop}%
\bibitem [{\citenamefont {Yang}(1989)}]{PhysRevLett.63.2144}%
  \BibitemOpen
  \bibfield  {author} {\bibinfo {author} {\bibfnamefont {C.~N.}\ \bibnamefont
  {Yang}},\ }\bibfield  {title} {\bibinfo {title} {\ensuremath{\eta} pairing
  and off-diagonal long-range order in a hubbard model},\ }\href
  {https://doi.org/10.1103/PhysRevLett.63.2144} {\bibfield  {journal} {\bibinfo
   {journal} {Phys. Rev. Lett.}\ }\textbf {\bibinfo {volume} {63}},\ \bibinfo
  {pages} {2144} (\bibinfo {year} {1989})}\BibitemShut {NoStop}%
\bibitem [{Note7()}]{Note7}%
  \BibitemOpen
  \bibinfo {note} {For simplicity, we have omitted here the possibility that
  the global fermion parity $\protect \hat P$ appears on the right-hand side,
  which would be the case if the chosen straightened arrow passes through the
  ``decorated link'' we fixed. In any case, since $\protect \hat P$ is
  invariant under conjugation by $\protect \hat U(A)$ our discussion is
  unaffected even if it is present.}\BibitemShut {Stop}%
\bibitem [{Note8()}]{Note8}%
  \BibitemOpen
  \bibinfo {note} {In fact, we can claim something slightly stronger: with the
  prescription in Sec.\ \ref {sec:global} we have the identification $\protect
  \hat \gamma _{\protect \mathbf {0}}^1 \protect \dot \DOTSB \leftarrow
  \protect \joinrel \rightarrow \protect \hat \Lambda ^{11}_{\protect \mathbf
  {0}}$, and $\Lambda ^{11}_{\protect \mathbf {0}}$ also transforms as it
  should under our definition of $\protect \hat \Xi $.}\BibitemShut {Stop}%
\bibitem [{Note9()}]{Note9}%
  \BibitemOpen
  \bibinfo {note} {One can also generate $T_y$ using $C_4$ and $T_x$, but we
  keep it here as a sanity check.}\BibitemShut {Stop}%
\bibitem [{Note10()}]{Note10}%
  \BibitemOpen
  \bibinfo {note} {In general, a spatial rotation symmetry should also act on
  the spin index. In our notation, one can construct such an operator by
  combing the ``spinless'' $C_4$ together with the corresponding internal
  unitary action $e^{- i \protect \frac {\pi }{4} \protect \hat \sigma ^3 }$.
  Alternatively, in a system with full ${\protect \rm SU}(2)$ spin rotation
  symmetry the ``spinless'' rotation considered here can also be a symmetry of
  the system.}\BibitemShut {Stop}%
\bibitem [{Note11()}]{Note11}%
  \BibitemOpen
  \bibinfo {note} {In case this may appear out of the blue, we comment that the
  two parts are actually generators for different ${\protect \rm SU}(2)$
  factors when we view the unitary as an element of ${\protect \rm Spin}(4)$,
  c.f. the corresponding definitions for $\protect \hat \theta $ in Eq.\
  \protect \textup {\hbox {\mathsurround \z@ \protect \normalfont
  (\ignorespaces \ref {eq:Spin_trans}\unskip \@@italiccorr )}}.}\BibitemShut
  {Stop}%
\bibitem [{\citenamefont {Wen}(2003)}]{PhysRevLett.90.016803}%
  \BibitemOpen
  \bibfield  {author} {\bibinfo {author} {\bibfnamefont {X.-G.}\ \bibnamefont
  {Wen}},\ }\bibfield  {title} {\bibinfo {title} {Quantum orders in an exact
  soluble model},\ }\href {https://doi.org/10.1103/PhysRevLett.90.016803}
  {\bibfield  {journal} {\bibinfo  {journal} {Phys. Rev. Lett.}\ }\textbf
  {\bibinfo {volume} {90}},\ \bibinfo {pages} {016803} (\bibinfo {year}
  {2003})}\BibitemShut {NoStop}%
\bibitem [{Note12()}]{Note12}%
  \BibitemOpen
  \bibinfo {note} {Not to be confused with $\protect \hat \Gamma _\pm $: in the
  physical Hilbert space we always have $\protect \hat \Gamma \protect \dot
  \DOTSB \mapstochar \rightarrow -1$, but here $\protect \hat P$ denotes the
  on-site fermion parity, which is dynamical and can take both $\pm 1$
  values.}\BibitemShut {Stop}%
\bibitem [{\citenamefont {Kitaev}(2003)}]{KITAEV20032}%
  \BibitemOpen
  \bibfield  {author} {\bibinfo {author} {\bibfnamefont {A.}~\bibnamefont
  {Kitaev}},\ }\bibfield  {title} {\bibinfo {title} {Fault-tolerant quantum
  computation by anyons},\ }\href
  {https://doi.org/https://doi.org/10.1016/S0003-4916(02)00018-0} {\bibfield
  {journal} {\bibinfo  {journal} {Annals of Physics}\ }\textbf {\bibinfo
  {volume} {303}},\ \bibinfo {pages} {2} (\bibinfo {year} {2003})}\BibitemShut
  {NoStop}%
\bibitem [{\citenamefont {Fu}\ \emph {et~al.}(2007)\citenamefont {Fu},
  \citenamefont {Kane},\ and\ \citenamefont {Mele}}]{PhysRevLett.98.106803}%
  \BibitemOpen
  \bibfield  {author} {\bibinfo {author} {\bibfnamefont {L.}~\bibnamefont
  {Fu}}, \bibinfo {author} {\bibfnamefont {C.~L.}\ \bibnamefont {Kane}},\ and\
  \bibinfo {author} {\bibfnamefont {E.~J.}\ \bibnamefont {Mele}},\ }\bibfield
  {title} {\bibinfo {title} {Topological insulators in three dimensions},\
  }\href {https://doi.org/10.1103/PhysRevLett.98.106803} {\bibfield  {journal}
  {\bibinfo  {journal} {Phys. Rev. Lett.}\ }\textbf {\bibinfo {volume} {98}},\
  \bibinfo {pages} {106803} (\bibinfo {year} {2007})}\BibitemShut {NoStop}%
\bibitem [{\citenamefont {Tantivasadakarn}(2020)}]{PhysRevResearch.2.023353}%
  \BibitemOpen
  \bibfield  {author} {\bibinfo {author} {\bibfnamefont {N.}~\bibnamefont
  {Tantivasadakarn}},\ }\bibfield  {title} {\bibinfo {title} {Jordan-wigner
  dualities for translation-invariant hamiltonians in any dimension: Emergent
  fermions in fracton topological order},\ }\href
  {https://doi.org/10.1103/PhysRevResearch.2.023353} {\bibfield  {journal}
  {\bibinfo  {journal} {Phys. Rev. Research}\ }\textbf {\bibinfo {volume}
  {2}},\ \bibinfo {pages} {023353} (\bibinfo {year} {2020})}\BibitemShut
  {NoStop}%
\bibitem [{\citenamefont {{Shirley}}(2020)}]{2020arXiv200212026S}%
  \BibitemOpen
  \bibfield  {author} {\bibinfo {author} {\bibfnamefont {W.}~\bibnamefont
  {{Shirley}}},\ }\bibfield  {title} {\bibinfo {title} {{Fractonic order and
  emergent fermionic gauge theory}},\ }\href@noop {} {\bibfield  {journal}
  {\bibinfo  {journal} {arXiv e-prints}\ ,\ \bibinfo {eid} {arXiv:2002.12026}}
  (\bibinfo {year} {2020})},\ \Eprint {https://arxiv.org/abs/2002.12026}
  {arXiv:2002.12026} \BibitemShut {NoStop}%
\bibitem [{\citenamefont {Pozo}\ \emph {et~al.}(2021)\citenamefont {Pozo},
  \citenamefont {Rao}, \citenamefont {Chen},\ and\ \citenamefont
  {Sodemann}}]{PhysRevB.103.035145}%
  \BibitemOpen
  \bibfield  {author} {\bibinfo {author} {\bibfnamefont {O.}~\bibnamefont
  {Pozo}}, \bibinfo {author} {\bibfnamefont {P.}~\bibnamefont {Rao}}, \bibinfo
  {author} {\bibfnamefont {C.}~\bibnamefont {Chen}},\ and\ \bibinfo {author}
  {\bibfnamefont {I.}~\bibnamefont {Sodemann}},\ }\bibfield  {title} {\bibinfo
  {title} {Anatomy of ${\mathbb{z}}_{2}$ fluxes in anyon fermi liquids and bose
  condensates},\ }\href {https://doi.org/10.1103/PhysRevB.103.035145}
  {\bibfield  {journal} {\bibinfo  {journal} {Phys. Rev. B}\ }\textbf {\bibinfo
  {volume} {103}},\ \bibinfo {pages} {035145} (\bibinfo {year}
  {2021})}\BibitemShut {NoStop}%
\bibitem [{\citenamefont {Rao}\ and\ \citenamefont
  {Sodemann}(2021)}]{PhysRevResearch.3.023120}%
  \BibitemOpen
  \bibfield  {author} {\bibinfo {author} {\bibfnamefont {P.}~\bibnamefont
  {Rao}}\ and\ \bibinfo {author} {\bibfnamefont {I.}~\bibnamefont {Sodemann}},\
  }\bibfield  {title} {\bibinfo {title} {Theory of weak symmetry breaking of
  translations in ${\mathbb{z}}_{2}$ topologically ordered states and its
  relation to topological superconductivity from an exact lattice
  ${\mathbb{z}}_{2}$ charge-flux attachment},\ }\href
  {https://doi.org/10.1103/PhysRevResearch.3.023120} {\bibfield  {journal}
  {\bibinfo  {journal} {Phys. Rev. Research}\ }\textbf {\bibinfo {volume}
  {3}},\ \bibinfo {pages} {023120} (\bibinfo {year} {2021})}\BibitemShut
  {NoStop}%
\bibitem [{\citenamefont {Li}\ and\ \citenamefont {Po}(2021)}]{Kangle}%
  \BibitemOpen
  \bibfield  {author} {\bibinfo {author} {\bibfnamefont {K.}~\bibnamefont
  {Li}}\ and\ \bibinfo {author} {\bibfnamefont {H.~C.}\ \bibnamefont {Po}},\
  }\href@noop {} {\bibinfo {title} {Higher-dimensional jordan-wigner
  transformation and auxiliary majorana fermions}} (\bibinfo {year} {2021}),\
  \Eprint {https://arxiv.org/abs/2107.14083} {arXiv:2107.14083} \BibitemShut
  {NoStop}%
\bibitem [{\citenamefont {Bochniak}\ \emph {et~al.}(2021)\citenamefont
  {Bochniak}, \citenamefont {Ruba},\ and\ \citenamefont
  {Wosiek}}]{bochniak2021bosonization}%
  \BibitemOpen
  \bibfield  {author} {\bibinfo {author} {\bibfnamefont {A.}~\bibnamefont
  {Bochniak}}, \bibinfo {author} {\bibfnamefont {B.}~\bibnamefont {Ruba}},\
  and\ \bibinfo {author} {\bibfnamefont {J.}~\bibnamefont {Wosiek}},\
  }\href@noop {} {\bibinfo {title} {Bosonization of majorana modes and edge
  states}} (\bibinfo {year} {2021}),\ \Eprint
  {https://arxiv.org/abs/2107.06335} {arXiv:2107.06335} \BibitemShut {NoStop}%
\end{thebibliography}%
\clearpage

\end{document}